\documentclass[superscriptaddress, showpacs, nofootinbib, amsmath, amssymb, aps, prd, notitlepage]{revtex4-1}
\usepackage{amsmath,amssymb,amsfonts,mathrsfs,bbold}
\usepackage{yfonts}

\usepackage[dvipsnames]{xcolor}

\usepackage{float}
\usepackage{graphics,graphicx}
\usepackage{dcolumn}
\usepackage{bm}
\usepackage{hyperref}
\usepackage{comment}
\usepackage{xcolor}
\usepackage{slashed}
\usepackage{color}
\usepackage{simplewick}
\usepackage{caption}
\usepackage{subcaption}
\usepackage{braket}

\usepackage[export]{adjustbox}

\usepackage{ragged2e} 
\DeclareCaptionJustification{justified}{\justifying}

\usepackage{multirow}
\usepackage{mathbbol}
\usepackage{mathtools}

\usepackage{slashed}
\usepackage{simpler-wick}

\usepackage[normalem]{ulem}

\usepackage{tikz}
\usetikzlibrary{decorations.pathmorphing, decorations.markings, shapes.geometric, shapes.misc, calc, math}
\tikzstyle{gluon}=[decorate, decoration={coil,aspect=0.8, amplitude=1.5pt,  segment length=3pt}]

\usepackage{stackrel}
\usepackage[most]{tcolorbox}

\allowdisplaybreaks[1]

\newcommand{\ben}{\begin{eqnarray*}}
\newcommand{\een}{\end{eqnarray*}}

\newcommand{\xx}{{\underline{x}}}
\newcommand{\yy}{\underline{y}}
\newcommand{\zz}{\underline{z}}
\newcommand{\pp}{\underline{p}}
\newcommand{\qq}{\underline{q}}
\newcommand{\bb}{\underline{b}}
\newcommand{\rr}{\underline{r}}
\newcommand{\ee}{\underline{\varepsilon}}

\newcommand{\kk}{\underline{k}}

\newcommand{\wg}{{\widetilde G}}
\newcommand{\dd}{\mathrm{d}}

\newcommand{\as}{\alpha_\mathrm{s}}

\makeatletter
\DeclareRobustCommand{\cev}[1]{%
  {\mathpalette\do@cev{#1}}%
}
\newcommand{\do@cev}[2]{%
  \vbox{\offinterlineskip
    \sbox\z@{$\m@th#1 x$}%
    \ialign{##\cr
      \hidewidth\reflectbox{$\m@th#1\vec{}\mkern4mu$}\hidewidth\cr
      \noalign{\kern-\ht\z@}
      $\m@th#1#2$\cr
    }%
  }%
}
\makeatother

\begin{document}

\title{Polarized Dipole Scattering Amplitudes meet the Valence Quark Model}

\author{Adrian Dumitru} 
         \email[Email: ]{adrian.dumitru23@login.cuny.edu}
         \affiliation{Department of Natural Sciences, Baruch College, CUNY, 17 Lexington Avenue, New York, NY 10010, USA}
         \affiliation{The Graduate School and University Center, The City University of New York, 365 Fifth Avenue, New York, NY 10016, USA}

\author{Heikki M\"antysaari} 
         \email[Email: ]{heikki.mantysaari@jyu.fi}
         \affiliation{Department of Physics, University of Jyv\"askyl\"a,  P.O. Box 35, 40014 University of Jyv\"askyl\"a, Finland}
         \affiliation{Helsinki Institute of Physics, P.O. Box 64, 00014 University of Helsinki, Finland}

\author{Yossathorn Tawabutr}
         \email[Email: ]{yossathorn.j.tawabutr@jyu.fi}
         \affiliation{Department of Physics, University of Jyv\"askyl\"a,  P.O. Box 35, 40014 University of Jyv\"askyl\"a, Finland}
         \affiliation{Helsinki Institute of Physics, P.O. Box 64, 00014 University of Helsinki, Finland}

\date{\today}

\begin{abstract}
The recently revised small-$x$ helicity evolution~\cite{Cougoulic:2022gbk}, resumming the double-logarithmic factor, $\as\ln^2(1/x)$, allows for the study of helicity distributions of quarks and gluons at small Bjorken $x$, corresponding to high center-of-mass energy. In this work, we calculate the moderate-$x$ initial conditions in the regime, $\as \ln^2(1/x)\sim 1$, for the small-$x$ helicity evolution using a light-front valence quark model of the proton, which provides additional physical information about the target. The perturbative emission and absorption of a gluon by the valence quarks are also included. The results, given in Eqs.~\eqref{IC_summary_final}, provide a new set of initial conditions with a significantly reduced number of free parameters than conventional models~\cite{Adamiak:2023yhz}. Consequently, the predictive power of small-$x$ helicity evolution is expected to improve once the initial conditions from this work are incorporated.
\end{abstract}

\maketitle
\tableofcontents


\section{Introduction}

Proton spin puzzle is a longstanding unsolved problem in particle physics~\cite{Ji:2020ena}, concerning the amount of proton spin that comes from the quarks and gluons inside. Based on the Jaffe-Manohar decomposition~\cite{Jaffe:1989jz}, each parton could contribute to the proton spin via its spin or orbital angular momentum, amounting to the sum rule,
\begin{align}
    \frac{1}{2} = S_q+S_G+L_q+L_G\, ,
\end{align}
where $S_q$ ($S_G$) refers to the spin and $L_q$ ($L_G$) the orbital angular momenta of the quarks (gluons) inside the spin-$\frac{1}{2}$ proton. In the helicity basis, the spin contributions can be written in terms of the integrals over Bjorken $x$ of helicity-dependent parton distribution functions (hPDFs),
\begin{subequations}\label{spin_hPDF}
\begin{align}
    S_q(Q^2) &= \frac{1}{2}\int\limits_0^1\dd x\,\Delta\Sigma(x,Q^2) \, , \\
    S_G(Q^2) &= \int\limits_0^1\dd x\,\Delta G(x,Q^2) \, ,
\end{align}
\end{subequations}
where $\Delta\Sigma$ is the flavor singlet quark hPDF, that is, with the flavors of quark and antiquark summed over, and $\Delta G$ is the gluon hPDF. Recent measurements and analyses by RHIC spin program show that at resolution $Q^2=10$ GeV$^2$ we have $S_q \in [0.15,0.20]$ for $x\geq 10^{-3}$ and $S_G\in [0.13,0.26]$ for $x\geq 0.05$~\cite{Aschenauer:2013woa,Aschenauer:2015eha}. The reason for a lower bound on $x$ in the integrals, c.f. Eqs.~\eqref{spin_hPDF}, is due to the high center-of-mass energy collisions required to probe partons at small Bjorken $x$. With the numbers shown above, there remains a missing contribution to proton helicity that could come from the orbital angular momenta or the small-$x$ regions of quark and gluon hPDFs. This work focuses on quantifying the latter.

In the past decade, there have been several theoretical developments in order to understand spins at small Bjorken $x$~\cite{Cougoulic:2022gbk,Kovchegov:2015pbl,Kovchegov:2016weo,Kovchegov:2016zex,Kovchegov:2017jxc,Kovchegov:2017lsr,Kovchegov:2018znm,Cougoulic:2019aja,Kovchegov:2020hgb,Cougoulic:2020tbc,Kovchegov:2021lvz,Adamiak:2023okq,Chirilli:2018kkw,Chirilli:2021lif}, particularly surrounding the objective of deriving a high-energy evolution that would allow one to write down hPDFs at small $x$ based on their values at moderate $x$, which could be determined from experimental measurements. The framework modifies the dipole formalism~\cite{Mueller:1989st,Nikolaev:1990ja} to account for helicity dependence in the scattering processes, resulting in energy-suppressed, i.e. sub-eikonal, contributions to the dipole amplitude~\cite{Kovchegov:2015pbl,Kovchegov:2018znm,Cougoulic:2022gbk}. At the end, with the inclusion of kinematic constraint~\cite{Beuf:2014uia,Iancu:2015joa,Iancu:2015vea,Ducloue:2019ezk}, the small-$x$ evolution resums double logarithms of energy, that is, each step of evolution yields the factor, $\as\ln^2(1/x)$~\cite{Kovchegov:2015pbl}. The evolution equation does not close in general, but becomes a closed system of linear integral equation upon taking the large-$N_c$~\cite{tHooft:1973alw} or the large-$N_c\&N_f$~\cite{Veneziano:1976wm} limit with the mean-field approximation. The discussion in this work focuses on the latter limit, as it is more realistic in the helicity evolution where the quark exchange becomes important relative to the gluon counterpart~\cite{Cougoulic:2022gbk,Adamiak:2023okq}. 

Recently, a global analysis has been performed in Ref.~\cite{Adamiak:2023yhz} using a generalized Born-level amplitude as initial condition at moderate $x$, which contains 24 free parameters to be fitted to the experimental data. The analysis employs the Jefferson Laboratory Angular Momentum (JAM) Monte Carlo Bayesian framework and includes all available polarized deep-inelastic scatter (DIS) and polarized semi-inclusive DIS (SIDIS) measurements at $0.005\leq x \leq 0.1$ and 1.69 GeV$^2 \leq Q^2 \leq 10.4~\text{GeV}^2$ with proton, deuteron and helium-3 targets. In total, with 226 data points, the initial condition model with the small-$x$ helicity evolution is able to describe the data excellently, with $\chi^2$ of 1.03 per data point~\cite{Adamiak:2023yhz}. Unfortunately, the analysis has a shortcoming when it comes to the prediction power. Due to the limited amount of available polarized scattering measurements at small $x$, the free parameters in the initial condition model, and hence also the physical predictions, are not sufficiently well-constrained. In particular, the predicted amount of total parton spin from the region $10^{-5}\leq x\leq 1$ comes out to be $S_q+S_G = -0.64\pm 0.60$. Although the uncertainty will reduce dramatically~\cite{Adamiak:2023yhz} upon the inclusion of measurements from the future Electron-Ion Collider (EIC)~\cite{Accardi:2012qut,AbdulKhalek:2021gbh,Abir:2023fpo}, several improvements are possible on the theoretical side to equip the small-$x$ helicity evolution with a higher prediction power given the currently available data.

In this work, we model the proton target at moderate $x$ as a bound state of three valence quarks, based on which the correlators of polarized Wilson line operators are evaluated~\cite{Dumitru:2018vpr,Dumitru:2020fdh}. With the help of the Yang-Mills equation, the gluon fields are related to the color charges or currents and subsequently expressed in terms of creation and annihilation operators, which are later applied to the three-valence-quark state. We also include perturbative corrections involving gluon emission and absorption both within the target and connecting the target and the projectile, in a similar fashion to~\cite{Dumitru:2020gla,Dumitru:2021tvw,Dumitru:2021tqp,Dumitru:2022ooz,Dumitru:2023sjd}. As a result, we obtain the leading-order moderate-$x$ expressions that serve as initial conditions for the polarized dipole amplitudes in the small-$x$ helicity evolution~\cite{Cougoulic:2022gbk}. Summarized in Eqs.~\eqref{IC_summary_final}, our results contain a linear power of transverse logarithm of the dipole size that is completely determined, together with the constant term accounting for the infrared (large-dipole) physics that we leave as a free parameter. Overall, the physical proton model employed in this work is capable of reducing the number of free parameters in the initial conditions for helicity evolution from the original 24 in Ref.~\cite{Adamiak:2023yhz} to 3--9 depending on the flexibility of the model used in the global analysis. 

In the past, there have been similar model calculations for gluon hPDF, including~\cite{Chen:2006ng,Pasquini:2008ax}, both of which employ three-valence-quark models for the polarized target. The latter work~\cite{Pasquini:2008ax} also incorporates the Melosh rotation, which takes into account the fact that the helicity of a particle with nonzero transverse momentum does not exactly correspond to the spin along the light-cone direction. In contrast, the current work is the first to employ a valence quark model for the polarized dipole operator relevant to the small-$x$ helicity evolution~\cite{Cougoulic:2022gbk}. Although relations exist between the polarized dipole degrees of freedom and the parton hPDFs, these relations are derived with the assumption that $x$ is small. Applying such relations to the polarized dipole amplitude obtained in this work requires caution, as our calculation is based on a model valid at moderate $x$ only.

Throughout this article, we use the light-cone coordinates in which $v^{\mu}=(v^+,v^-,\underline{v})$ with $v^{\pm}=(v^0\pm v^3)/\sqrt{2}$. Transverse vectors are denoted by $\underline{v}=(v^1,v^2)$ with $v_{\perp} = |\underline{v}|$. The light-cone directions are chosen such that the projectile has a large light-cone minus momentum, $k^-$, while the target proton has  a large light-cone plus momentum, $P^+$. The paper begins with Section~\ref{sec:polarized}, which provides a quick introduction to the formalism employed in the small-$x$ helicity evolution~\cite{Cougoulic:2022gbk}. Subsequently, Section~\ref{sec:main} specifies the valence quark model we employ for the proton target, explains the main calculation steps, then proceeds to summarize and discuss the main results, with detailed derivations provided in the appendices. Finally, we conclude in Section~\ref{sec:conclusion} and discuss future projects that can be built up on the results and development in this work.

\section{Polarized Scattering Processes at High Energy}
\label{sec:polarized}

Before we discuss helicity-dependent scattering processes, it is helpful to review the dipole formalism for unpolarized scattering processes, c.f. \cite{Gelis:2010nm,Gelis:2012ri,Kovchegov:2012mbw} for more complete reviews. Eikonal propagation of a high-energy quark through the target color field at transverse coordinate $\xx$  is described by Wilson line $V_\xx$ defined as
\begin{equation}\label{V_eik}
    V_\xx = \mathcal{P} \exp\left[ ig \int \dd x^- A^+(0^+,x^-,\xx) \right].
\end{equation}
Similarly, the antiquark propagation is described by $V^\dagger_\xx$. These objects serve as building blocks within the dipole picture \cite{Mueller:1989st,Nikolaev:1990ja} of the deep-inelastic scattering at small $x$. In particular, the unpolarized parton distribution functions (PDF), transverse-momentum-dependent (TMD) PDF's and structure functions can be written in terms of the \emph{dipole amplitude,}
\begin{align}\label{unpol_dip_amp}
    N(\rr,\bb,Y) = 1-\frac{1}{N_c} \left\langle\text{tr}\left[V_{\bb-\rr/2}V^{\dagger}_{\bb+\rr/2}\right] \right\rangle_{\text{unpol}}.
\end{align}
Here, $\left\langle\cdots\right\rangle_{\text{unpol}}$ gives the forward light-front matrix element defined formally as 
\begin{align}\label{brackets_unpol}
    &\left\langle \hat{\mathcal{O}} \right\rangle_{\text{unpol}} = \frac{1}{2}\sum_{\mathcal{S}_L} \frac{\bra{P^+,\underline{P},\mathcal{S}_L} \hat{\mathcal{O}} \ket{P^+,\underline{P},\mathcal{S}_L}}{\braket{P^+,\underline{P},\mathcal{S}_L|P^+,\underline{P},\mathcal{S}_L}} \, ,
\end{align}
where $\mathcal{S}_L$ is the proton's longitudinal spin and $P^\mu$ is the proton's momentum. Note that here, and throughout this work, we are ultimately interested in quantities integrated over the impact parameter $\bb$, and as such only the forward matrix element is necessary. The transverse size of the dipole is denoted by $\rr$. This setup allows one to employ the Wilson line as the main degree of freedom to study unpolarized scattering at small $x$.

\subsection{Polarized Wilson lines}
\label{sect:pol_Wil_line}

In order to study the helicity of quarks and gluons inside the proton, the latter has to be probed through a helicity-dependent scattering process. This requires the inclusion of helicity-dependent interactions between a high-energy quark and the target. In~\cite{Kovchegov:2015pbl,Kovchegov:2018znm,Cougoulic:2022gbk}, such interactions are shown to come in as sub-eikonal corrections to the eikonal Wilson line, $V_{\xx}$. It is convenient to group these corrections into three types: (i) \emph{quark exchange}, (ii) \emph{type-1 gluon exchange} and (iii) \emph{type-2 gluon exchange.} In particular, consider an incoming quark at transverse position $\yy$ and helicity $\sigma$ that interacts with the target and results in an outgoing quark at transverse position $\xx$ and helicity $\sigma'$. Its interaction with the target shockwave at $x^-\to 0$ can be written as
\begin{align}\label{V_subeik_gen}
    &V_{\xx,\yy;\,\sigma',\sigma} = V_{\xx}\,\delta^2(\xx-\yy) \, \delta_{\sigma'\sigma} + \left[V_{\xx}^{\text{q}[1]} + V_{\xx}^{\text{G}[1]}\right]\,\delta^2(\xx-\yy) \, \sigma\delta_{\sigma'\sigma} + V_{\xx,\yy}^{\text{G}[2]}\, \delta_{\sigma'\sigma} + \text{etc} \, ,
\end{align}
where ``etc'' contains other terms that do not contribute to helicity-dependent scattering amplitudes. In the right-hand side of Eq.~\eqref{V_subeik_gen}, the first term corresponds to the eikonal Wilson line in Eq.~\eqref{V_eik}, while the next three objects correspond respectively to the three types of sub-eikonal corrections listed above that are relevant to helicity. Explicitly, they are given by 
\begin{subequations}\label{V_subeik}
\begin{align}
    &V_{\xx}^{\text{q}[1]} = \frac{g^2P^+}{2zs}\int\limits_{-\infty}^{\infty}\dd x_1^-\int\limits_{x_1^-}^{\infty}\dd x_2^- \, V_{\xx}[\infty,x_2^-] \, t^b \psi^f_{\beta}(x_2^-,\xx) \, U^{ba}_{\xx}[x_2^-,x_1^-] \left(\gamma^+\gamma_5\right)_{\alpha\beta} \bar{\psi}^f_{\alpha}(x_1^-,\xx) \, t^a V_{\xx}[x_1^-,-\infty]      \, , \label{V_subeik_q_1} \\
    &V_{\xx}^{\text{G}[1]} = \frac{igP^+}{zs} \int\limits_{-\infty}^{\infty}\dd x^- \, V_{\xx}[\infty,x^-]\,F^{12}(x^-,\xx)\,V_{\xx}[x^-,-\infty]     \, , \label{V_subeik_G_1} \\
    &V_{\xx,\yy}^{\text{G}[2]} = -\frac{iP^+}{zs}\int\limits_{-\infty}^{\infty}\dd z^-\int \dd^2 \zz \; V_{\xx}[\infty,z^-] \, \delta^2(\xx-\zz)\,\cev{D}^i(z^-,\zz)\,\vec{D}^i(z^-,\zz)\,\delta^2(\zz-\yy) \, V_{\yy}[z^-,-\infty]      \, , \label{V_subeik_G_2}
\end{align}
\end{subequations}
where $zs$ is the center-of-mass energy of the collision between the polarized (anti)quark and the proton. Here, the covariant derivatives are defined such that $\vec{D}^{\mu} = \vec{\partial}^{\mu} - igA^{\mu}$ and $\cev{D}^{\mu} = \cev{\partial}^{\mu} + igA^{\mu}$. The signs of the covariant terms are such that the gauge transformation property is consistent with that of the Wilson line~\eqref{V_eik}. In Eqs.~\eqref{V_subeik}, the notation for partial Wilson lines is such that
\begin{align}\label{V_eikonal}
    &V_{\xx}[a^-,b^-] = \mathcal{P} \exp\left[ ig \int\limits_{b^-}^{a^-} \dd x^- A^+(0^+,x^-,\xx) \right],
\end{align}
and similarly for the adjoint representation, $U_{\xx}[a^-,b^-]$.

\subsection{Parton Helicity TMDs}

Polarized Wilson lines introduced in Section~\ref{sect:pol_Wil_line} arise naturally in the calculation of the quark helicity TMD, $g_{1L}^{\text{q}}(x,k_{\perp})$. Upon propagating the (anti)quark from one end of the gauge link to the other under the shockwave picture, one obtains~\cite{Kovchegov:2018znm,Cougoulic:2022gbk}
\begin{align}\label{quark_TMD}
    &g_{1L}^{\text{q}}(x,k_{\perp}) = - \frac{4iP^+}{(2\pi)^5}  \int \dd^2\xx\,\dd^2\yy \int \dd k_1^- \\
    &\;\;\;\;\times \left\{ e^{-i\kk\cdot(\xx-\yy)} \frac{(\xx-\yy)}{|\xx-\yy|^2}\cdot\frac{\kk}{k^2_{\perp}} \left\langle \text{T\,tr}\left[V_{\xx}V_{\yy}^{\text{q}[1]\dagger}\right] + \bar{\text{T}}\,\text{tr}\left[V_{\xx}^{\text{q}[1]}V_{\yy}^{\dagger}\right] + \text{T\,tr}\left[V_{\xx}V_{\yy}^{\text{G}[1]\dagger}\right] + \bar{\text{T}}\,\text{tr}\left[V_{\xx}^{\text{G}[1]}V_{\yy}^{\dagger}\right] \right\rangle   \right. \notag \\
    &\;\;\;\;\;\;\;\;+ \left. i\,\frac{(\xx-\yy)}{|\xx-\yy|^2}\times\frac{\kk}{k^2_{\perp}}  \int\dd^2\zz  \left\langle e^{i\kk\cdot(\zz-\xx)}\,\text{T\,tr}\left[V_{\xx}V_{\zz,\yy}^{\text{G}[2]\dagger}\right] + e^{-i\kk\cdot(\zz-\xx)} \,\bar{\text{T}}\,\text{tr}\left[V_{\zz,\yy}^{\text{G}[2]}V_{\xx}^{\dagger}\right]   \right\rangle  \right\} , \notag
\end{align}
where T and $\bar{\text{T}}$ refer to (anti)time ordering of the traces that depend on whether the Wilson lines arise from the amplitude or the complex-conjugate amplitude in the propagation of the quark field. In Eq.~\eqref{quark_TMD}, the angle brackets denote the \emph{helicity-dependent CGC averaging} defined as
\begin{align}\label{exp_heli}
    &\left\langle\hat{\mathcal{O}}\right\rangle = \frac{1}{2}\sum_{\mathcal{S}_L}\mathcal{S}_L\frac{\bra{P^+,\underline{P},\mathcal{S}_L} \hat{\mathcal{O}} \ket{P^+,\underline{P},\mathcal{S}_L}}{\braket{P^+,\underline{P},\mathcal{S}_L|P^+,\underline{P},\mathcal{S}_L}}\,,
\end{align}
which will be used throughout this work. Notice that this averaging differs from the unpolarized counterpart in Eq.~\eqref{brackets_unpol} in the extra factor of $\mathcal{S}_L$ within the summation. Finally, $k_1^-$ refers to the momentum of the polarized (anti)quark in the dipole. Once we sum over the quark and antiquark flavors, we arrive at the flavor singlet quark helicity TMD, while the difference between the quark and the antiquark yields the ``flavor non-singlet'' counterparts \cite{Kovchegov:2016zex}. Explicitly,
\begin{subequations}\label{quark_TMD_singlet_nonsinglet}
\begin{align}
    g^{\text{S}}_{1L}(x,k_{\perp}) &= \sum_f\left[g^{\text{q}}_{1L}(x,k_{\perp}) + g^{\bar{\text{q}}}_{1L}(x,k_{\perp})\right], \label{quark_TMD_singlet} \\
    g^{\text{NS}}_{1L}(x,k_{\perp}) &= g^{\text{q}}_{1L}(x,k_{\perp}) - g^{\bar{\text{q}}}_{1L}(x,k_{\perp}) \, . \label{quark_TMD_nonsinglet}
\end{align}
\end{subequations}
Subsequently, we integrate Eq.~\eqref{quark_TMD_singlet} over transverse momentum, $k_{\perp}$, to obtain the flavor-singlet quark helicity PDF, $\Delta\Sigma(x,Q^2)$, which is most conveniently written as~\cite{Cougoulic:2022gbk}
\begin{align}\label{quark_singlet_PDF}
    \Delta\Sigma(x,Q^2) &= \int\limits^{Q^2}\dd^2 \kk \; g^{\text{S}}_{1L}(x,k_{\perp}) = -\frac{N_c}{2\pi^3}\sum_f\int\limits_{\Lambda^2/s}^1\frac{\dd z}{z} \int\limits_{1/zs}^{\min\{1/zQ^2,1/\Lambda^2\}}\frac{\dd x^2_{10}}{x^2_{10}} \left[Q_f(r_{\perp},zs) + 2G_2(r_{\perp},zs)\right] ,   
\end{align}
where $z$ is the minus momentum fraction of the polarized quark within the dipole. Here, we have imposed a kinematic constraint, $x\ll 1/zsx^2_{10}$, and employed $\Lambda$ as the infrared cutoff. Furthermore, we defined the \emph{type-1 polarized dipole amplitude} as 
\begin{align}\label{def_Q}
    &Q_f(r_{\perp},zs) = \int \dd^2\left(\frac{\xx+\yy}{2}\right) Q_f(\xx,\yy,zs)\Big|_{r_\perp = |\yy-\xx|} \\
    &= \frac{zs}{2N_c}\int \dd^2\left(\frac{\xx+\yy}{2}\right) \text{Re}\left\langle\text{T\,tr}\left[V_{\xx}V_{\yy}^{\text{q}[1]\dagger}\right] + \text{T\,tr}\left[V_{\xx}V_{\yy}^{\text{G}[1]\dagger}\right] + \text{T\,tr}\left[V_{\yy}^{\text{q}[1]}V_{\xx}^{\dagger}\right] + \text{T\,tr}\left[V_{\yy}^{\text{G}[1]}V_{\xx}^{\dagger}\right] \right\rangle \Big|_{r_\perp = |\yy-\xx|} \, , \notag
\end{align}
where the flavor, $f$, corresponds physically to the flavors of the quark and antiquark in the Wilson line traces. In deriving Eq.~\eqref{quark_singlet_PDF}, integration by-part has been performed on the covariant derivatives in the type-2 Wilson line, $V_{\zz,\yy}^{\text{G}[2]}$, from Eq.~\eqref{quark_TMD}, eventually leading to a convenient expression in term of the \emph{type-2 polarized dipole amplitude,}
\begin{align}\label{def_G2}
    G_2(r_{\perp},zs) &= \frac{\epsilon^{ij}\rr^j}{r^2_{\perp}}\int \dd^2\left(\frac{\xx+\yy}{2}\right) G^{i}(\xx,\yy,zs)\Big|_{r_\perp = |\yy-\xx|} \\
    &= \frac{zs}{2N_c}\, \frac{\epsilon^{ij}\rr^j}{r^2_{\perp}} \int \dd^2\left(\frac{\xx+\yy}{2}\right)  \left\langle\text{tr}\left[V_{\xx}^{\dagger}V_{\yy}^{i\text{G}[2]}\right] + \text{tr}\left[V_{\yy}^{i\text{G}[2]\dagger}V_{\xx} \right] \right\rangle \Big|_{r_\perp = |\yy-\xx|} \, , \notag
\end{align}
where
\begin{align}\label{ViG2}
    &V_{\xx}^{i\text{\,G}[2]}=\frac{P^+}{2s}\int\limits_{-\infty}^{\infty}\dd x^-\,V_{\xx}[\infty,x^-] \left[\vec{D}^i(x^-,\xx) - \cev{D}^i(x^-,\xx)\right] V_{\xx}[x^-,-\infty] \, .
\end{align}
In most cases, it is more convenient to express the contribution originated from the type-2 polarized Wilson line in terms of $G_2(r_{\perp},zs)$ or $V_{\xx}^{i\text{\,G}[2]}$. 

The flavor non-singlet quark helicity TMD and PDF are important for polarized semi-inclusive deep-inelastic scattering (SIDIS)~\cite{Kovchegov:2016zex,Adamiak:2023yhz}. Integrating Eq.~\eqref{quark_TMD_nonsinglet} over $k_{\perp}$ yields
\begin{align}\label{quark_nonsinglet_PDF}
    &\Delta q^-_f(x,Q^2) = \int^{Q^2}\dd^2\kk\; g_{1L}^{\text{NS}}(x,k_{\perp}) = \frac{N_c}{2\pi^3}\int\limits_{\Lambda^2/s}^1\frac{\dd z}{z} \int\limits_{1/zs}^{\min\{1/zQ^2,1/\Lambda^2\}}\frac{\dd x^2_{10}}{x^2_{10}} \,Q_f^{\text{NS}}(r_{\perp},zs) \, ,
\end{align}
where
\begin{align}\label{QfNS}
    &Q_f^{\text{NS}}(r_{\perp},zs) = \int \dd^2\left(\frac{\xx+\yy}{2}\right) Q_f^{\text{NS}}(\xx,\yy,zs)\Big|_{r_\perp = |\yy-\xx|} \\
    &= \frac{zs}{2N_c}\int \dd^2\left(\frac{\xx+\yy}{2}\right) \text{Re}\left\langle\text{T\,tr}\left[V_{\xx}V_{\yy}^{\text{q}[1]\dagger}\right] + \text{T\,tr}\left[V_{\xx}V_{\yy}^{\text{G}[1]\dagger}\right] - \text{T\,tr}\left[V_{\yy}^{\text{q}[1]}V_{\xx}^{\dagger}\right] - \text{T\,tr}\left[V_{\yy}^{\text{G}[1]}V_{\xx}^{\dagger}\right] \right\rangle \Big|_{r_\perp = |\yy-\xx|} \, , \notag
\end{align}
where the flavor, $f$, corresponds again to the flavor of the Wilson lines.

Next, we consider the gluon helicity TMD. The type-2 gluon exchange arises naturally in term of $V_{\xx}^{i\text{\,G}[2]}$ from the sub-eikonal term in the expansion of dipole gluon helicity TMD, $g^{\text{G\,dip}}_{1L}(x,k_{\perp})$, via the Lipatov vertex~\cite{Kovchegov:2017lsr,Cougoulic:2022gbk}. Ultimately, the calculation gives
\begin{align}\label{gluon_TMD}
    &g_{1L}^{\text{G\,dip}}(x,k_{\perp}) = \frac{4iN_c}{\alpha_s(2\pi)^4}\int \dd^2\rr\;e^{-i\kk\cdot\rr} \, (\kk\cdot\rr) \, G_2\left(r_{\perp},zs=\frac{Q^2}{x}\right) ,
\end{align}
which integrates over $\kk$ to the gluon helicity PDF,
\begin{align}\label{gluon_PDF}
    &\Delta G(x,Q^2) = \frac{2N_c}{\alpha_s\pi^2} \left[\left(1+r^2_{\perp}\frac{\partial}{\partial r^2_{\perp}} \right) G_2\left(r_{\perp},zs=\frac{Q^2}{x}\right) \right]\Big|_{r^2_{\perp}=1/Q^2} \, .
\end{align}
From Eqs.~\eqref{quark_singlet_PDF}, \eqref{gluon_TMD} and \eqref{gluon_PDF}, we see that both quark and gluon helicity distributions at small $x$ can be expressed in terms of the sub-eikonal polarized dipole amplitudes, which trace over a product of an unpolarized and a polarized Wilson line.

\subsection{Small-$x$ Helicity Evolution}

In Refs.~\cite{Kovchegov:2015pbl,Kovchegov:2017lsr,Kovchegov:2018znm,Cougoulic:2022gbk}, the small-$x$ evolution equation has been derived for the flavor singlet polarized dipole amplitudes. In the mean-field Veneziano's limit of large $N_c$ and $N_f$~\cite{Veneziano:1976wm}, where we take $N_c\gg 1$ while keeping the ratio $N_f/N_c$ finite, the evolution equation reduces to a system of linear integral equations involving $Q_f(r_{\perp},zs)$, $G_2(r_{\perp},zs)$ and $\wg(r_{\perp},zs)$, with the last object being the adjoint representation of $Q(r_{\perp},zs)$. It describes the type-1 quark and gluon exchanges by a gluon dipole at large $N_c$ and $N_f$. Its explicit form is
\begin{align}\label{G_tilde}
    &\wg(r_{\perp},zs) = \int \dd^2\left(\frac{\xx+\yy}{2}\right) \wg(\xx,\yy,zs)\Big|_{r_\perp = |\yy-\xx|} \\
    &= \frac{zs}{2N_c} \sum_f \int \dd^2\left(\frac{\xx+\yy}{2}\right) \text{Re}\left\langle\text{T\,tr}\left[V_{\xx}W_{\yy}^{\text{q}[1]\dagger}\right] + \text{T\,tr}\left[W_{\yy}^{\text{q}[1]}V_{\xx}^{\dagger}\right] + \text{T\,tr}\left[V_{\xx}V_{\yy}^{\text{G}[1]\dagger}\right] + \text{T\,tr}\left[V_{\yy}^{\text{G}[1]}V_{\xx}^{\dagger}\right] \right\rangle \Big|_{r_\perp = |\yy-\xx|} \, , \notag
\end{align}
where 
\begin{align}\label{W_q1}
    &W_{\yy}^{\text{q}[1]} = \frac{g^2P^+}{8s} \int\limits_{-\infty}^{\infty}\dd x_1^-\int\limits_{x_1^-}^{\infty}\dd x_2^- \, V_{\yy}[\infty,x_2^-] \, \psi^f_{\beta}(x_2^-,\yy) \left[\gamma^+\gamma_5\right]_{\alpha\beta}\bar{\psi}^f_{\alpha}(x_1^-,\yy) \, V_{\yy}[x_1^-,-\infty]  
\end{align}
describes the quark exchange by a gluon line at large $N_c$ and $N_f$. As for the flavor non-singlet sector, the evolution equation in the mean-field large-$N_c\& N_f$ limit~\cite{Veneziano:1976wm} is an integral equation of $Q_f^{\text{NS}}(r_{\perp},zs)$, separately for each flavor, $f$~\cite{Kovchegov:2016zex}.

For both flavor singlet and non-singlet sectors, the small-$x$ helicity evolution resums the double-logarithmic factor of $\alpha_s\ln^2(1/x)$, which is in contrast to the single-logarithmic factor, $\alpha_s\ln(1/x)$, resummed by the unpolarized small-$x$ evolution. As a result, the helicity evolution becomes essential to study polarized scattering physics starting from $x_0\simeq 0.1$ and below, as this is roughly the point where the resummation parameter, $\alpha_s\ln^2(1/x)\sim 1$. Furthermore, the starting point of the evolution at $x=0.1$ is supported through global analyses~\cite{Adamiak:2021ppq,Adamiak:2023yhz} that the evolution equation is capable of describing all polarized DIS and SIDIS data at $x\leq 0.1$. Altogether, this justifies the main purpose of this work: to calculate the initial condition of the helicity evolution at $x\sim 0.1$ using the valence-quark picture, $\ket{qqq}$, of the proton target. As discussed below, we will also include corrections from gluon emissions so that the final results contain all the terms up to $\mathcal{O}(\alpha_s^2)$ accuracy.

Finally, it is important to note that in practice our calculation requires $N_c$ and $N_f$ to be taken explicitly to 3. On the surface, this seems to contradict the large-$N_c\& N_f$ limit imposed above in obtaining the helicity evolution equation. However, the corrections to these limits would come in at order $\mathcal{O}(1/N_c^2)$ or higher~\cite{Kovchegov:2012mbw,Kovchegov:2015pbl,Lappi:2020srm}. This is already smaller than typical values of the strong coupling constant, $\as$, which we also assume to be sufficiently small in order for our perturbative calculation to be valid.


\section{Initial conditions for helicity evolution from a light-front quark model}
\label{sec:main}

\subsection{The Proton Model}
\label{sec:proton_model}

In our moderate-$x$ calculation, we adopt the light-front valence quark model of Refs.~\cite{Schlumpf:1992vq,Brodsky:1994fz} which has been applied to unpolarized small-$x$ dipole - proton 
scattering~\cite{Dumitru:2018vpr,Dumitru:2020fdh,Dumitru:2020gla,Dumitru:2021tqp,Dumitru:2021tvw,Dumitru:2022ooz,Dumitru:2023sjd}. There, the proton state with (predominantly plus) momentum $P^{\mu}\simeq(P^+,0^-,\underline{P})$ and helicity $\mathcal{S}_L$ is written as
\begin{align}\label{proton}
\ket{P^+,\underline{P},\mathcal{S}_L} &= \frac{1}{\sqrt{6}} \int\frac{\dd x_1  \dd x_2  \dd x_3}{(4\pi)^3\sqrt{x_1x_2x_3}}\,4\pi\delta(1-x_1-x_2-x_3) \int\frac{\dd^2\qq_1  \dd^2\qq_2  \dd^2\qq_3}{(2\pi)^6}\,(2\pi)^2\delta^2(\qq_1+\qq_2+\qq_3) \\
&\;\;\;\;\times \sum_{\{f_1,f_2,f_3\}=\{u,u,d\}} \sum_{\sigma_1,\sigma_2,\sigma_3} \psi_{\mathcal{S}_L}(x_1,\qq_1,\sigma_1,f_1;x_2,\qq_2,\sigma_2,f_2;x_3,\qq_3,\sigma_3,f_3) \notag \\
&\;\;\;\;\times \sum_{i_1,i_2,i_3}\epsilon_{i_1i_2i_3}
\ket{q(p_1,i_1,\sigma_1,f_1)}\otimes\ket{q(p_2,i_2,\sigma_2,f_2)}\otimes \ket{q(p_3,i_3,\sigma_3,f_3)} , \notag
\end{align}
where  $p_i^+ = x_i P^+$ and $\pp_i = x_i\underline{P} + \qq_i$ denote, respectively, the longitudinal and transverse
momenta of the quarks. In particular, $\qq_i$ is the transverse momentum of the $i$-th quark relative to the proton.
The light-cone wave function, $\psi_{\mathcal{S}_L}$, is invariant under both longitudinal and transverse boosts, and hence it depends only on the $x_i$ and $\qq_i$.
It is also symmetric under exchange of any pair of quarks. The indices, $i_1\cdots i_3$ and $\sigma_1\cdots\sigma_3$, denote the
colors and spins of the valence quarks, respectively. Note that the proton state as written in Eq.~\eqref{proton} assumes $N_c=3$ colors. Hence, throughout this work, upon plugging in the explicit proton state and/or wave function, $N_c=3$ should be automatically assumed.

As written in Eq.~\eqref{proton}, the wave function is normalized such that
\begin{equation}\label{psi_norm}
\int[\dd x_i] \int[\dd^2 \qq_i]\sum_{\{f_1,f_2,f_3\}=\{u,u,d\}} \sum_{\sigma_1,\sigma_2,\sigma_3} \left|
\psi_{\mathcal{S}_L}(x_1,\qq_1,\sigma_1,f_1;x_2,\qq_2,\sigma_2,f_2;x_3,\qq_3,\sigma_3,f_3) \right|^2 = 1\,, 
\end{equation}
where for convenience we have defined the integration measures,
\begin{subequations}
\begin{align}
    &[\dd x_i] = \frac{\dd x_1 \dd x_2 \dd x_3}{(4\pi)^3}\,4\pi\delta(1-x_1-x_2-x_3) \, , \\
    &[\dd^2 \qq_i] = \frac{\dd^2 \qq_1\dd^2 \qq_2\dd^2 \qq_3}{(2\pi)^6} \, (2\pi)^2\delta^2(\qq_1+\qq_2+\qq_3) \, .
\end{align}
\end{subequations}
This is consistent with the following normalization convention,
\begin{align}\label{norm_quark}
    &\braket{q(p',i',\sigma',f') | q(p,i,\sigma,f)} = \delta_{i'i}\delta_{\sigma'\sigma}\delta^{f',f}\,2p^+\,2\pi \delta(p^+-p'^+) \, (2\pi)^2\delta^2(\pp-\pp') \, ,
\end{align}
for the quark state, together with the convention,
\begin{align}\label{denom}
&\braket{K^+,\underline{K},\mathcal{S}_L | P^+,\underline{P},\mathcal{S}_L} = 2P^+\,2\pi\delta(P^+-K^+)\,(2\pi)^2\delta^2(\underline{P}-\underline{K}) \, ,
\end{align}
for the proton state.

Throughout the paper we assume a factorized wave function of the form
\begin{equation}  \label{eq:factorized-LCwf}
    \psi_{\mathcal{S}_L}(x_1,\qq_1,\sigma_1,f_1;x_2,\qq_2,\sigma_2,f_2;x_3,\qq_3,\sigma_3,f_3) = \Phi(x_i,\qq_i) \, S_{\mathcal{S}_L}(f_i, \sigma_i)\,,
\end{equation}
where $\Phi(x_i,\qq_i) = \Phi(x_1,\qq_1;x_2,\qq_2;x_3,\qq_3)$ is a momentum-space wave function and 
$S_{\mathcal{S}_L}(f_i, \sigma_i) = S_{\mathcal{S}_L}(f_1, \sigma_1;f_2, \sigma_2;f_3, \sigma_3)$ is the normalized spin-flavor wave function of the non-relativistic
quark model,
\begin{align}\label{qk11} 
    S_{\mathcal{S}_L}(f_i, \sigma_i) &=  \frac{1}{\sqrt{18}}\left\{ \left[ 2\delta_{\sigma_1,\mathcal{S}_L}\delta_{\sigma_2,\mathcal{S}_L}\delta_{\sigma_3,-\mathcal{S}_L} - \delta_{\sigma_1,\mathcal{S}_L}\delta_{\sigma_2,-\mathcal{S}_L}\delta_{\sigma_3,\mathcal{S}_L} - \delta_{\sigma_1,-\mathcal{S}_L}\delta_{\sigma_2,\mathcal{S}_L}\delta_{\sigma_3,\mathcal{S}_L} \right] \delta^{f_1,u} \delta^{f_2,u} \delta^{f_3,d}  \right.   \\
    &\;\;\;\;\;\;+ \left[  2\delta_{\sigma_1,\mathcal{S}_L}\delta_{\sigma_2,-\mathcal{S}_L}\delta_{\sigma_3,\mathcal{S}_L} - \delta_{\sigma_1,\mathcal{S}_L}\delta_{\sigma_2,\mathcal{S}_L}\delta_{\sigma_3,-\mathcal{S}_L} - \delta_{\sigma_1,-\mathcal{S}_L}\delta_{\sigma_2,\mathcal{S}_L}\delta_{\sigma_3,\mathcal{S}_L} \right] \delta^{f_1,u} \delta^{f_2,d} \delta^{f_3,u}   \notag \\
    &\;\;\;\;\;\;+ \left. \left[  2\delta_{\sigma_1,-\mathcal{S}_L}\delta_{\sigma_2,\mathcal{S}_L}\delta_{\sigma_3,\mathcal{S}_L} - \delta_{\sigma_1,\mathcal{S}_L}\delta_{\sigma_2,-\mathcal{S}_L}\delta_{\sigma_3,\mathcal{S}_L} - \delta_{\sigma_1,\mathcal{S}_L}\delta_{\sigma_2,\mathcal{S}_L}\delta_{\sigma_3,-\mathcal{S}_L} \right] \delta^{f_1,d} \delta^{f_2,u} \delta^{f_3,u}   \right\} . \notag
\end{align}
By using this spin wave function we are neglecting the Melosh rotation of the non-relativistic spins
to light-cone helicities, c.f.~\cite{Schlumpf:1992vq,Brodsky:1994fz,Pasquini:2010af}. However, as will be shown below, this simplification enables us to obtain simple analytic expressions for the polarized dipole amplitudes at moderate $x$. 
We intend to investigate
quantitative corrections due to the Melosh rotation in a future work.

The momentum-space wave function is taken to be the ``harmonic oscillator'' model by Schlumpf~\cite{Schlumpf:1992vq,Brodsky:1994fz},
\begin{align}\label{Phi}
    &\Phi(x_i,\qq_i) = \mathcal{N}\,\exp\left[-\frac{1}{2\beta^2}\sum_{i=1}^3\frac{q^2_{i\perp}+M^2}{x_i}\right] .
\end{align}
This is a minimal model for the effective valence quark light-cone wave function, in that fluctuations of the squared invariant mass of the three quarks, i.e. the sum of their light-cone energies, about the extremal point are suppressed exponentially. The width of the Gaussian exponential is controlled by the parameter $\beta$ which is of order $N_c$ times the QCD confinement scale. Specifically, the proton electromagnetic form factor require $\beta=0.55$ GeV and $M=0.26$ GeV in order to achieve the realistic proton charge radius of about 0.75 fm~\cite{Schlumpf:1992vq}. Without loss of generality, we take $\mathcal{N}$ to be real and positive in this work. This form of the wave function allows for relatively simple way to analytically evaluate the integrals over $\qq_i$'s, as will be clear below.

\subsection{Polarized Dipole Amplitudes at Moderate $x$}
\label{sec:main_results}

We calculate the expectation values of the dipole operators defined in Eqs.~\eqref{def_Q}, \eqref{def_G2}, \eqref{QfNS} and \eqref{G_tilde} from Section~\ref{sec:polarized} using the proton model specified in Section~\ref{sec:proton_model}. Except for the type-2 dipole amplitude, $G_2(r_{\perp},zs)$, from Eq.~\eqref{def_G2} that only contains the gluon-exchange term, the three remaining polarized dipole amplitudes -- $Q_f(r_{\perp},zs)$, $Q_f^{\text{NS}}(r_{\perp},zs)$ and $\wg(r_{\perp},zs)$ -- contain quark- and gluon-exchange contributions, each of which is more conveniently evaluated separately. Each of the two contributions involves the respective sub-eikonal operator, c.f. Eqs.~\eqref{V_subeik_q_1}, \eqref{V_subeik_G_1}, \eqref{ViG2} and \eqref{W_q1}, sandwiched between eikonal Wilson lines. Hence, the dominant contribution that preserves helicity dependence follows from taking all the eikonal Wilson lines to identity while keeping the sub-eikonal operator. 

For the quark-exchange term, taking all the unpolarized Wilson lines to identity yields a contribution of order $\mathcal{O}(\as)$. In Appendix~\ref{sec:O-g2-quark-Xchange}, the expansion is performed in detail. An important step involves writing the quark fields in the sub-eikonal quark-exchange operator in terms of quark creation and annihilation operators, c.f. Eq.~\eqref{qkfield}. The latter operators then act straightforwardly to the proton state~\eqref{proton}. With the valence quark model employed in this work, we see that the expression vanishes, and we need to look into the $\mathcal{O}(\as^2)$ corrections to find the leading nonzero quark-exchange contribution. Considered in Appendix~\ref{sec:O-g4-quark-Xchange}, the corrections include (i) a gluon connecting the projectile dipole with the target, corresponding to expanding one of the eikonal Wilson lines~\eqref{V_eikonal} to order $\mathcal{O}(g)$ and evaluating the resulting gluon field, $A^+$, within the target proton state, together with (ii) a gluon emitted and absorbed within the target, which corresponds to expanding the valence quark state~\eqref{proton} to include an extra perturbative gluon, c.f. Eq.~\eqref{r4}. As discussed in Appendix~\ref{sec:O-g4-quark-Xchange}, only the first type of corrections contributes perturbatively to the quark-exchange term at $\mathcal{O}(\as^2)$, while the remaining terms do not depend on the dipole size, $r_{\perp}$. The latter raises a potential question regarding the perturbative nature of the calculation, as the insensitivity to $r_{\perp}$ requires the strong coupling constant to run with an alternative scale inherent to the problem.\footnote{In Appendix~\ref{sec:r-indep}, we briefly discuss these $r_{\perp}$-independent terms and present the results that must be taken with caution as elaborated in the text.} To evaluate the healthy $r_{\perp}$-dependent contributions, we employ the Poisson equation~\eqref{qk21} to relate the eikonal gluon field, $A^+$, to the light-cone color charge density~\cite{Dumitru:2018vpr}, which can subsequently be written in terms of the quark fields and eventually the quark ladder operators. As a result, the dominant quark-exchange contribution for each of the type-1 dipole amplitudes reads
\begin{subequations}\label{qk_xchng_interm}
\begin{align}
    &Q_f^{\text{q}}(r_{\perp},zs) = Q_f^{\text{q},\text{NS}}(r_{\perp},zs) = \frac{2\alpha_s^2 \pi^2 P^+}{9} \, \text{Im} \left\{ (t^a)_{ji} \, (t^a)_{\ell m} \int\dd^2\left(\frac{\xx+\yy}{2}\right) \int\limits_{-\infty}^{\infty}\dd x_1^-\int\limits_{x_1^-}^{\infty}\dd x_2^-  \int\limits_{-\infty}^{\infty}\dd x_3^- \int\frac{\dd^2 \pp}{(2\pi)^2} \, \frac{1}{p^2_{\perp}} \right.   \label{qk_xchng_interm_Q}  \\ 
    &\;\;\;\;\times \int \dd^2 \xx' \int\frac{\dd^2 \pp_1 \, \dd p_1^+}{(2\pi)^3\sqrt{2p_1^+}} \int\frac{\dd^2 \pp_2 \, \dd p_2^+}{(2\pi)^3\sqrt{2p_2^+}} \int\frac{\dd^2 \pp_3 \, \dd p_3^+}{(2\pi)^3\sqrt{2p_3^+}} \int\frac{\dd^2 \pp_4 \, \dd p_4^+}{(2\pi)^3\sqrt{2p_4^+}} \, e^{-ip_1^+x_1^- + ip_2^+x_2^- - i(p_3^+ - p_4^+)\,x_3^-}   \notag \\ 
    &\;\;\;\;\times \left. e^{i\pp\cdot\xx + i(\pp_1-\pp_2)\cdot\yy - i(\pp-\pp_3+\pp_4)\cdot\xx'} \, \sum_{f'} \sum_{S,S'} S \left\langle \hat{b}^{f'\dagger}_{p_4,\ell,S'} \hat{b}^{f'}_{p_3,m,S'} \hat{b}^f_{p_1,i,S} \hat{b}^{f\dagger}_{p_2,j,S}\right\rangle \right\}\Bigg|_{r_\perp = |\yy-\xx|} + (r_{\perp}\text{- independent terms})  \, ,  \notag \\
    &\wg^{\text{q}}(r_{\perp},zs) = - \frac{3}{2}\sum_f Q_f^{\text{q}}(r_{\perp},zs)  + (r_{\perp}\text{- independent terms}) \, , \label{qk_xchng_interm_GT}
\end{align}
\end{subequations}
c.f. Eq.~\eqref{qk26}. From the expressions, we evaluate the ladder operators via the valence quark state~\eqref{proton}. At the end, we integrate over the impact parameter and other kinematic variables to obtain the final results~\eqref{qk28} and \eqref{qk30} at the end of Appendix~\ref{sec:O-g4-quark-Xchange}.

Next, the gluon-exchange term is worked out in Appendix~\ref{sec:O-g4-gluon-Xchange}. There, the leading contribution that results from putting all eikonal Wilson lines to identity is shown to be nonzero and of order $\mathcal{O}(\as^2)$. As can be seen from Eqs.~\eqref{V_subeik_G_1} and \eqref{ViG2}, the sub-eikonal gluon-exchange operators involve transverse gluon field, $A^i$, that must be evaluated in the valence quark state. To do so, we employ the Poisson equation~\eqref{gl2} relating $A^i$ to transverse color current, which can then be written in terms of the quark fields and subsequently the quark ladder operators, c.f. Eq.~\eqref{gl3}. This results in the gluon-exchange contributions given by
\begin{subequations}\label{gl_xchng_interm}
\begin{align}
    &Q_f^{\text{G}}(r_{\perp},zs) = \frac{1}{3}\,\wg^{\text{G}}(r_{\perp},zs) = \frac{2\as^2\pi^2P^+}{3} \, \text{Re} \left\{ (t^a)_{\ell m} \, (t^a)_{pq} \int\dd^2\left(\frac{\xx+\yy}{2}\right) \int\frac{\dd^2 \pp}{(2\pi)^2} \int\frac{\dd^2 \pp'}{(2\pi)^2} \, \frac{\pp'^i}{p_{\perp}^2p'^2_{\perp}} \, e^{i(\pp+\pp')\cdot\yy}      \right.\label{gl_xchng_interm_Q} \\
    &\;\;\;\;\times \left[1-e^{-i\pp\cdot(\yy-\xx)}\right] \int \dd^2 \xx' \int \dd^2 \yy' \int\frac{\dd p_1^+ \dd^2 \pp_1}{(2\pi)^3 \sqrt{2p_1^+}} \int\frac{\dd p_2^+ \dd^2 \pp_2}{(2\pi)^3 \sqrt{2p_2^+}} \int\frac{\dd p_3^+ \dd^2 \pp_3}{(2\pi)^3 \sqrt{2p_3^+}} \int\frac{\dd p_4^+ \dd^2 \pp_4}{(2\pi)^3 \sqrt{2p_4^+}} \, e^{i(\pp_1-\pp_2-\pp')\cdot\yy' + i(\pp_3-\pp_4-\pp)\cdot\xx'}  \notag \\
    &\;\;\;\;\times \left. \int\limits_{-\infty}^{\infty}\dd x_1^- \int\limits_{-\infty}^{\infty}\dd x_2^- \, e^{-i(p_1^+ - p_2^+)x_1^- - i(p_3^+ - p_4^+)x_2^-} \left[\frac{\pp_1^i}{p_1^+} - \frac{\pp_2^i}{p_2^+} \right] \sum_{f,f'} \sum_{S,S'} S \left\langle \left\{ \hat{b}^{f\dagger}_{p_2,\ell,S} \hat{b}^{f}_{p_1,m,S} , \hat{b}^{f'\dagger}_{p_4,p,S'} \hat{b}^{f'}_{p_3,q,S'} \right\} \right\rangle   \right\}\Bigg|_{r_\perp = |\yy-\xx|}  \, ,  \notag \\
    &G_2(r_{\perp},zs) = \frac{2i\as^2\pi^2P^+}{3} \,  (t^a)_{\ell m} \, (t^a)_{pq} \int\dd^2\left(\frac{\xx+\yy}{2}\right) \int\frac{\dd^2\pp}{(2\pi)^2} \int\frac{\dd^2\pp'}{(2\pi)^2} \, \frac{\rr^i}{p_{\perp}^2p'^2_{\perp}r^2_{\perp}} \, e^{i(\pp+\pp')\cdot\yy} \left[1-e^{-i\pp\cdot(\yy-\xx)}\right]    \label{gl_xchng_interm_GT}\\
    &\;\;\;\;\times \int \dd^2 \xx' \int \dd^2 \yy' \int\frac{\dd p_1^+ \dd^2 \pp_1}{(2\pi)^3 \sqrt{2p_1^+}} \int\frac{\dd p_2^+ \dd^2 \pp_2}{(2\pi)^3 \sqrt{2p_2^+}} \int\frac{\dd p_3^+ \dd^2 \pp_3}{(2\pi)^3 \sqrt{2p_3^+}} \int\frac{\dd p_4^+ \dd^2 \pp_4}{(2\pi)^3 \sqrt{2p_4^+}} \, e^{i(\pp_1-\pp_2-\pp')\cdot\yy' + i(\pp_3-\pp_4-\pp)\cdot\xx'}  \notag \\
    &\;\;\;\;\times  \int\limits_{-\infty}^{\infty}\dd x_1^- \int\limits_{-\infty}^{\infty}\dd x_2^- \, e^{-i(p_1^+ - p_2^+)x_1^- - i(p_3^+ - p_4^+)x_2^-} \left[\frac{\pp_1^i}{p_1^+} - \frac{\pp_2^i}{p_2^+} \right] \sum_{f,f'} \sum_{S,S'} S \left\langle \left\{ \hat{b}^{f\dagger}_{p_2,\ell,S} \hat{b}^{f}_{p_1,m,S} , \hat{b}^{f'\dagger}_{p_4,p,S'} \hat{b}^{f'}_{p_3,q,S'} \right\} \right\rangle  \bigg|_{r_\perp = |\yy-\xx|}  \, ,  \notag
\end{align}
\end{subequations}
c.f. Eq~\eqref{gl5}, while $Q_f^{\text{G},\text{NS}}(r_{\perp},zs)$ is proportional to the commutator of the ladder operators and eventually vanishes. Then, we similarly evaluate each of these expressions via the valence quark state~\eqref{proton}. Integrating over the impact parameter and other kinematic variables yields the final results~\eqref{gl7} and \eqref{gl15} at the end of Appendix~\ref{sec:O-g4-gluon-Xchange}.

Putting together the quark-exchange terms -- Eqs.~\eqref{qk28} and \eqref{qk30} -- and gluon-exchange terms -- Eqs.~\eqref{gl7}, \eqref{gl9} and \eqref{gl15} -- we obtain the following expressions for the polarized dipole amplitudes at moderate $x$,
\begin{subequations}\label{IC_summary}
\begin{align}
    Q_f(r_{\perp},zs) &= - \frac{8\pi^2\alpha_s^2}{81}\left[18 + 4\delta^{f,u}-\delta^{f,d}\right] \int [\dd x_i] \, \frac{1}{x_1} \int\frac{\dd^2 \pp}{(2\pi)^2} \, \frac{1}{p^2_{\perp}} \, e^{-i\pp\cdot\rr}  \, F(x_i;\pp)  + (r_{\perp}\text{- independent terms}) \, ,  \label{Q_IC} \\
    Q^{\text{NS}}_f(r_{\perp},zs) &= - \frac{8\pi^2\alpha_s^2}{81}\left[4\delta^{f,u}-\delta^{f,d}\right] \int [\dd x_i] \, \frac{1}{x_1} \int\frac{\dd^2 \pp}{(2\pi)^2} \, \frac{1}{p^2_{\perp}} \, e^{-i\pp\cdot\rr}  \, F(x_i;\pp)  + (r_{\perp}\text{- independent terms}) \, ,  \label{QNS_IC} \\
    \wg(r_{\perp},zs) &= - \frac{44\pi^2\alpha_s^2}{9} \int [\dd x_i] \, \frac{1}{x_1} \int\frac{\dd^2 \pp}{(2\pi)^2} \, \frac{1}{p^2_{\perp}} \, e^{-i\pp\cdot\rr}  \, F(x_i;\pp)  + (r_{\perp}\text{- independent terms}) \, , 
    \label{GT_IC}   \\
    G_2(r_{\perp},zs) &=  \frac{16\pi^2\as^2}{9} \int [\dd x_i] \, \frac{1}{x_1}  \int\frac{\dd^2 \pp}{(2\pi)^2} \, \frac{i(\pp\cdot\rr)}{p^4_{\perp}r^2_{\perp}} \, e^{- i\pp\cdot\rr}  \, F(x_i;\pp)  + (r_{\perp}\text{- independent terms}) \, ,   \label{G2_IC} 
\end{align}
\end{subequations}
where we have $Q_f(r_{\perp},zs)=Q_f^{\text{q}}(r_{\perp},zs)+Q_f^{\text{G}}(r_{\perp},zs)$ and so on. Here, we only kept track of the terms sensitive to the dipole size, $r_{\perp}$, which guaranteed the calculation of such terms to be perturbative when the dipole size is perturbatively small. In contrast, the $r_{\perp}$-independent term is left unknown. However, as shown in Appendix~\ref{sec:r-indep}, the $r_{\perp}$-independent term for $G_2(r_{\perp},zs)$ in Eq.~\eqref{G2_IC} vanishes, and hence we will drop them below. In Eqs.~\eqref{IC_summary}, for convenience, we have also defined
\begin{align}\label{IC2}
    &F(x_i;\pp) = \int [\dd^2q_i]  \left[  |\Phi(x_i,\qq_i)|^2 - \Phi^*(x_1,\qq_1 + \pp;x_2,\qq_2 - \pp;x_3,\qq_3) \, \Phi(x_i,\qq_i) \right] .
\end{align}
Note that this function vanishes in the $p_{\perp} \to 0$ limit, so that the integrals over $\pp$ in 
Eqs.~\eqref{IC_summary} are free of infrared divergences. This is a consequence
of the Ward identity, $\mathcal{J}^+(\pp=0) \ket{P} = 0$, for any color singlet state, $|P\rangle$, where
$\mathcal{J}^+(\pp)$ is the transverse Fourier pair of $\int \dd x^-J^{+}(x^-,\xx)$ with the color charge density, $J^+$, defined in Eq.~\eqref{qk21}.

Then, with the momentum-space wave function from Eq.~\eqref{Phi}, it is straightforward to analytically evaluate the Gaussian integrals in Eq.~\eqref{IC2}, which yields
\begin{align}\label{IC3}
    &F(x_i;\pp) = \frac{\mathcal{N}^2 \beta^4}{16\pi^2}\,x_1x_2x_3\,\exp\left[- \frac{M^2}{\beta^2}\sum_{i=1}^3 \frac{1}{x_i}\right] \left\{ 1 - \exp\left[- \frac{p^2_{\perp}}{4\beta^2}\left(\frac{1}{x_1}+\frac{1}{x_2}\right)\right] \right\} .
\end{align}
Note that $F(x_i;\pp)$ actually depends only on $p_{\perp} = |\pp|$, that is, the dependence on the azimuthal directions of $\pp$ is limited to the Fourier factor and, in the case of Eq.~\eqref{G2_IC}, the factor $i(\pp\cdot\rr)$. As a result, the integrals over $\pp$ can be written in terms of the incomplete gamma function which allows us to numerically evaluate the integrals over $x_1$, $x_2$ and $x_3$. The details of these numerical integrals are given in Appendix~\ref{sec:HO-LCwf}.

Ultimately, for small dipoles, we arrive at the following approximate analytic parametrizations,
\begin{subequations}\label{IC_summary_final}
\begin{align}
    Q_f(r_{\perp},zs) &= - \frac{2\pi\alpha_s^2}{81}\left[18 + 4\delta^{f,u}-\delta^{f,d}\right] \, 
    \overline{x^{-1}}  \, \ln\left(\frac{1}{r^2_{\perp}\Lambda^2_1}+c_1\right)   + (r_{\perp}\text{- independent terms})\, ,  \label{Q_IC_final} \\
    Q^{\text{NS}}_f(r_{\perp},zs) &=  - \frac{2\pi\alpha_s^2}{81}\left[4\delta^{f,u}-\delta^{f,d}\right] \, \overline{x^{-1}}  \, \ln\left(\frac{1}{r^2_{\perp}\Lambda^2_1}+c_1\right)   + (r_{\perp}\text{- independent terms}) \, ,  \label{QNS_IC_final} \\
    \wg(r_{\perp},zs) &=  - \frac{11\pi\alpha_s^2}{9} \; \overline{x^{-1}}  \, \ln\left(\frac{1}{r^2_{\perp}\Lambda^2_1}+c_1\right)   + (r_{\perp}\text{- independent terms}) \, , 
    \label{GT_IC_final}   \\
    G_2(r_{\perp},zs) &=  \frac{2\pi\as^2}{9} \; \overline{x^{-1}}  \, \ln\left(\frac{1}{r^2_{\perp}\Lambda^2_2}+c_2\right)    ,   \label{G2_IC_final} 
\end{align}
\end{subequations}
where the parameters are given by $\Lambda_1 = 0.267$ GeV, $\Lambda_2 = 0.161$ GeV, $c_1=0.668$ and $c_2=0.832$. In particular, $\Lambda_1$ and $c_1$ are fitted from the numerical integrals performed on the exact expression for the type-1 polarized dipole amplitudes shown in Eqs.~\eqref{IC_summary}, and similarly $\Lambda_2$ and $c_2$ from the type-2. Here, $\overline{x^{-1}} = {\cal O}(N_c)$ 
denotes the average of the inverse light-cone momentum fraction of a valence quark in the proton. It is given by 
\begin{equation}\label{x-1bar}
\overline{x^{-1}} = \int[\dd x_i] \, \frac{1}{x_1} \int[\dd^2 \qq_i]\sum_{\{f_1,f_2,f_3\}=\{u,u,d\}} \sum_{\sigma_1,\sigma_2,\sigma_3} \left|
\psi_{\mathcal{S}_L}(x_1,\qq_1,\sigma_1,f_1;x_2,\qq_2,\sigma_2,f_2;x_3,\qq_3,\sigma_3,f_3) \right|^2 , 
\end{equation}
c.f. Eq.~\eqref{psi_norm}. Numerically, for the momentum space
wave function described above, we have $\overline{x^{-1}} =3.64$. As evident from Figs.~\ref{fig:I1_fit} and \ref{fig:I2_fit} shown in Appendix.~\ref{sec:HO-LCwf}, the approximate parametrizations~\eqref{IC_summary_final} agree very closely with the exact integral expressions~\eqref{IC_summary} in the perturbative region up to the dipole size of $r_{\perp} \sim 5$ GeV$^{-1}$.

Eqs.~\eqref{IC_summary} and \eqref{IC_summary_final} are the main results of this work. They provide the moderate-$x$ expressions for the polarized dipole amplitudes, which can serve as initial conditions for the small-$x$ helicity evolution~\cite{Kovchegov:2016zex,Cougoulic:2022gbk}. The evolved dipoles then describe polarized dipole-target scattering at small $x$, corresponding to large center-of-mass energy. With Eqs.~\eqref{quark_singlet_PDF}, \eqref{quark_nonsinglet_PDF}, \eqref{gluon_TMD} and \eqref{gluon_PDF}, which were derived in the small-$x$ regime~\cite{Cougoulic:2022gbk}, these evolved polarized dipole amplitudes completely determine the quark and gluon hPDFs. 

Furthermore, Eqs.~\eqref{IC_summary_final} can be generalized to a more flexible model given by
\begin{subequations}\label{IC_summary_approx}
\begin{align}
    Q_f(r_{\perp},zs) &= - \frac{2\pi\alpha_s^2}{81}\left[18 + 4\delta^{f,u}-\delta^{f,d}\right] \, 
    \overline{x^{-1}}  \, \ln\left(\frac{1}{r^2_{\perp}\Lambda^2_{\text{IR}}}\right) + C_f \, ,  \label{Q_IC_approx} \\
    Q^{\text{NS}}_f(r_{\perp},zs) &=  - \frac{2\pi\alpha_s^2}{81}\left[4\delta^{f,u}-\delta^{f,d}\right] \, \overline{x^{-1}}  \, \ln\left(\frac{1}{r^2_{\perp}\Lambda^2_{\text{IR}}}\right) + C_f^{\text{NS}} \, ,  \label{QNS_IC_approx} \\
    \wg(r_{\perp},zs) &=  - \frac{11\pi\alpha_s^2}{9} \; \overline{x^{-1}}  \, \ln\left(\frac{1}{r^2_{\perp}\Lambda^2_{\text{IR}}}\right) + {\widetilde C} \, , 
    \label{GT_IC_approx}   \\
    G_2(r_{\perp},zs) &=  \frac{2\pi\as^2}{9} \; \overline{x^{-1}}  \, \ln\left(\frac{1}{r^2_{\perp}\Lambda^2_{\text{IR}}}\right) + C_2 \, ,   \label{G2_IC_approx} 
\end{align}
\end{subequations}
where the $r_{\perp}$-independent terms are now absorbed into $C_f$, $C_f^{\text{NS}}$ and ${\widetilde C}$. These parameters could be treated as free parameters that will be fitted to small-$x$ data in a future global analysis. Furthermore, we also shifted $\Lambda_1$ and $\Lambda_2$ to a common infrared scale, $\Lambda_{\text{IR}}$, with the resulting constants in Eqs.~\eqref{Q_IC_approx}--\eqref{GT_IC_approx} absorbed into $C_f$, $C_f^{\text{NS}}$ and ${\widetilde C}$, respectively, while the constant in Eq.~\eqref{G2_IC_approx} yielded the constant, $C_2$. These constant terms were also made to account for the behavior at moderately large dipole, allowing for the infrared-regulating parameters, $c_1$ and $c_2$, inside the logarithms in Eqs.~\eqref{IC_summary_final} to be neglected. We emphasize that the coefficients of the  $r_\perp$-dependent terms remain completely fixed by the perturbative calculation.

As the $C$-parameters encode the physics of the $r_{\perp}$-independent terms, their expressions could also include single ultraviolet logarithms, which at small $x$ usually come from requiring the squared dipole size to be bounded below by the scale, $1/zs$, dictated by the center-of-mass energy. Such contributions cannot be captured in this work due to the fact that our formalism does not guarantee perturbativity for the terms independent of the dipole size, $r_{\perp}$. In Appendix~\ref{sec:r-indep}, we sketch the results for these $r_{\perp}$-independent terms assuming that the caveat about perturbative nature of the problem could be set aside. There, we also discuss how the ultraviolet condition $r^2_{\perp}\gg 1/zs$ would come into play and eventually yield single energy logarithms already at moderate $x$. 

Finally, it is important to remark that single logarithms in the initial conditions at moderate $x$ are not redundant with the small-$x$ helicity evolution because the latter is a double-logarithmic evolution, that is, the dominant terms of its kernel yield $\as\ln^2(1/x)$ per step of the evolution. This is in direct contrast with the unpolarized counterpart, e.g. Balitsky-Kovchegov (BK) equation~\cite{Balitsky:1995ub,Kovchegov:1999yj}, whose evolution kernel only yields a single power of high-energy logarithm per iteration, that is, the latter evolution instead resums $\as\ln(1/x)$.

\subsection{Comparison to Previously Employed Parametrizations}

In Refs.~\cite{Kovchegov:2016zex,Kovchegov:2017lsr,Kovchegov:2018znm,Cougoulic:2022gbk,Adamiak:2023okq}, the initial conditions for polarized dipole amplitudes are calculated based on the tree-level diagrams that are dominant in the Regge limit. In that calculation, not only did they have the dipole projectile, but the target was also taken to be a $q\bar{q}$ dipole. As a result, the polarized dipole amplitudes at moderate $x$ can be written as
\begin{subequations}\label{Born_IC}
\begin{align}
    Q_f(r_{\perp},zs) &= \frac{1}{N_f}\wg(r_{\perp},zs) = \frac{\as^2 \pi C_F}{2N_c}\left[(C_F-2)\ln\left(\frac{zs}{\Lambda^2_{\text{IR}}}\right) + 2\ln\left(\frac{1}{r^2_{\perp}\Lambda^2_{\text{IR}}}\right)\right] , \\
    Q_f^{\text{NS}}(r_{\perp},zs) &= \frac{\as^2 \pi C_F^2}{N_c}\,\ln\left(\frac{zs}{\Lambda^2_{\text{IR}}}\right) , \\
    G_2(r_{\perp},zs) &= -\frac{\as^2 \pi C_F}{2N_c} \ln\left(\frac{1}{r^2_{\perp}\Lambda^2_{\text{IR}}}\right) .
\end{align}
\end{subequations}
The only possible comparison between Eqs.~\eqref{IC_summary_approx} and \eqref{Born_IC} is on the coefficients of the transverse logarithms. As it turns out, all the coefficients based on the valence quark model are of the opposite signs to their Born-level counterparts, with the former consistently being 3.3--4.5 times greater in magnitude. However, it is difficult to draw definite conclusions at the current stage. For instance, in Ref.~\cite{Adamiak:2023yhz}, a global analysis was performed under the JAM framework based on a generalized version of initial condition~\eqref{Born_IC}, namely
\begin{subequations}\label{generalized_Born}
\begin{align}
    Q_f(r_{\perp},zs) &= a_f\ln\left(\frac{zs}{\Lambda^2_{\text{IR}}}\right) + b_f\ln\left(\frac{1}{r^2_{\perp}\Lambda^2_{\text{IR}}}\right) + c_f \, ,  \label{Q_gb} \\
    Q^{\text{NS}}_f(r_{\perp},zs) &=  a^{\text{NS}}_f\ln\left(\frac{zs}{\Lambda^2_{\text{IR}}}\right) + b^{\text{NS}}_f\ln\left(\frac{1}{r^2_{\perp}\Lambda^2_{\text{IR}}}\right) + c^{\text{NS}}_f \, ,  \label{QNS_gb} \\
    \wg(r_{\perp},zs) &= \tilde{a}\ln\left(\frac{zs}{\Lambda^2_{\text{IR}}}\right) + \tilde{b}\ln\left(\frac{1}{r^2_{\perp}\Lambda^2_{\text{IR}}}\right) + \tilde{c} \, , 
    \label{GT_gb}   \\
    G_2(r_{\perp},zs) &= a_2\ln\left(\frac{zs}{\Lambda^2_{\text{IR}}}\right) + b_2\ln\left(\frac{1}{r^2_{\perp}\Lambda^2_{\text{IR}}}\right) + c_2 \, .   \label{G2_gb} 
\end{align}
\end{subequations}
This results in the coefficients, $b_f$, $b_f^{\text{NS}}$, $\tilde{b}$ and $b_2$, of the transverse logarithm being 5--20 times greater in magnitude than the respective coefficients from our calculation, i.e. Eqs.~\eqref{IC_summary_approx}, with 4 out of 7 nonzero coefficients having the same sign. Most importantly, all the coefficients from this work fall with the 95\% confidence interval of the respective JAM coefficients. Qualitatively, our results from the valence quark model lead to the coefficients that are roughly in the middle between the Born-level calculation~\cite{Kovchegov:2016zex,Adamiak:2023okq} and the JAM global analysis~\cite{Adamiak:2023yhz}. 

Ultimately, to quantitatively test the viability of our initial conditions, a new global analysis of small-$x$ helicity data must be performed starting from initial conditions~\eqref{IC_summary_approx} with $\as$ together with $C_f$, $C_f^{\text{NS}}$, ${\widetilde C}$ and $C_2$ as free parameters. The strictest valence-quark model would take the $C$-parameters of all the type-1 amplitudes in Eqs.~\eqref{IC_summary_approx} to be identical regardless of flavors, $C_f=C_f^{\text{NS}}={\widetilde C}$, resulting in 3 free parameters: $\as$, $C_u$ and $C_2$. On the other hand, a more flexible valence-quark model could allow for each $C$-parameter to be distinct, resulting in 9 free parameters: $\as$, $C_u$, $C_d$, $C_s$, $C_u^{\text{NS}}$, $C_d^{\text{NS}}$, $C_s^{\text{NS}}$, ${\widetilde C}$ and $C_2$. Furthermore, as discussed in more detail in Appendix~\ref{sec:r-indep}, the energy-logarithmic terms could also be added to Eqs.~\eqref{IC_summary_approx} without additional free parameters. In any case, the initial conditions derived in this work based on the physical proton model of three valence quarks provides a significant improvement to the generalized Born-level model~\eqref{generalized_Born} employed in JAM analysis~\cite{Adamiak:2023yhz}, reducing the number of free parameters from 24 in the latter to 3--9 in this work. As a result, we expect a significant reduction in the uncertainty of the physical predictions, including the parton helicity PDFs and $g_1$ structure function at small Bjorken $x$, that result from a future global analysis based on this work.

\section{Conclusion and Outlook}
\label{sec:conclusion}

In this work, we employ the valence quark model of the proton target to calculate the polarized dipole amplitudes at moderate values of Bjorken $x$ to the order $\mathcal{O}(\as^2)$ in perturbation theory. These dipoles are the key ingredients for the small-$x$ helicity evolution and allow for parton helicity PDFs, $g_1$ structure functions and other physical observables in polarized DIS and SIDIS processes to be calculated at small $x$. Particularly, the final results~\eqref{IC_summary} of our calculation serves as a set of initial conditions for the small-$x$ helicity evolution at large $N_c\& N_f$~\cite{Cougoulic:2022gbk}. We also reported a simple analytical parametrization, Eq.~\eqref{IC_summary_final}, that approximates the full result very accurately and can be conveniently employed in fits. In contrast to previous initial condition calculations~\cite{Kovchegov:2016zex,Cougoulic:2020tbc,Adamiak:2023okq}, our results encode physical characteristics of the proton target at moderate $x$ and provide explicit flavor structure that reflects the valence quark content of the proton. The framework employed here can be generalized to other hadronic targets. 

Our results contain free non-perturbative parameters to be fixed by an upcoming global analysis, which will include at least the polarized DIS and SIDIS data at small $x\leq 0.1$~\cite{Adamiak:2021ppq}. In contrast to the recent global analysis~\cite{Adamiak:2023yhz} of polarized small-$x$ data based on the generalized Born-level initial conditions~\eqref{generalized_Born}, the new analysis based on the initial condition presented in this work will automatically encode part of the physical information about the target state through the valence quark model, leaving us with fewer free parameters to be fitted to the data. In particular, the dipole-size dependence of the polarized dipole amplitude at moderate $x$ is completely determined by our perturbative calculation. We are optimistic that the upcoming global analysis will improve on the uncertainties in the predictions of hPDFs and $g_1$ structure functions, contributing to an improved understanding of the proton spin puzzle even before future EIC measurements. Furthermore, the resulting gluon hPDF will allow for a direct comparison to the results of~\cite{Chen:2006ng} where similar valence quark models are employed directly to the gluon hPDF operator without the polarized dipole framework considered here.

With the framework established in this paper, a potential future direction towards higher precision is to perform the calculation of the polarized dipole amplitudes with a more realistic model that includes the Melosh rotation. Physically, this takes into account that the helicity of each individual valence quark, which generally has nonzero transverse momentum, is not exactly the same as its spin along the light-cone direction. Finally, the developments in this work are also 
relevant for other calculations of correlators involving sub-eikonal Wilson lines, including the polarized dipole amplitudes pertaining to other TMDs~\cite{Kovchegov:2021iyc,Kovchegov:2022kyy,Santiago:2023rfl,qkTMD}, the orbital angular momentum~\cite{Kovchegov:2019rrz,Kovchegov:2023yzd,Manley:2024pcl} and other sub-eikonal corrections to high-energy scattering processes~\cite{Altinoluk:2020oyd,Altinoluk:2021lvu,Altinoluk:2022jkk,Altinoluk:2023qfr,Chirilli:2018kkw,Chirilli:2021lif}.


\begin{acknowledgments}

\label{sec:acknowledgement}

YT would like to thank Risto Paatelainen, Yuri Kovchegov, Feng Yuan, Florian Cougoulic and Daniel Adamiak
for useful discussions about various aspects of this work.

AD acknowledges support by the DOE Office of Nuclear Physics through
Grant DE-SC0002307.

HM and YT are supported by the Academy of Finland, the Centre of Excellence in Quark Matter and projects 338263 and 346567, under the European Union’s Horizon 2020 research and innovation programme by the European Research Council (ERC, grant agreements No. ERC-2023-101123801 GlueSatLight and No. ERC-2018-ADG-835105 YoctoLHC) and by the STRONG-2020 project (grant agreement No. 824093). The content of this article does not reflect the official opinion of the European Union and responsibility for the information and views expressed therein lies entirely with the authors. 

\end{acknowledgments}


\appendix

\section{Calculation of The Matrix Elements}
\label{sec:matrix_element_details}

In this Appendix, we present some details of the calculation of the polarized dipole amplitudes. The general method is similar to the one previously used to calculate matrix elements of the unpolarized dipole $S$-matrix in Refs.~\cite{Dumitru:2018vpr,Dumitru:2020gla,Dumitru:2021tqp}. As discussed in Section~\ref{sec:polarized}, sub-eikonal corrections to the light-cone Wilson line that are relevant to helicity come in two types. While the type-2 dipole amplitude is pure-glue, containing only gluon-exchange terms, the type-1 dipole amplitudes involve both quark and gluon exchanges. In our calculation, it is convenient to define
\begin{subequations}\label{type1_qG}
\begin{align}
    &Q_f^{\text{q}}(r_{\perp},zs) = \frac{zs}{2N_c}\int\dd^2\left(\frac{\xx+\yy}{2}\right) \text{Re}\left\langle\text{T\,tr}\left[V_{\xx}V_{\yy}^{\text{q}[1]\dagger}\right] + \text{T\,tr}\left[V_{\yy}^{\text{q}[1]}V_{\xx}^{\dagger}\right]\right\rangle \Big|_{r_\perp=|\yy-\xx|} \, , \label{Q_q} \\
    &Q_f^{\text{G}}(r_{\perp},zs) = \frac{1}{N_f}\wg^{\text{G}}(r_{\perp},zs) = \frac{zs}{2N_c}\int\dd^2\left(\frac{\xx+\yy}{2}\right) \text{Re}\left\langle\text{T\,tr}\left[V_{\xx}V_{\yy}^{\text{G}[1]\dagger}\right] + \text{T\,tr}\left[V_{\yy}^{\text{G}[1]}V_{\xx}^{\dagger}\right]\right\rangle \Big|_{r_\perp=|\yy-\xx|} \, , \label{Q_G} \\
    &Q_f^{\text{NS},\text{q}}(r_{\perp},zs) = \frac{zs}{2N_c}\int\dd^2\left(\frac{\xx+\yy}{2}\right) \text{Re}\left\langle\text{T\,tr}\left[V_{\xx}V_{\yy}^{\text{q}[1]\dagger}\right] - \text{T\,tr}\left[V_{\yy}^{\text{q}[1]}V_{\xx}^{\dagger}\right]\right\rangle \Big|_{r_\perp=|\yy-\xx|} \, , \label{QNS_q} \\
    &Q_f^{\text{NS},\text{G}}(r_{\perp},zs) = \frac{zs}{2N_c}\int\dd^2\left(\frac{\xx+\yy}{2}\right) \text{Re}\left\langle\text{T\,tr}\left[V_{\xx}V_{\yy}^{\text{G}[1]\dagger}\right] - \text{T\,tr}\left[V_{\yy}^{\text{G}[1]}V_{\xx}^{\dagger}\right]\right\rangle \Big|_{r=|\yy-\xx|} \, , \label{QNS_G} \\
    &\wg^{\text{q}}(r_{\perp},zs) = \frac{zs}{2N_c}\int\dd^2\left(\frac{\xx+\yy}{2}\right) \text{Re}\left\langle\text{T\,tr}\left[V_{\xx}W_{\yy}^{\text{q}[1]\dagger}\right] + \text{T\,tr}\left[W_{\yy}^{\text{q}[1]}V_{\xx}^{\dagger}\right]\right\rangle \Big|_{r=|\yy-\xx|} \, , \label{GT_q} 
\end{align}
\end{subequations}
for the quark- and gluon-exchange terms of the flavor singlet, flavor non-singlet and adjoint type-1 dipole amplitudes, respectively. In Eq.~\eqref{Q_G}, we denote by $N_f$ the number of quark flavors, which throughout this work is taken to be 3, for up, down and strange quarks. This is because the gluon-exchange terms in the polarized Wilson lines do not depend on the flavor of the (anti)quark that moves along the light cone. Recall that the polarized Wilson lines, $V_{\yy}^{\text{q}[1]}$, $V_{\yy}^{\text{G}[1]}$ and $W_{\yy}^{\text{q}[1]}$, are defined in Eqs.~\eqref{V_subeik} and \eqref{W_q1}. The type-2 dipole amplitude, $G_2(r_{\perp},zs)$, which only contains gluon-exchange terms, can be calculated as a whole starting from Eqs.~\eqref{def_G2} and \eqref{ViG2}.

\subsection{Quark Exchange Diagrams at ${\cal O}(\as)$}
\label{sec:O-g2-quark-Xchange}

We start with the quark-exchange terms contained in $Q_f^{\text{q}}(r_{\perp},zs)$, $Q_f^{\text{NS},\text{q}}(r_{\perp},zs)$ and $\wg^{\text{q}}(r_{\perp},zs)$. At order $\alpha_s$, there is no extra vertex outside of the ones contained in the polarized Wilson lines themselves. Since the proton only contains the three valence quarks and no antiquark, the terms $\sim \text{Re}\left\langle \text{T\,tr}\left[V_{\yy}^{\text{q}[1]}V_{\xx}^{\dagger}\right]\right\rangle,\text{Re}\left\langle \text{T\,tr}\left[W_{\yy}^{\text{q}[1]}V_{\xx}^{\dagger}\right]\right\rangle$ with polarized quark in the dipole would result in disconnected diagrams and therefore vanish. As for the remaining terms, we will begin with the calculation for $\text{Re}\left\langle \text{T\,tr}\left[V_{\xx}V_{\yy}^{\text{q}[1]\dagger}\right]\right\rangle$, which contributes to $Q_f^{\text{q}}(r_{\perp},zs)$ and $Q_f^{\text{NS},\text{q}}(r_{\perp},zs)$, then later generalize the results to the counterpart with $W_{\yy}^{\text{q}[1]\dagger}$, which contributes to $\wg^{\text{q}}(r_{\perp},zs)$. 

\begin{figure}
    \centering
    \includegraphics[width=0.35\textwidth]{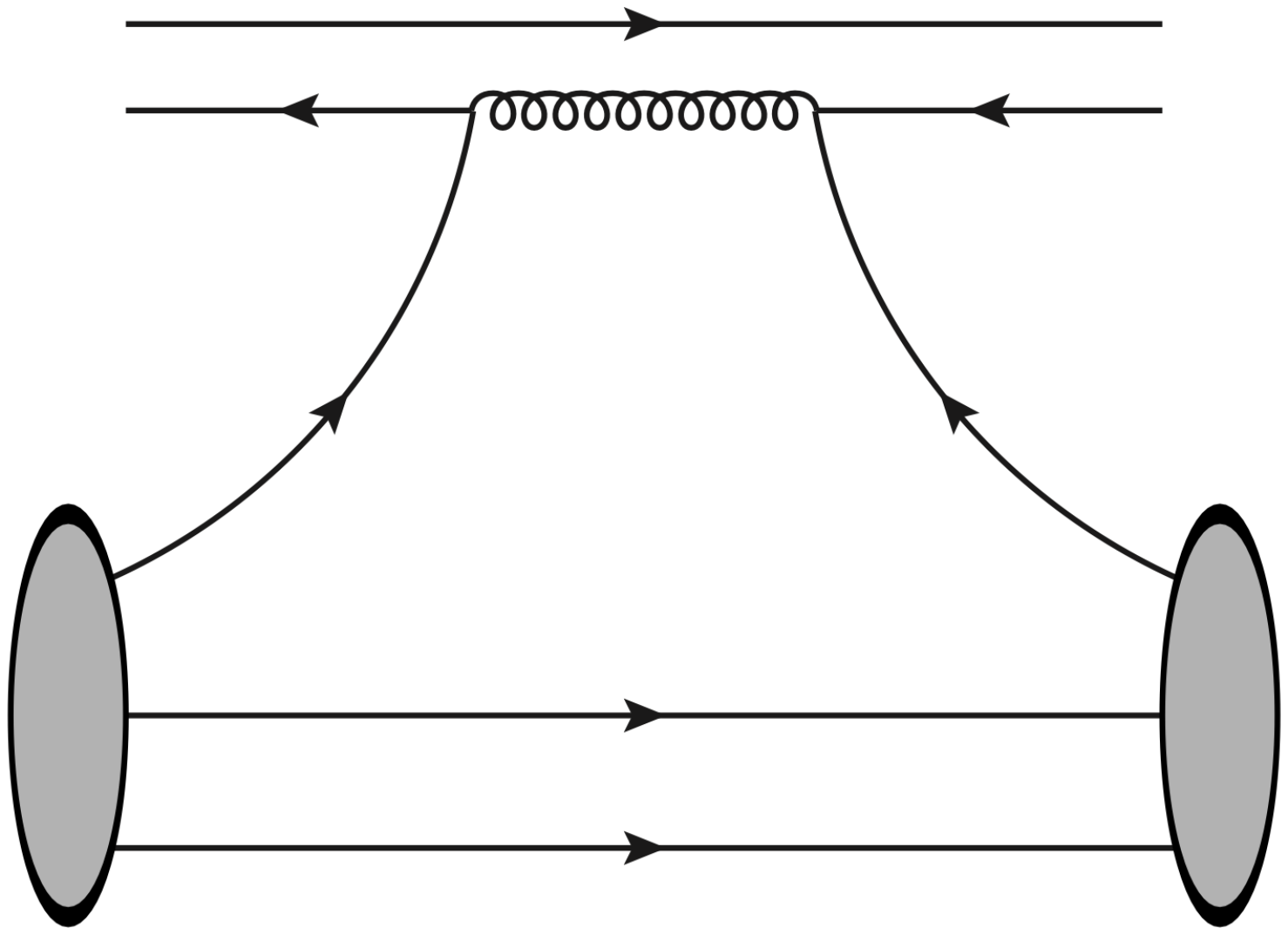}
    \caption{The diagram contributing to the quark-exchange term at order $\mathcal{O}(\alpha_s)$. The upper half of the diagram corresponds to the projectile dipole that makes up the polarized dipole trace, while the lower half corresponds to the interactions within the proton target.}
    \label{fig:qk_Oas1}
\end{figure}

With the help of Eq.~\eqref{V_subeik_q_1}, we have that
\begin{align}\label{qk0}
\frac{zs}{2N_c} &\left\langle\text{T\,tr}\left[V_{\xx}V_{\yy}^{\text{q}[1]\dagger}\right]\right\rangle  = \frac{\alpha_s \pi P^+}{N_c}\int\limits_{-\infty}^{\infty}\dd x_1^-\int\limits_{x_1^-}^{\infty}\dd x_2^- \\ 
&\;\;\;\;\times \left\langle \text{T\,tr}\left[ V_{\xx} V_{\yy}[-\infty,x_1^-]\,t^a\psi^f_{\alpha}(x_1^-,\yy)\,U_{\yy}^{ab}[x_1^-,x_2^-]\,(\gamma^+\gamma_5)_{\beta\alpha} \bar{\psi}^f_{\beta}(x_2^-,\yy)\,t^b\,V_{\yy}[x_2^-,\infty] \right] \right\rangle ,  \notag
\end{align}
where $P^+$ is the proton's longitudinal momentum. In Eq.~\eqref{qk0}, $\alpha$ and $\beta$ are Dirac indices, and $f$ is the flavor of the dipole that would finally go into $Q_f^{\text{q}}$.  At order $\mathcal{O}(\alpha_s)$, we must keep only the non-interacting term of each Wilson line, which gives
\begin{align}\label{qk1}
&\frac{zs}{2N_c}\left\langle\text{T\,tr}\left[V_{\xx}V_{\yy}^{\text{q}[1]\dagger}\right]\right\rangle  = \frac{\alpha_s \pi C_F P^+}{N_c}\int\limits_{-\infty}^{\infty}\dd x_1^-\int\limits_{x_1^-}^{\infty}\dd x_2^-   \left\langle  \psi^f_{i\alpha}(x_1^-,\yy)\, (\gamma^+\gamma_5)_{\beta\alpha} \bar{\psi}^f_{i\beta}(x_2^-,\yy)   \right\rangle + \mathcal{O}(\alpha_s^2)\, ,  
\end{align}
where $i$ is a color index. Throughout the rest of this Section, we will drop $\mathcal{O}(\alpha_s^2)$ at the end for brevity, as it should be clear that our current calculation concerns the terms of order $\mathcal{O}(\alpha_s)$ only. The operator form in Eq.~\eqref{qk1} implies that the contribution we are calculating here corresponds to the diagram shown in Fig.~\ref{fig:qk_Oas1}.

To proceed, we plug in the quark field in terms of the quark creation and annihilation operators,
\begin{align}\label{qkfield}
\psi^f_{i\alpha}(x^-,\xx) &= \int\frac{\dd p^+\dd^2 \pp}{(2\pi)^32p^+}\sum_{S}\left[\hat{b}^f_{p,i,S}u_{S}^{\alpha}(p)\,e^{-ip^+x^- + i\pp\cdot\xx} + \hat{d}^{f\dagger}_{p,i,S}v_{S}^{\alpha}(p)\,e^{ip^+x^- - i\pp\cdot \xx} \right],
\end{align}
together with its Hermitian conjugate. Here, one can choose the basis spinor solutions, $u_S(p)$ and $v_S(p)$ to be the Brodsky-Lepage spinor \cite{Lepage:1980fj}. Since both the incoming and outgoing states in $\left\langle\cdots\right\rangle$ contain the three valence quarks, we are left with only one term, which reads
\begin{align}\label{qk2}
\frac{zs}{2N_c}\left\langle\text{T\,tr}\left[V_{\xx}V_{\yy}^{\text{q}[1]\dagger}\right]\right\rangle  &= - \frac{\alpha_s \pi C_F P^+}{N_c}\int\limits_{-\infty}^{\infty}\dd x_1^-\int\limits_{x_1^-}^{\infty}\dd x_2^- \int\frac{\dd p^+\dd^2\pp}{(2\pi)^32p^+}\int\frac{\dd p'^+\dd^2\pp'}{(2\pi)^32p'^+}\sum_{S_1,S_2} \\
&\;\;\;\;\;\times \left[ \bar{u}_{S_2}(p')\gamma^+\gamma_5u_{S_1}(p) \right] e^{-ip^+x^-_1 +ip'^+x^-_2 + i(\pp-\pp')\cdot\yy}\left\langle \hat{b}^{f\dagger}_{p',i,S_2} \hat{b}^f_{p,i,S_1} \right\rangle   . \notag
\end{align}
Along the way, we also anticommuted the two ladder operators using the fact that the anticommutator would yield a number times $\sum_{\mathcal{S}_L}\mathcal{S}_L=0$. To further simplify, we employ the spinor matrix elements from Brodsky-Lepage~\cite{Lepage:1980fj},
\begin{align}\label{BL1}
&\bar{u}_{S_2}(p'_2)\,\gamma^+\gamma_5u_{S_1}(p_2) = 2\sqrt{p_2^+p'^+_2}\,S_1\delta_{S_1,S_2} \, .
\end{align}
Plugging this result into Eq.~\eqref{qk2}, we obtain
\begin{align}\label{qk3}
\frac{zs}{2N_c}\left\langle\text{T\,tr}\left[V_{\xx}V_{\yy}^{\text{q}[1]\dagger}\right]\right\rangle  &= - \frac{\alpha_s \pi C_F P^+}{N_c}\int\limits_{-\infty}^{\infty}\dd x_1^-\int\limits_{x_1^-}^{\infty}\dd x_2^- \int\frac{\dd p^+\dd^2 \pp}{(2\pi)^3\sqrt{2p^+}}\int\frac{\dd p'^+\dd^2 \pp'}{(2\pi)^3\sqrt{2p'^+}} \, e^{-ip^+x^-_1 +ip'^+x^-_2 + i(\pp-\pp')\cdot\yy} \\
&\;\;\;\;\;\times \sum_SS \left\langle \hat{b}^{f\dagger}_{p',i,S} \hat{b}^f_{p,i,S} \right\rangle   . \notag
\end{align}

The second line of Eq.~\eqref{qk3} can be written as follows based on Eq.~\eqref{exp_heli},
\begin{align}\label{qk4}
    &\sum_SS \left\langle \hat{b}^{f\dagger}_{p',i,S} \hat{b}^f_{p,i,S} \right\rangle = \lim_{K\to P}\frac{1}{2}\sum_{S,\mathcal{S}_L} S\mathcal{S}_L \, \frac{\bra{K^+,\underline{K},\mathcal{S}_L} \hat{b}^{f\dagger}_{p',i,S} \hat{b}^f_{p,i,S} \ket{P^+,\underline{P},\mathcal{S}_L}}{\braket{K^+,\underline{K},\mathcal{S}_L | P^+,\underline{P},\mathcal{S}_L}} \, .
\end{align}
Then, we plug the proton state from Eq.~\eqref{proton} into the numerator to get
\begin{align}\label{qk5}
    &\frac{1}{2}\sum_{S,\mathcal{S}_L} S\mathcal{S}_L \bra{K^+,\underline{K},\mathcal{S}_L} \hat{b}^{f\dagger}_{p',i,S} \hat{b}^f_{p,i,S} \ket{P^+,\underline{P},\mathcal{S}_L} = \frac{1}{9} \left[4\delta^{f,u} - \delta^{f,d}\right] \int[\dd x_i] \int[\dd^2 \qq_i] \,  
    \sqrt{\frac{(P^+)^3}{K^+}p^+p'^+}  \\
    &\;\;\;\;\times \Phi^*\left(\frac{p'^+}{K^+},\pp'-\frac{p'^+}{K^+}\underline{K}; x_2\frac{P^+}{K^+},\qq_2+x_2\underline{P}-x_2\frac{P^+}{K^+}\underline{K}; x_3\frac{P^+}{K^+},\qq_3+x_3\underline{P}-x_3\frac{P^+}{K^+}\underline{K}\right) \Phi(x_i,\qq_i)   \notag \\
    &\;\;\;\;\times 4\pi\delta(p^+-x_1P^+) (2\pi)^2\delta^2(\pp-x_1\underline{P}-\qq_1) \, 4\pi\delta(p'^+ - p^+ + P^+ - K^+ ) (2\pi)^2\delta^2(\pp'-\pp+\underline{P}-\underline{K}) \,  , \notag
\end{align}
where we also made use of Eqs.~\eqref{eq:factorized-LCwf} and \eqref{qk11}. The latter implies that
\begin{align}\label{wfS_sum1}
    &\sum_{\sigma_2,\sigma_3}\sum_{f_2,f_3} |S_{\mathcal{S}_L}(S,f;\sigma_2,f_2;\sigma_3,f_3)|^2 = \frac{1}{9}\left\{\left[5\delta_{S,\mathcal{S}_L} + \delta_{S,-\mathcal{S}_L}\right]\delta^{f,u} + \left[\delta_{S,\mathcal{S}_L} + 2\delta_{S,-\mathcal{S}_L}\right]\delta^{f,d}\right\} . 
\end{align}
Furthermore, we multiply to the initial result the symmetry factor of $3$ to account for the fact that any of the three valence quarks could be the one interacting with the projectile, instead of ``quark 1'' as depicted in Fig.~\ref{fig:qk_Oas1}. Due to the symmetric nature of the proton wave function, the results remain unchanged no matter which valence quark interacts with the projectile.

Finally, plugging Eqs.~\eqref{qk3}--\eqref{qk5} into Eqs.~\eqref{Q_q} and \eqref{QNS_q} while keeping in mind that the polarized quark term vanishes, we have that
\begin{align}\label{qk6}
    &Q^{\text{q}}_{f}(r_{\perp},zs) = Q^{\text{NS},\text{q}}_{f}(r_{\perp},zs) = -  \frac{\alpha_s\pi C_F}{18N_c} \left[4\delta^{f,u} - \delta^{f,d}\right]  \int[\dd x_i] \int[\dd^2\qq_i]  \, 2\pi\delta(x_1) \, |\Phi(x_i,\qq_i)|^2 \, , 
\end{align}
where along the way we picked out the real part of the expression. Physically, owing to the factor of $\delta(x_1)$, this diagram corresponds exclusively to the edge case where the interacting valence quark (``quark 1'' in Fig.~\ref{fig:qk_Oas1}) carries no longitudinal momentum. This is consistent with the fact that the diagram we are considering requires the interacting valence quark to have the same kinematics as those of the exchange quarks, whose momenta are dominated by their transverse components. Ultimately, the whole expression vanishes because the proton momentum-space wave function, $\Phi(x_i,\qq_i)$, has no support at $x_1=0$, implying that the dominant quark-exchange contribution is at least of order $\mathcal{O}(\as^2)$.

Before we proceed to calculate the $\mathcal{O}(\as^2)$ contributions, we consider the quark-exchange term for the adjoint dipole amplitude, $\wg^{\text{q}}(r_{\perp},zs)$. The Wilson line trace that contains a connected diagram can be written as
\begin{align}\label{qk7}
    \frac{zs}{2N_c} \sum_f \left\langle\text{T\,tr}\left[V_{\xx}W_{\yy}^{\text{q}[1]\dagger}\right]\right\rangle  &= \frac{\alpha_s \pi P^+}{4N_c} \int\limits_{-\infty}^{\infty}\dd x_1^-\int\limits_{x_1^-}^{\infty}\dd x_2^- \sum_f \left\langle \text{tr}\left[ V_{\xx} V_{\yy}[-\infty,x_1^-]\,\psi^f_{\alpha}(x_1^-,\yy)\,(\gamma^+\gamma_5)_{\beta\alpha} \bar{\psi}^f_{\beta}(x_2^-,\yy)\,V_{\yy}[x_2^-,\infty] \right] \right\rangle \notag \\
    &= \frac{\alpha_s \pi P^+}{4N_c} \int\limits_{-\infty}^{\infty}\dd x_1^-\int\limits_{x_1^-}^{\infty}\dd x_2^- \sum_f \left\langle \psi^f_{i\alpha}(x_1^-,\yy)\,(\gamma^+\gamma_5)_{\beta\alpha} \bar{\psi}^f_{i\beta}(x_2^-,\yy) \right\rangle  \\
    &= \frac{1}{4C_F}\times \frac{zs}{2N_c} \sum_f\left\langle\text{T\,tr}\left[V_{\xx}V_{\yy}^{\text{q}[1]\dagger}\right]\right\rangle   \, .   \notag
\end{align}
This implies exactly the same qualitative features for $\wg^{\text{q}}(r_{\perp},zs)$ at order $\mathcal{O}(\as)$ as those of the $Q^{\text{q}}(r_{\perp},zs)$ and $Q^{\text{NS},\text{q}}(r_{\perp},zs)$ counterparts. Most importantly, the contribution to the moderate-$x$ initial condition of $\wg^{\text{q}}(r_{\perp},zs)$ also begins at $\mathcal{O}(\as^2)$.

\subsection{ Quark Exchange Diagrams at ${\cal O}(\as^2)$}
\label{sec:O-g4-quark-Xchange}

Since the $\mathcal{O}(\as)$ contribution to the quark-exchange term vanishes, we proceed to consider the $\mathcal{O}(\as^2)$ expressions. Such corrections come from two sources: (i) the $\mathcal{O}(g)$ corrections to the Wilson lines within the polarized dipole trace, and (ii) the emission and absorption of a gluon by valence quarks inside the proton. 

\begin{figure}[]
    \centering
    \begin{subfigure}[t]{0.2\textwidth}
        \centering
        \adjincludegraphics[width = \textwidth,trim={{0.07\width} {0.3\height} {0.07\width} {0.3\height}},clip]{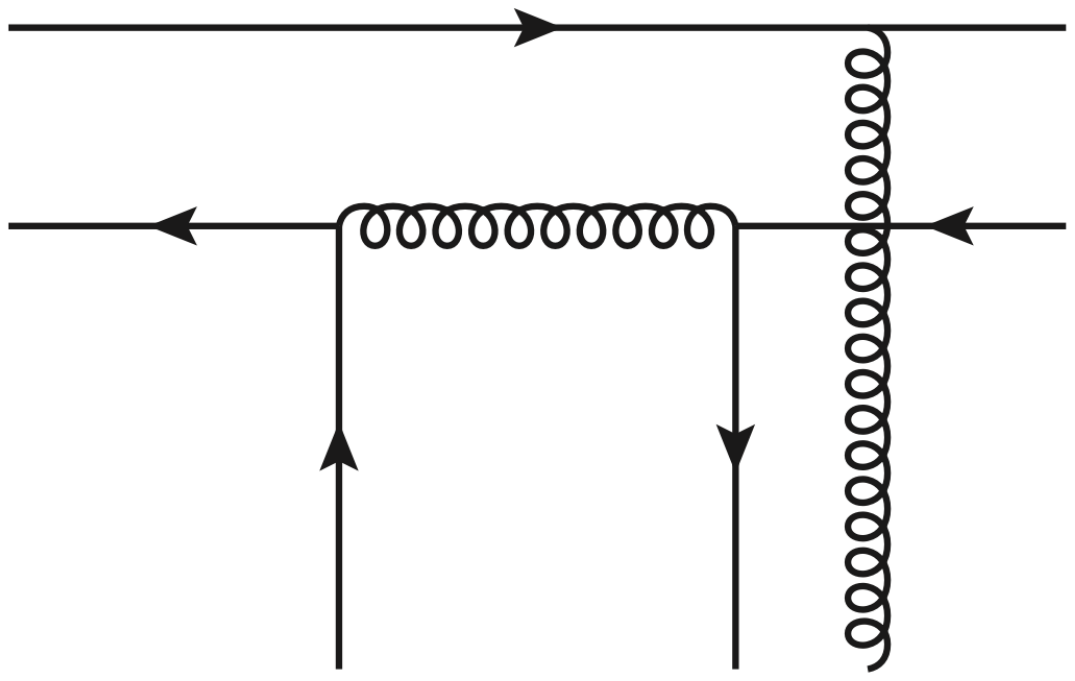}
        \caption{Correction to $V_{\xx}$}
        \label{fig:Vx_Og}
    \end{subfigure}
    ~ 
    \begin{subfigure}[t]{0.24\textwidth}
        \centering
        \adjincludegraphics[width = 0.8333333\textwidth,trim={{0.07\width} {0.3\height} {0.07\width} {0.3\height}},clip]{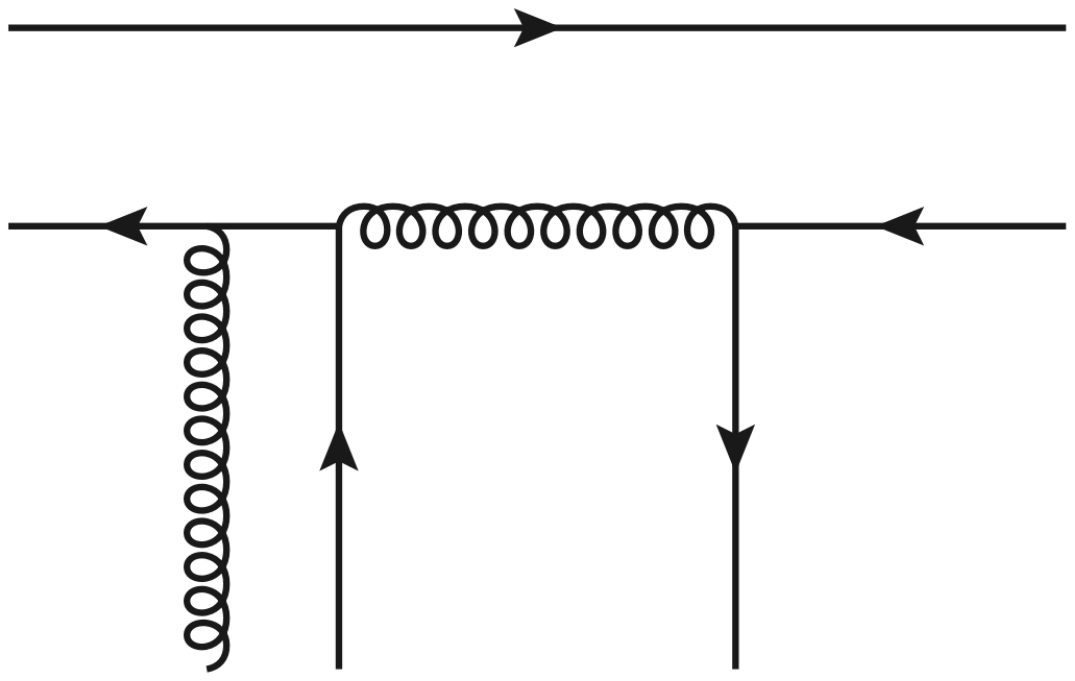}
        \caption{Correction to $V_{\yy}[-\infty,x_1^-]$}
        \label{fig:Vyminus_Og}
    \end{subfigure}
    ~ 
    \begin{subfigure}[t]{0.23\textwidth}
        \centering
        \adjincludegraphics[width = 0.869565217\textwidth,trim={{0.07\width} {0.3\height} {0.07\width} {0.3\height}},clip]{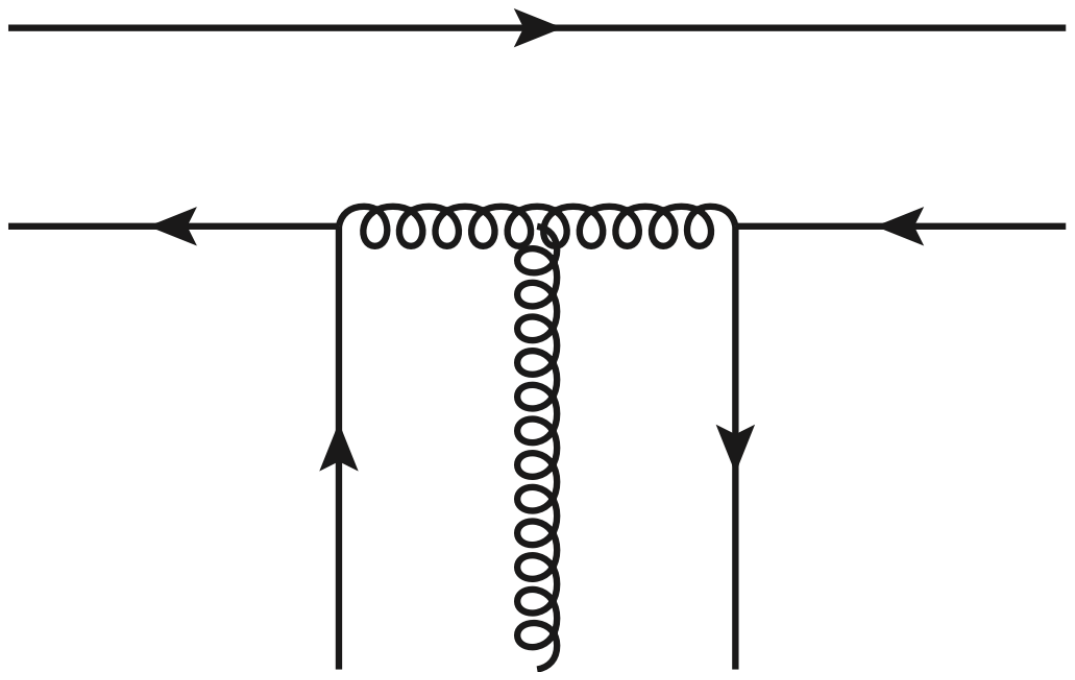}
        \caption{Correction to $U_{\yy}[x_1^-,x_2^-]$}
        \label{fig:Uy_Og}
    \end{subfigure}
    ~ 
    \begin{subfigure}[t]{0.22\textwidth}
        \centering
        \adjincludegraphics[width = 0.909090909\textwidth,trim={{0.07\width} {0.3\height} {0.07\width} {0.3\height}},clip]{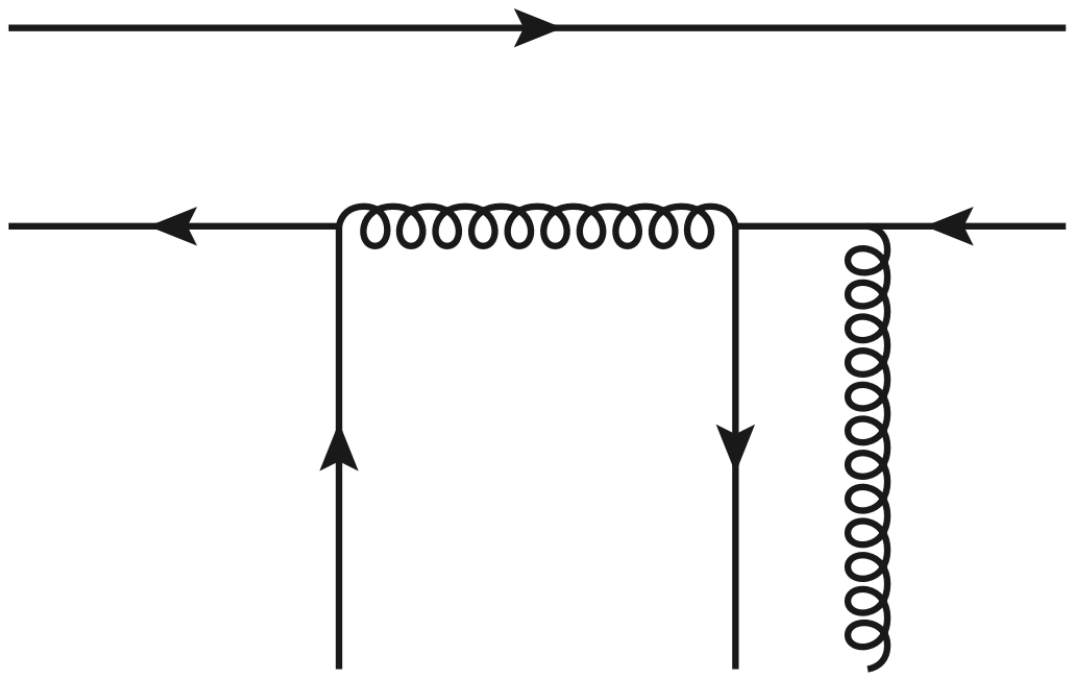}
        \caption{Correction to $V_{\yy}[x_2^-,\infty]$}
        \label{fig:Vyplus_Og}
    \end{subfigure}
    \caption{Corrections of order $\mathcal{O}(g)$ to the quark-exchange term of the polarized dipole, resulting in eikonal gluon emissions from one of the parton lines that receives the correction.}
    \label{fig:WilsonLine_Og}
\end{figure}

Starting with the first contribution, we revisit Eq.~\eqref{qk0} to notice that there are four Wilson lines that could receive an $\mathcal{O}(g)$ correction. Each of these correction terms corresponds physically to an eikonal gluon emission by the respective parton line. Diagrammatically, they corresponds to the four diagrams shown in Figs.~\ref{fig:WilsonLine_Og}. Respectively from Fig.~\ref{fig:Vx_Og} to \eqref{fig:Vyplus_Og}, the corrections to Eq.~\eqref{qk1} read
\begin{subequations}\label{qk20}
\begin{align}
    \frac{zs}{2N_c} \left\langle\text{T\,tr}\left[V_{\xx}V_{\yy}^{\text{q}[1]\dagger}\right]\right\rangle\Big|_{V_{\xx}} &= - \frac{ig\,\alpha_s \pi P^+}{2N_c^2} \, (t^a)_{ji} \, (\gamma^+\gamma_5)_{\beta\alpha} \int\limits_{-\infty}^{\infty}\dd x_1^-\int\limits_{x_1^-}^{\infty}\dd x_2^-  \int\limits_{-\infty}^{\infty}\dd x_3^-    \label{qk20_Vx} \\ 
    &\;\;\;\;\times \left\langle A^{+a}(x_3^-,\xx)\, \psi^f_{i\alpha}(x_1^-,\yy) \,\bar{\psi}^f_{j\beta}(x_2^-,\yy)  \right\rangle ,  \notag \\
    \frac{zs}{2N_c} \left\langle\text{T\,tr}\left[V_{\xx}V_{\yy}^{\text{q}[1]\dagger}\right]\right\rangle\Big|_{V_{\yy}[-\infty,x_1^-]} &= \frac{ig\,\alpha_s \pi P^+}{2N_c^2} \, (t^a)_{ji} \, (\gamma^+\gamma_5)_{\beta\alpha} \int\limits_{-\infty}^{\infty}\dd x_1^-\int\limits_{x_1^-}^{\infty}\dd x_2^- \int\limits_{-\infty}^{x_1^-}\dd x_3^-    \label{qk20_Vyminus} \\ 
    &\;\;\;\;\times \left\langle A^{+a}(x_3^-,\yy)\,\psi^f_{i\alpha}(x_1^-,\yy) \,\bar{\psi}^f_{j\beta}(x_2^-,\yy)  \right\rangle ,  \notag \\
    \frac{zs}{2N_c} \left\langle\text{T\,tr}\left[V_{\xx}V_{\yy}^{\text{q}[1]\dagger}\right]\right\rangle\Big|_{U_{\yy}[x_1^-,x_2^-]} &=  \frac{ig\,\alpha_s \pi P^+}{2} \, (t^a)_{ji}  \, (\gamma^+\gamma_5)_{\beta\alpha} \int\limits_{-\infty}^{\infty}\dd x_1^-\int\limits_{x_1^-}^{\infty}\dd x_2^-\int\limits_{x_1^-}^{x_2^-}\dd x_3^-    \label{qk20_Uy} \\ 
    &\;\;\;\;\times \left\langle \psi^f_{i\alpha}(x_1^-,\yy)\,A^{+a}(x_3^-,\yy)\, \bar{\psi}^f_{j\beta}(x_2^-,\yy)  \right\rangle ,  \notag \\
    \frac{zs}{2N_c} \left\langle\text{T\,tr}\left[V_{\xx}V_{\yy}^{\text{q}[1]\dagger}\right]\right\rangle\Big|_{V_{\yy}[x_2^-,\infty]} &=  \frac{ig\,\alpha_s \pi P^+}{2N_c^2} \, (t^a)_{ji} \, (\gamma^+\gamma_5)_{\beta\alpha}  \int\limits_{-\infty}^{\infty}\dd x_1^-\int\limits_{x_1^-}^{\infty}\dd x_2^- \int\limits_{x_2^-}^{\infty}\dd x_3^-    \label{qk20_Vyplus} \\ 
    &\;\;\;\;\times \left\langle \psi^f_{i\alpha}(x_1^-,\yy)\, \bar{\psi}^f_{j\beta}(x_2^-,\yy) A^{+a}(x_3^-,\yy) \right\rangle ,  \notag
\end{align}
\end{subequations}
where along the way we used
\begin{align}
    &t^at^bt^a = -\frac{1}{2N_c}\,t^b \;\;\;\;\; \text{and} \;\;\;\;\; t^bt^a(T^c)^{ab} = -\frac{N_c}{2}\,t^c\,.
\end{align}

A close look at Eqs.~\eqref{qk20} allows us to realize that only Eq.~\eqref{qk20_Vx} (corresponding to Fig.~\ref{fig:Vx_Og}) is sensitive to the dipole's transverse separation, $\rr = \yy-\xx$. The three remaining expression depend only on $\yy$. Upon the integration over impact parameter, $\bb=(\xx+\yy)/2$, the results would not contain a transverse scale that could guarantee the validity of perturbative physics employed in our calculation. In fact, the results at $\mathcal{O}(\alpha_s)$ suffer the same symptom, as the diagram contains no parton exchange between the target and the unpolarized quark in the dipole. Furthermore, the other type of $\mathcal{O}(\as^2)$ corrections that involve gluon emission and absorption within the proton would include exactly the same pair of quark exchanges, resulting in the expression independent of $\rr$. Hence, for the rest of this Section, we simply focus on Eq.~\eqref{qk20_Vx}. We will revisit the terms indedendent of $\rr$ in Appendix~\ref{sec:r-indep}.

\begin{figure}[t!]
    \centering
    \begin{subfigure}[t]{0.45\textwidth}
        \centering
        \includegraphics[width = 0.6\textwidth]{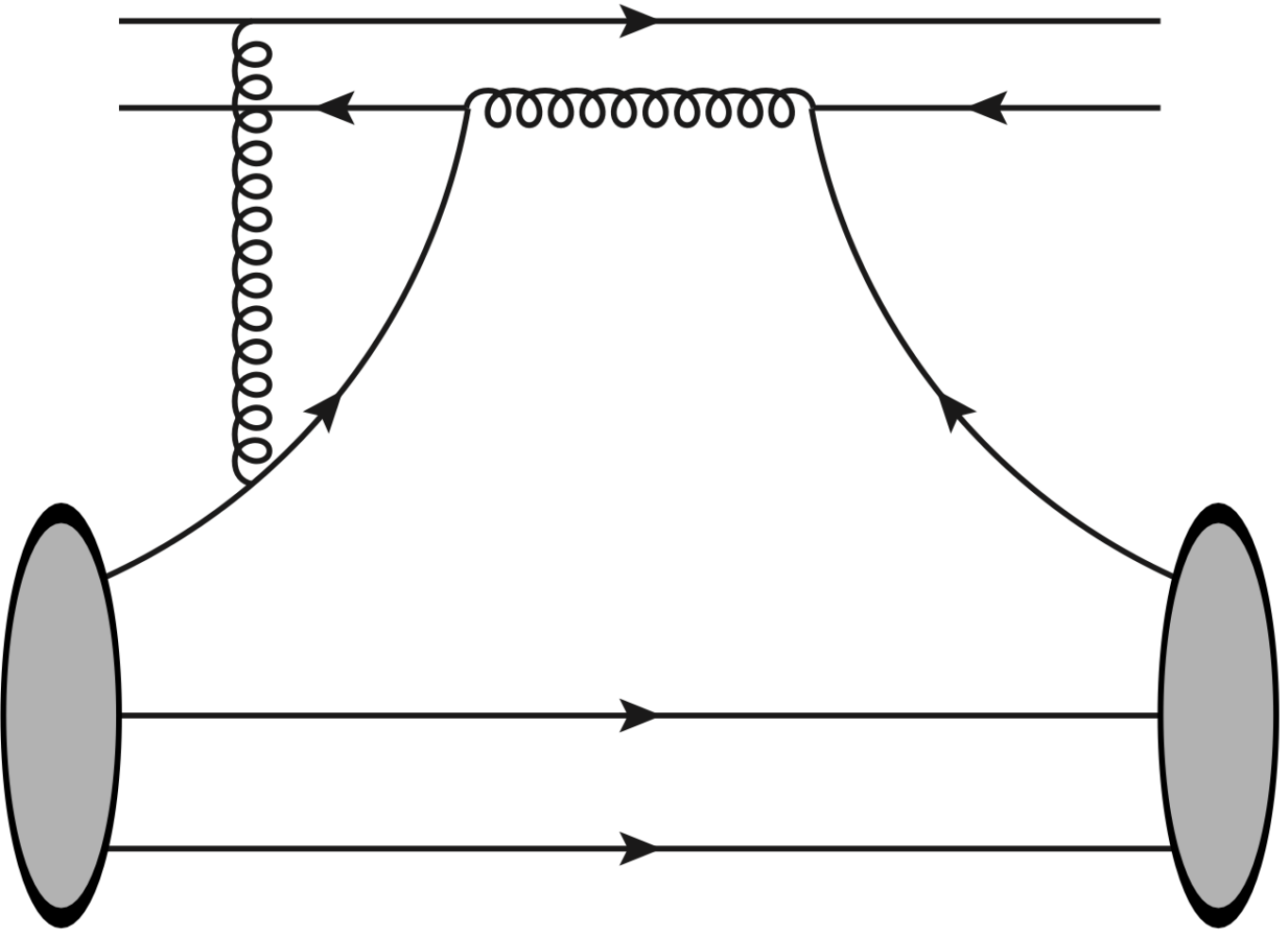}
        \caption{All the fields interact with one valence quark.}
        \label{fig:gqq_11}
    \end{subfigure}
    \;\;\;\;\;
    \begin{subfigure}[t]{0.45\textwidth}
        \centering
        \includegraphics[width = 0.6\textwidth]{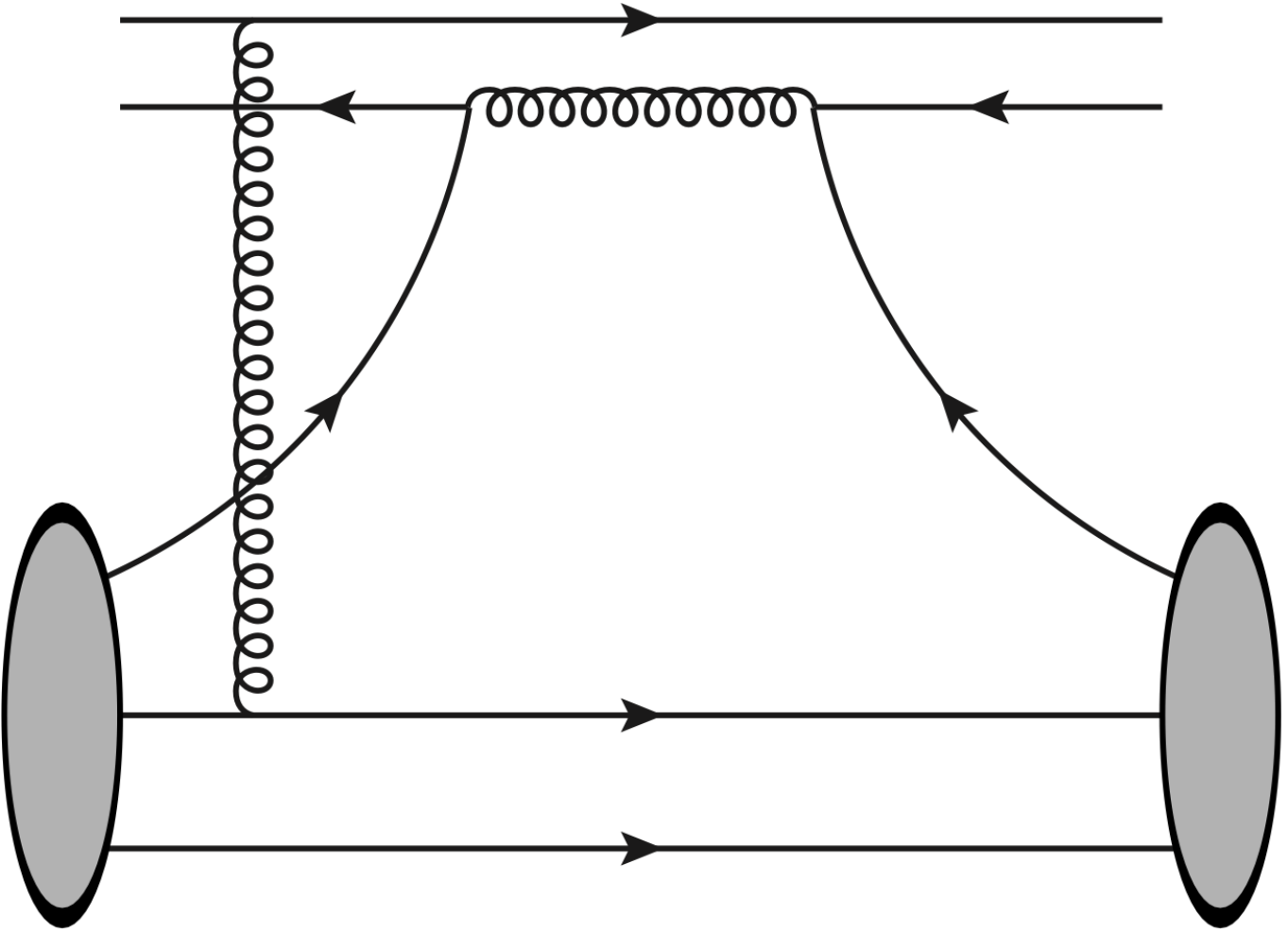}
        \caption{The quark and gluon fields interact with distinct valence quarks.}
        \label{fig:gqq_12}
    \end{subfigure}
    \caption{Diagrams illustrating possible ways the two quark fields from $V_{\yy}^{\text{q}[1]\dagger}$ and one gluon field from $V_{\xx}$ could interact with valence quarks in the proton target. The diagrams are drawn for the case where $x_3^-<x_1^-<x_2^-$, while the generalization to all the other values of $x_3^-$ is straightforward.}
    \label{fig:gqq}
\end{figure}

To calculate the contribution in Eq.~\eqref{qk20_Vx}, we employ the Poisson Equation for the covariant gauge
Yang-Mills field to trade the gluon field for the light-cone color charge density~\cite{Dumitru:2018vpr},
\begin{align}\label{qk21}
    \rho^a(x^-,\xx) & \equiv J^{+a}(x^-,\xx) = - \nabla^2A^{+a}(x^-,\xx) \, .
\end{align}
In the Fourier space, this gives
\begin{align}\label{qk22}
    A^{+a}(x^-,\xx) &= \int\frac{\dd^2 \pp}{(2\pi)^2} \, e^{i\pp\cdot\xx} \tilde{A}^{+a}(x^-,\pp) = \int\frac{\dd^2 \pp}{(2\pi)^2} \, e^{i\pp\cdot\xx} \frac{1}{p^2_{\perp}} \, \tilde{\rho}^a(x^-,\pp) \, .
\end{align}
Furthermore, the charge density can be written in terms of the quark fields as
\begin{align}\label{qk23}
    \rho^a(x^-,\xx) &= g\sum_{f'}\bar{\psi}^{f'}_{\ell\eta}(x^-,\xx) \, (\gamma^+)_{\eta\zeta}(t^a)_{\ell m}\psi^{f'}_{m\zeta}(x^-,\xx) \, .
\end{align}
Putting Eqs.~\eqref{qk22} and \eqref{qk23} together, we have
\begin{align}\label{qk24}
    A^{+a}(x^-,\xx) &= g\sum_{f'} \int\frac{\dd^2 \pp}{(2\pi)^2} \, \frac{1}{p^2_{\perp}} \int \dd^2 \xx' \, e^{i\pp\cdot(\xx-\xx')} \bar{\psi}^{f'}_{\ell\eta}(x^-,\xx') \, (\gamma^+)_{\eta\zeta}(t^a)_{\ell m}\psi^{f'}_{m\zeta}(x^-,\xx') \, . 
\end{align}
Then, plugging this result into Eq.~\eqref{qk20_Vx} yields
\begin{align}\label{qk25}
    &\frac{zs}{2N_c} \left\langle\text{T\,tr}\left[V_{\xx}V_{\yy}^{\text{q}[1]\dagger}\right]\right\rangle\Big|_{V_{\xx}} = - \frac{2i\alpha_s^2 \pi^2 P^+}{N_c^2} \, (t^a)_{ji} \, (t^a)_{\ell m} \, (\gamma^+\gamma_5)_{\beta\alpha} \, (\gamma^+)_{\eta\zeta}  \int\limits_{-\infty}^{\infty}\dd x_1^-\int\limits_{x_1^-}^{\infty}\dd x_2^-  \int\limits_{-\infty}^{\infty}\dd x_3^-      \\ 
    &\;\;\;\;\times \int\frac{\dd^2 \pp}{(2\pi)^2} \, \frac{1}{p^2_{\perp}} \int \dd^2\xx' \, e^{i\pp\cdot(\xx-\xx')} \sum_{f'} \left\langle \bar{\psi}^{f'}_{\ell\eta}(x_3^-,\xx')\, \psi^{f'}_{m\zeta}(x_3^-,\xx') \,\psi^f_{i\alpha}(x_1^-,\yy) \,\bar{\psi}^f_{j\beta}(x_2^-,\yy)  \right\rangle .  \notag
\end{align}
When evaluating this expression, one needs to be careful that the two quark fields with flavor $f'$ stemmed from one single gluon field. As such, they must act on the same valence quark inside the proton target. In total, as the proton only contains three valence quarks, this leaves us with two contributions shown in Figs.~\ref{fig:gqq}: (i) all four quark fields in Eq.~\eqref{qk25} act on the same valence quark, c.f. Fig.~\ref{fig:gqq_11}, and (ii) two quark fields with flavor $f$ act on one quark, while the remaining two quark fields with flavor $f'$ act on another distinct valence quark, c.f. Fig.~\ref{fig:gqq_12}. 

To proceed, we recall from Eq.~\eqref{qkfield} that the quark field is a linear combination of quark and antiquark ladder operators. With a close look at Eq.~\eqref{qk25}, any terms involving $\hat{d}$ and $\hat{d}^{\dagger}$ lead to either a disconnected diagram or a term proportional to $\text{tr}(t^a)=0$ for both case (i) and case (ii). Thus, we are left with the terms involving two quark creation operators and two quark annihilation operators.
\begin{align}\label{qk26}
    &\frac{zs}{2N_c} \left\langle\text{T\,tr}\left[V_{\xx}V_{\yy}^{\text{q}[1]\dagger}\right]\right\rangle\Big|_{V_{\xx}} = - \frac{2i\alpha_s^2 \pi^2 P^+}{N_c^2} \, (t^a)_{ji} \, (t^a)_{\ell m} \int\limits_{-\infty}^{\infty}\dd x_1^-\int\limits_{x_1^-}^{\infty}\dd x_2^-  \int\limits_{-\infty}^{\infty}\dd x_3^- \int\frac{\dd^2 \pp}{(2\pi)^2} \, \frac{1}{p^2_{\perp}} \int \dd^2 \xx'      \\ 
    &\;\;\;\;\times \int\frac{\dd^2 \pp_1 \, \dd p_1^+}{(2\pi)^3\sqrt{2p_1^+}} \int\frac{\dd^2 \pp_2 \, \dd p_2^+}{(2\pi)^3\sqrt{2p_2^+}} \int\frac{\dd^2 \pp_3 \, \dd p_3^+}{(2\pi)^3\sqrt{2p_3^+}} \int\frac{\dd^2 \pp_4 \, \dd p_4^+}{(2\pi)^3\sqrt{2p_4^+}} \, e^{-ip_1^+x_1^- + ip_2^+x_2^- - i(p_3^+ - p_4^+)\,x_3^- + i\pp\cdot\xx + i(\pp_1-\pp_2)\cdot\yy - i(\pp-\pp_3+\pp_4)\cdot\xx'}  \notag \\ 
    &\;\;\;\;\times \sum_{f'} \sum_{S,S'} S \left\langle \hat{b}^{f'\dagger}_{p_4,\ell,S'} \hat{b}^{f'}_{p_3,m,S'} \hat{b}^f_{p_1,i,S} \hat{b}^{f\dagger}_{p_2,j,S}\right\rangle ,  \notag
\end{align}
where we also used Eq.~\eqref{BL1}, together with another Brodsky-Lepage spinor matrix element~\cite{Lepage:1980fj},
\begin{align}\label{BL2}
&\bar{u}_{S_2}(p'_2)\,\gamma^+u_{S_1}(p_2) = 2\sqrt{p_2^+p'^+_2}\,\delta_{S_1,S_2} \, .
\end{align}
Then, with the proton state specified by Eqs.~\eqref{proton}, \eqref{eq:factorized-LCwf} and \eqref{qk11}, we obtain
\begin{align}\label{qk27}
    &\frac{zs}{2N_c} \,\text{Re}\left\langle\text{T\,tr}\left[V_{\xx}V_{\yy}^{\text{q}[1]\dagger}\right]\right\rangle\Big|_{V_{\xx}} = - \lim_{K\to P} \frac{2\alpha_s^2 \pi^2 C_F}{3N_c^2}\left[4\delta^{f,u}-\delta^{f,d}\right] \frac{4\pi P^+\delta(P^+-K^+)}{\braket{K|P}}  \; e^{i(\underline{P}-\underline{K})\cdot\yy}  \\
    &\;\;\;\;\;\;\;\;\times \int\frac{\dd^2 \pp}{(2\pi)^2} \, \frac{1}{p^2_{\perp}} \, e^{i\pp\cdot(\xx-\yy)} \int[\dd x_i] \int[\dd^2 \qq_i] \, \frac{1}{x_1}   \notag \\
    &\;\;\;\;\;\;\;\;\times \Phi^*(x_1,\qq_1-(1-x_1)\underline{P}+(1-x_1)\underline{K}; x_2,\qq_2+x_2\underline{P}-x_2\underline{K}; x_3,\qq_3+x_3\underline{P}-x_3\underline{K}) \, \Phi(x_i,\qq_i)      \notag \\
    &\;\;\;\;+ \lim_{K\to P} \frac{2\alpha_s^2 \pi^2 C_F}{9N_c} \left[4\delta^{f,u}-\delta^{f,d}\right] \frac{4\pi P^+\delta(P^+-K^+) }{\braket{K|P}}  \; e^{i(\underline{P}-\underline{K})\cdot\yy}  \int\frac{\dd^2 \pp}{(2\pi)^2} \, \frac{1}{p^2_{\perp}} \, e^{i\pp\cdot(\xx-\yy)} \int[\dd x_i] \int[\dd^2 \qq_i] \, \frac{1}{x_1}    \notag  \\  
    &\;\;\;\;\;\;\;\;\times  \Phi^*(x_1,\qq_1 - (1-x_1)\underline{P} + (1 - x_1)\underline{K} + \pp;x_2,\qq_2 + x_2\underline{P} - x_2\underline{K} - \pp;x_3,\qq_3 + x_3\underline{P} - x_3\underline{K}) \, \Phi(x_i,\qq_i) \, .  \notag
\end{align}
Finally, we plug this result into Eqs.~\eqref{Q_q} and \eqref{QNS_q} for the fundamental dipole amplitudes to get
\begin{align}\label{qk28}
    &Q_f^{\text{q}}(r_{\perp},zs) = Q_f^{\text{NS},\text{q}}(r_{\perp},zs) = - \frac{8\pi^2\alpha_s^2}{81}\left[4\delta^{f,u}-\delta^{f,d}\right] \int\frac{\dd^2\pp}{(2\pi)^2} \, \frac{1}{p^2_{\perp}} \, e^{-i\pp\cdot\rr} \int[\dd x_i] \int[\dd^2\qq_i] \, \frac{1}{x_1}    \\
    &\;\;\;\;\times \left[ |\Phi(x_i,\qq_i)|^2 - \Phi^*(x_1,\qq_1 + \pp;x_2,\qq_2 - \pp;x_3,\qq_3) \, \Phi(x_i,\qq_i) \right] + (r_{\perp} \text{- independent terms}) \, ,  \notag
\end{align}
where we also put $N_c=3$ explicitly. This gives the leading nonzero contribution to the quark-exchange term of these dipole amplitudes, which gives one of the terms in Eqs.~\eqref{Q_IC} and \eqref{QNS_IC}.

As for the adjoint dipole amplitude, $\wg^{\text{q}}(r_{\perp},zs)$, we revisit its operator form~\eqref{qk7} and expand each of the Wilson lines to order $\mathcal{O}(g)$. This gives
\begin{subequations}\label{qk29}
\begin{align}
    \frac{zs}{2N_c} \sum_f \left\langle\text{T\,tr}\left[V_{\xx}W_{\yy}^{\text{q}[1]\dagger}\right]\right\rangle\Big|_{V_{\xx}} &= \frac{ig\,\alpha_s \pi P^+}{4N_c} \, (t^a)_{ji} \, (\gamma^+\gamma_5)_{\beta\alpha} \int\limits_{-\infty}^{\infty}\dd x_1^-\int\limits_{x_1^-}^{\infty}\dd x_2^- \int\limits_{-\infty}^{\infty}\dd x_3^-      \label{qk29_Vx} \\ 
    &\;\;\;\;\times \sum_f \left\langle A^{+a}(x_3^-,\xx)\, \psi^f_{i\alpha}(x_1^-,\yy)\, \bar{\psi}^f_{j\beta}(x_2^-,\yy)  \right\rangle ,  \notag \\ 
    \frac{zs}{2N_c} \sum_f \left\langle\text{T\,tr}\left[V_{\xx}W_{\yy}^{\text{q}[1]\dagger}\right]\right\rangle\Big|_{V_{\yy}[-\infty,x_1^-]} &= - \frac{ig\,\alpha_s \pi P^+}{4N_c} \, (t^a)_{ji} \, (\gamma^+\gamma_5)_{\beta\alpha}  \int\limits_{-\infty}^{\infty}\dd x_1^-\int\limits_{x_1^-}^{\infty}\dd x_2^- \int\limits_{-\infty}^{x_1^-}\dd x_3^-   \label{qk29_Vyminus} \\ 
    &\;\;\;\;\times \sum_f \left\langle A^{+a}(x_3^-,\yy)\, \psi^f_{i\alpha}(x_1^-,\yy) \, \bar{\psi}^f_{j\beta}(x_2^-,\yy)  \right\rangle ,  \notag \\ 
    \frac{zs}{2N_c} \sum_f \left\langle\text{T\,tr}\left[V_{\xx}W_{\yy}^{\text{q}[1]\dagger}\right]\right\rangle\Big|_{V_{\yy}[x_2^-,\infty]} &= - \frac{ig\,\alpha_s \pi P^+}{4N_c} \, (t^a)_{ji} \, (\gamma^+\gamma_5)_{\beta\alpha}  \int\limits_{-\infty}^{\infty}\dd x_1^-\int\limits_{x_1^-}^{\infty}\dd x_2^- \int\limits_{x_2^-}^{\infty}\dd x_3^-     \label{qk29_Vyplus}  \\ 
    &\;\;\;\;\times \sum_f \left\langle \psi^f_{i\alpha}(x_1^-,\yy) \, \bar{\psi}^f_{j\beta}(x_2^-,\yy)\,A^{+a}(x_3^-,\yy)  \right\rangle .  \notag
\end{align}
\end{subequations}
This leads to exactly the same conclusion as that of $Q_f^{\text{q}}(r_{\perp},zs)$ and $Q_f^{\text{NS},\text{q}}(r_{\perp},zs)$, except for a constant prefactor. Hence, the term sensitive to the dipole separation, $\rr=\yy-\xx$, can be written as
\begin{align}\label{qk30}
    &\wg^{\text{q}}(r_{\perp},zs) = \frac{4\pi^2\alpha_s^2}{9} \int\frac{\dd^2\pp}{(2\pi)^2} \, \frac{1}{p^2_{\perp}} \, e^{-i\pp\cdot\rr} \int[\dd x_i] \int[\dd^2\qq_i] \, \frac{1}{x_1}    \\
    &\;\;\;\;\times \left[ |\Phi(x_i,\qq_i)|^2 - \Phi^*(x_1,\qq_1 + \pp;x_2,\qq_2 - \pp;x_3,\qq_3) \, \Phi(x_i,\qq_i) \right] + (r_{\perp} \text{- independent terms}) \, .  \notag
\end{align}
This gives the quark-exchange term in Eq.~\eqref{GT_IC}.

\subsection{ Gluon Exchange Diagrams at ${\cal O}(\as^2)$}
\label{sec:O-g4-gluon-Xchange}

In this Section, we consider the gluon-exchange contribution to each of the polarized dipole amplitudes. Here, both polarized-antiquark and polarized-quark terms are nonzero. Furthermore, there are gluon-exchange terms in both types of polarized Wilson lines. We will look at each case one-by-one.

\begin{figure}[t!]
    \centering
    \begin{subfigure}[t]{0.23\textwidth}
        \centering
        \includegraphics[width = \textwidth]{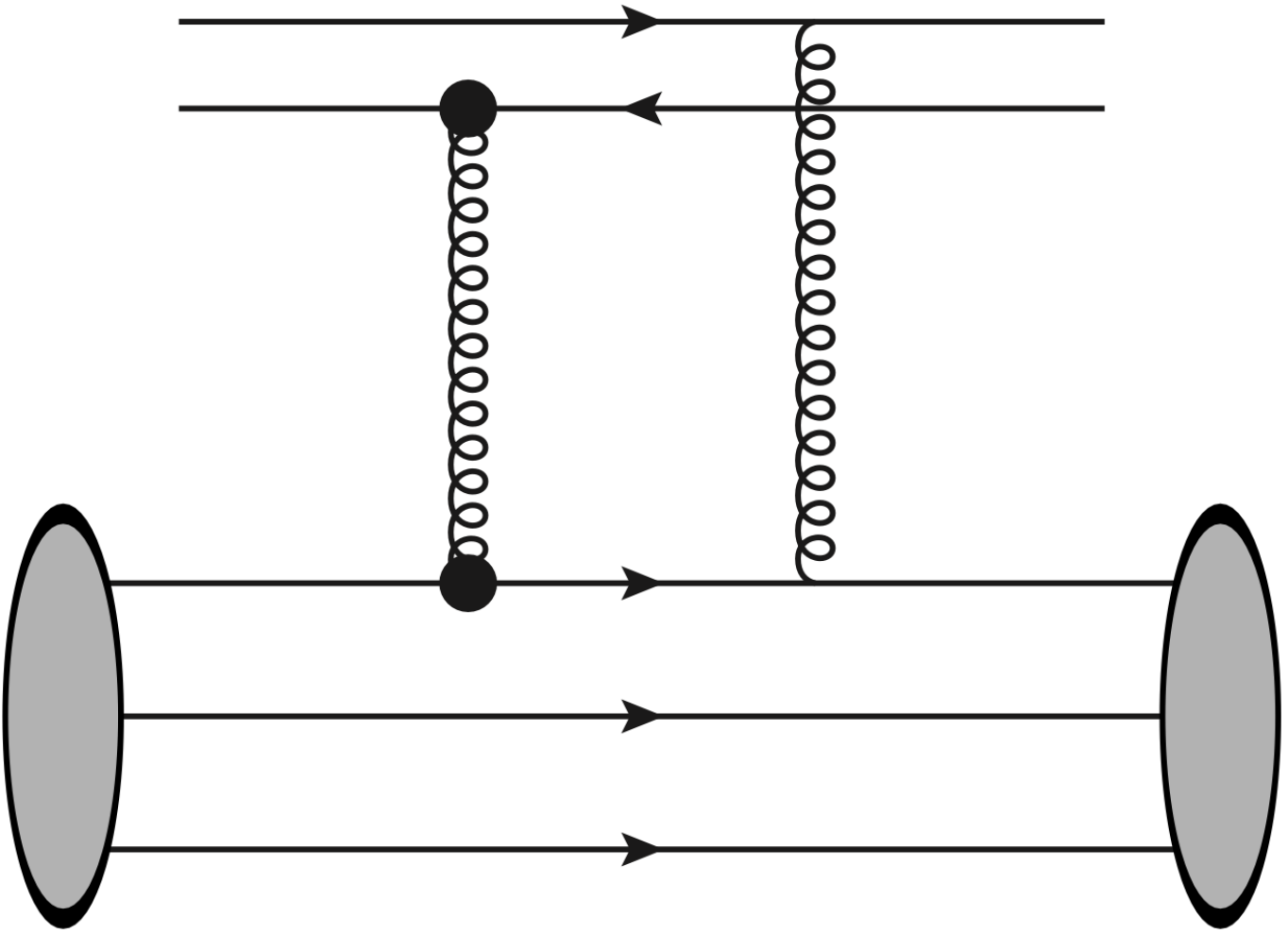}
        \caption{Emission from $V_{\xx}$, interacting with the same quark}
        \label{fig:F12_proj10_tg11}
    \end{subfigure}
    \;
    \begin{subfigure}[t]{0.23\textwidth}
        \centering
        \includegraphics[width = \textwidth]{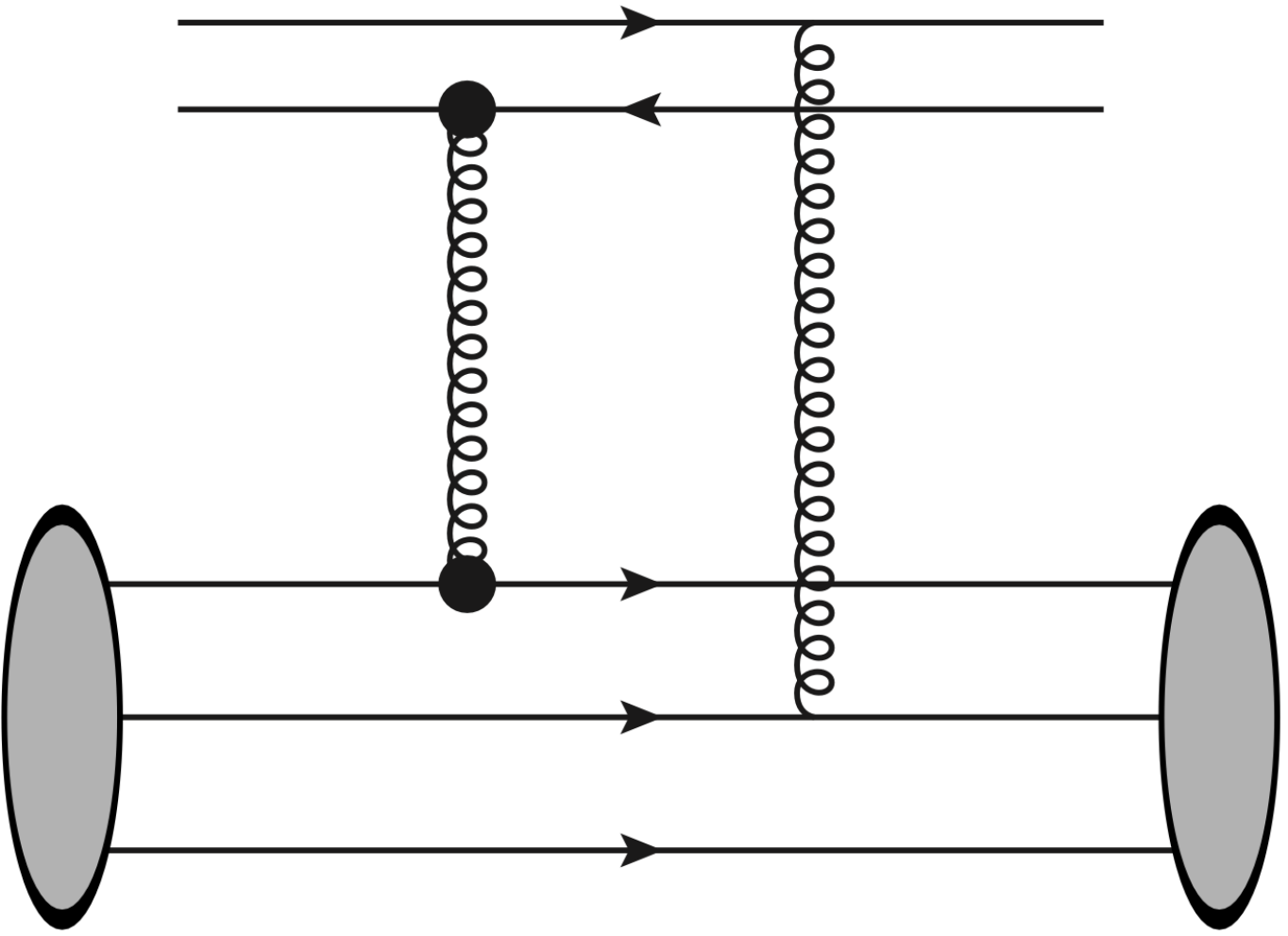}
        \caption{Emission from $V_{\xx}$, interacting with a distinct quark}
        \label{fig:F12_proj10_tg12}
    \end{subfigure}
    \;
    \begin{subfigure}[t]{0.23\textwidth}
        \centering
        \includegraphics[width = \textwidth]{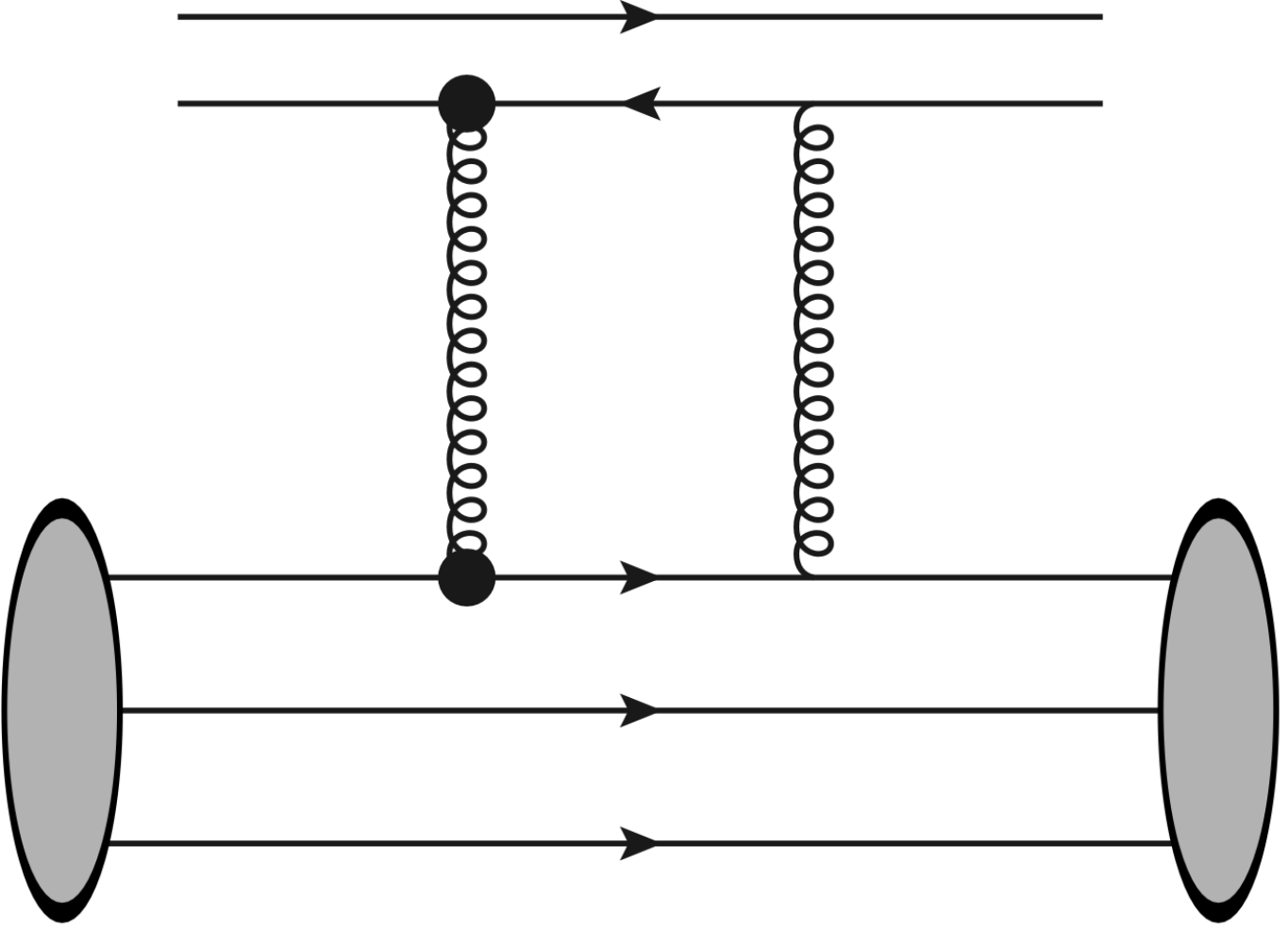}
        \caption{Emission from $V_{\yy}^{\text{G}[1]\dagger}$, interacting with the same quark}
        \label{fig:F12_proj11_tg11}
    \end{subfigure}
    \;
    \begin{subfigure}[t]{0.23\textwidth}
        \centering
        \includegraphics[width = \textwidth]{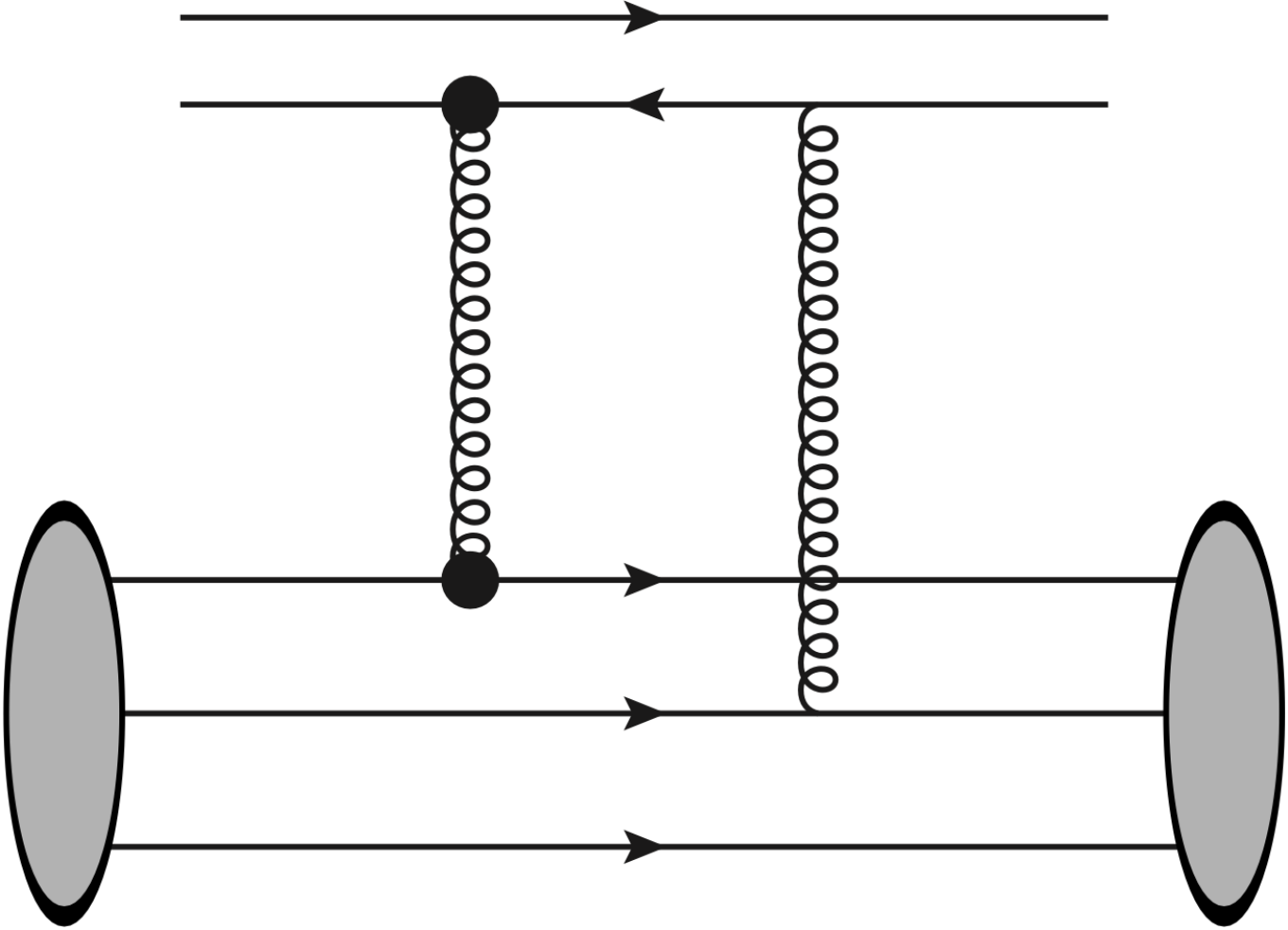}
        \caption{Emission from $V_{\yy}^{\text{G}[1]\dagger}$, interacting with a distinct quark}
        \label{fig:F12_proj11_tg12}
    \end{subfigure}
    \caption{Diagrams illustrating all possible gluon exchanges at order $\mathcal{O}(\as^2)$ from $V_{\yy}^{\text{G}[1]\dagger}$, with the extra power of $\as$ coming from an extra eikonal gluon emission from the dipole. In each diagram, the eikonal gluon is either emitted from $V_{\xx}$ or one of the semi-infinite Wilson lines that are parts of $V_{\yy}^{\text{G}[1]\dagger}$, and the eikonal gluon interacts with either the same valence quark or a distinct valence quark that interacts with the sub-eikonal gluon. The diagrams are drawn for the case where $x_2^->x_1^-$, while the generalization to the other ordering of light-cone times is straightforward.}
    \label{fig:F12}
\end{figure}

First, we start with the type-1 Wilson line, which contributes to $Q_f^{\text{G}}(r_{\perp},zs)$, $Q_f^{\text{NS},\text{G}}(r_{\perp},zs)$ and $\wg^{\text{G}}(r_{\perp},zs)$. With the help of Eq.~\eqref{V_subeik_G_1}, the term with polarized antiquark is 
\begin{align}\label{gl0}
    \frac{zs}{2N_c} \left\langle\text{T\,tr}\left[V_{\xx}V_{\yy}^{\text{G}[1]\dagger}\right]\right\rangle &= - \frac{igP^+}{2N_c} \int\limits_{-\infty}^{\infty}\dd x_1^- \left\langle\text{T\,tr}\left[V_{\xx} V_{\yy}[-\infty,x_1^-] \, F^{12}(x_1^-,\yy) \, V_{\yy}[x_1^-,\infty] \right]\right\rangle  \\
    &= \frac{igP^+}{2N_c} \, \epsilon^{ij} \int\limits_{-\infty}^{\infty}\dd x_1^- \left\langle\text{T\,tr}\left[V_{\xx} V_{\yy}[-\infty,x_1^-] \left( \frac{\partial}{\partial\yy^i}A^{ja}(x_1^-,\yy) \right) t^a \, V_{\yy}[x_1^-,\infty] \right]\right\rangle , \notag
\end{align}
where we neglected the four-gluon vertex. Naïvely, the leading-order contribution would follow from taking all the Wilson lines in Eq.~\eqref{gl0} to identity color matrices. However, such terms would be proportional to tr$(t^a)=0$. Thus, we need to expand one of the Wilson lines to order $\mathcal{O}(g)$. This gives
\begin{align}\label{gl1}
    \frac{zs}{2N_c} \left\langle\text{T\,tr}\left[V_{\xx}V_{\yy}^{\text{G}[1]\dagger}\right]\right\rangle &= - \frac{g^2P^+}{4N_c} \, \epsilon^{ij} \int\limits_{-\infty}^{\infty}\dd x_1^- \int\limits_{-\infty}^{\infty}\dd x_2^- \left\langle A^{+a}(x_2^-,\xx) \left( \frac{\partial}{\partial\yy^i}A^{ja}(x_1^-,\yy) \right) \right\rangle \\
    &\;\;\;\;+ \frac{g^2P^+}{4N_c} \, \epsilon^{ij} \int\limits_{-\infty}^{\infty}\dd x_1^- \int\limits_{-\infty}^{x_1^-}\dd x_2^- \left\langle A^{+a}(x_2^-,\yy) \left( \frac{\partial}{\partial\yy^i}A^{ja}(x_1^-,\yy) \right) \right\rangle \notag \\
    &\;\;\;\;+ \frac{g^2P^+}{4N_c} \, \epsilon^{ij} \int\limits_{-\infty}^{\infty}\dd x_1^- \int\limits_{x_1^-}^{\infty}\dd x_2^- \left\langle \left( \frac{\partial}{\partial\yy^i}A^{ja}(x_1^-,\yy) \right) A^{+a}(x_2^-,\yy)  \right\rangle + \mathcal{O}(g^3) \, , \notag
\end{align}
where for now we took the terms written explicitly to be of order $\mathcal{O}(g^2)$. However, below, once we write down the gluon fields in terms of color currents we will gain some additional factors of $g$. In Eq.~\eqref{gl1}, the first term comes from the $\mathcal{O}(g)$ correction to the unpolarized quark's Wilson line, $V_{\xx}$ (Figs.~\ref{fig:F12_proj10_tg11} and \ref{fig:F12_proj10_tg12}), while the two other terms come from the correction to one of the semi-infinite Wilson lines, $V_{\yy}[-\infty,x_1^-]$ or $V_{\yy}[x_1^-,\infty]$ (Figs.~\ref{fig:F12_proj11_tg11} and \ref{fig:F12_proj11_tg12}). Similarly, for the polarized quark term, we have
\begin{align}\label{gl1_cc}
    \frac{zs}{2N_c} \left\langle\text{T\,tr}\left[V_{\yy}^{\text{G}[1]}V_{\xx}^{\dagger}\right]\right\rangle &= \frac{igP^+}{2N_c} \int\limits_{-\infty}^{\infty}\dd x_1^- \left\langle\text{T\,tr}\left[V_{\yy}[\infty,x_1^-] \, F^{12}(x_1^-,\yy) \, V_{\yy}[x_1^-,-\infty] \, V_{\xx}^{\dagger} \right]\right\rangle  \\
    &= - \frac{g^2P^+}{4N_c} \, \epsilon^{ij} \int\limits_{-\infty}^{\infty}\dd x_1^- \int\limits_{-\infty}^{\infty}\dd x_2^- \left\langle \left( \frac{\partial}{\partial\yy^i}A^{ja}(x_1^-,\yy) \right) A^{+a}(x_2^-,\xx) \right\rangle \notag \\
    &\;\;\;\;+ \frac{g^2P^+}{4N_c} \, \epsilon^{ij} \int\limits_{-\infty}^{\infty}\dd x_1^- \int\limits_{-\infty}^{x_1^-}\dd x_2^-\left\langle  \left( \frac{\partial}{\partial\yy^i}A^{ja}(x_1^-,\yy) \right) A^{+a}(x_2^-,\yy) \right\rangle \notag \\
    &\;\;\;\;+ \frac{g^2P^+}{4N_c} \, \epsilon^{ij} \int\limits_{-\infty}^{\infty}\dd x_1^- \int\limits_{x_1^-}^{\infty}\dd x_2^-\left\langle A^{+a}(x_2^-,\yy)  \left( \frac{\partial}{\partial\yy^i}A^{ja}(x_1^-,\yy) \right)  \right\rangle  , \notag
\end{align}
again neglecting the four-gluon vertex. Then, altogether, we have that
\begin{align}\label{gl1_sum}
    &\frac{zs}{2N_c} \left\langle\text{T\,tr}\left[V_{\xx}V_{\yy}^{\text{G}[1]\dagger}\right] + \text{T\,tr}\left[V_{\yy}^{\text{G}[1]}V_{\xx}^{\dagger}\right]\right\rangle  \\
    &= \frac{g^2P^+}{4N_c} \, \epsilon^{ij} \int\limits_{-\infty}^{\infty}\dd x_1^- \int\limits_{-\infty}^{\infty}\dd x_2^- \left\langle \left\{ A^{+a}(x_2^-,\yy) - A^{+a}(x_2^-,\xx) , \frac{\partial}{\partial\yy^i}A^{ja}(x_1^-,\yy) \right\} \right\rangle . \notag
\end{align}

To proceed, we employ the Poisson field equation to relate, in the weak field limit, 
the covariant gauge transverse fields to color currents,
\begin{align}\label{gl2}
    &J^{ia}(x^-,\xx) = -\nabla^2 A^{ia}(x^-,\xx) \, .
\end{align}
Following the same steps outline in Eqs.~\eqref{qk21}--\eqref{qk24}, we obtain the following relation for transverse gluon fields,
\begin{align}\label{gl3}
    \frac{\partial}{\partial\yy^i}A^{ja}(y^-,\yy) &= ig\sum_{f} \int\frac{\dd^2 \pp}{(2\pi)^2} \, \frac{\pp^i}{p^2_{\perp}} \int \dd^2y' \, e^{i\pp\cdot(\yy-\yy')} \bar{\psi}^{f}_{\ell\eta}(y^-,\yy') \, (\gamma^j)_{\eta\zeta}(t^a)_{\ell m}\psi^{f}_{m\zeta}(y^-,\yy') \, . 
\end{align}
Then, together with Eq.~\eqref{qk24}, we rewrite Eq.~\eqref{gl1_sum} as
\begin{align}\label{gl4}
    &\frac{zs}{2N_c} \left\langle\text{T\,tr}\left[V_{\xx}V_{\yy}^{\text{G}[1]\dagger}\right] + \text{T\,tr}\left[V_{\yy}^{\text{G}[1]}V_{\xx}^{\dagger}\right]\right\rangle  \\
    &=  \frac{ig^4P^+}{4N_c} \, \epsilon^{ij} \, (\gamma^+)_{\alpha\beta} \, (\gamma^j)_{\eta\zeta} \, (t^a)_{\ell m} \, (t^a)_{pq}  \int\frac{\dd^2 \pp}{(2\pi)^2} \int\frac{\dd^2 \pp'}{(2\pi)^2} \, \frac{\pp^i}{p_{\perp}^2p'^2_{\perp}}  \int\limits_{-\infty}^{\infty}\dd x_1^- \int\limits_{-\infty}^{\infty}\dd x_2^- \sum_{f,f'} \notag \\
    &\;\;\;\;\times \left[ \int \dd^2 \xx' \int \dd^2 \yy' \, e^{i\pp\cdot(\yy-\yy') + i\pp'\cdot(\yy-\xx')}  \left\langle \left\{ \bar{\psi}^f_{\ell\eta}(x_1^-,\yy') \, \psi^f_{m\zeta}(x_1^-,\yy') , \bar{\psi}^{f'}_{p\alpha}(x_2^-,\xx') \, \psi^{f'}_{q\beta}(x_2^-,\xx') \right\} \right\rangle  \right.  \notag \\
    &\;\;\;\;\;\;\;\;- \left. \int \dd^2 \xx' \int \dd^2 \yy' \, e^{i\pp\cdot(\yy-\yy') + i\pp'\cdot(\xx-\xx')} \left\langle \left\{ \bar{\psi}^f_{\ell\eta}(x_1^-,\yy') \, \psi^f_{m\zeta}(x_1^-,\yy') , \bar{\psi}^{f'}_{p\alpha}(x_2^-,\xx') \, \psi^{f'}_{q\beta}(x_2^-,\xx') \right\} \right\rangle  \right] + \mathcal{O}(g^5) \, , \notag
\end{align}
where the quark fields of the same flavor, $f$ or $f'$, must act on the same valence quark in the proton. Similar to the quark case, all the nonzero terms that do not involve disconnected diagrams only contain quark ladder operators, $\hat{b}$ and $\hat{b}^{\dagger}$. By Eq.~\eqref{qkfield}, we have that
\begin{align}\label{gl5}
    &\frac{zs}{2N_c} \left\langle\text{T\,tr}\left[V_{\xx}V_{\yy}^{\text{G}[1]\dagger}\right] + \text{T\,tr}\left[V_{\yy}^{\text{G}[1]}V_{\xx}^{\dagger}\right]\right\rangle \\ 
    &=  \frac{ig^4P^+}{8N_c} \, \epsilon^{ij} \, (t^a)_{\ell m} \, (t^a)_{pq}  \int\frac{\dd^2 \pp}{(2\pi)^2} \int\frac{\dd^2 \pp'}{(2\pi)^2} \, \frac{\pp'^i}{p_{\perp}^2p'^2_{\perp}} \, e^{i(\pp+\pp')\cdot\yy} \left[1-e^{-i\pp\cdot(\yy-\xx)}\right]  \int\limits_{-\infty}^{\infty}\dd x_1^- \int\limits_{-\infty}^{\infty}\dd x_2^- \int \dd^2 \xx' \int \dd^2 \yy'   \notag \\
    &\;\;\times  \int\frac{\dd p_1^+ \dd^2 \pp_1}{(2\pi)^3 \sqrt{2p_1^+}} \int\frac{\dd p_2^+ \dd^2 \pp_2}{(2\pi)^3 \sqrt{2p_2^+}} \int\frac{\dd p_3^+ \dd^2 \pp_3}{(2\pi)^3 \sqrt{2p_3^+}} \int\frac{\dd p_4^+ \dd^2 \pp_4}{(2\pi)^3 \sqrt{2p_4^+}} \, e^{-i(p_1^+ - p_2^+)x_1^- - i(p_3^+ - p_4^+)x_2^- + i(\pp_1-\pp_2-\pp')\cdot\yy' + i(\pp_3-\pp_4-\pp)\cdot\xx'}  \notag \\
    &\;\;\times \sum_{f,f'} \sum_{S,S'} \left[\frac{\pp_1^j+iS\epsilon^{jk}\pp_1^k}{p_1^+} + \frac{\pp_2^j-iS\epsilon^{jk}\pp_2^k}{p_2^+} \right] \left\langle \left\{ \hat{b}^{f\dagger}_{p_2,\ell,S} \hat{b}^{f}_{p_1,m,S} , \hat{b}^{f'\dagger}_{p_4,p,S'} \hat{b}^{f'}_{p_3,q,S'} \right\} \right\rangle + \mathcal{O}(g^5) \, , \notag
\end{align}
where along the way we made use of another spinor product~\cite{Lepage:1980fj},
\begin{align}
    &\bar{u}_{S'}(p')\,\gamma^{j}u_{S}(p) = \delta_{SS'}\sqrt{p^+p'^+}\left[\frac{\pp^j+iS\epsilon^{jk}\pp^k}{p^+} + \frac{\pp'^j-iS\epsilon^{jk}\pp'^k}{p'^+}\right] .
\end{align}

Then, with the proton state specified in Eqs.~\eqref{proton}, \eqref{eq:factorized-LCwf} and \eqref{qk11}, we have that
\begin{align}\label{gl6}
    \frac{zs}{2N_c} \,&\text{Re}\left\langle\text{T\,tr}\left[V_{\xx}V_{\yy}^{\text{G}[1]\dagger}\right] + \text{T\,tr}\left[V_{\yy}^{\text{G}[1]}V_{\xx}^{\dagger}\right]\right\rangle = \lim_{K\to P} \frac{16\pi^2\as^2}{9} \int\frac{\dd^2 \pp}{(2\pi)^2} \int\frac{\dd^2 \pp'}{(2\pi)^2} \, \frac{1}{p^2_{\perp}} \, e^{i(\pp+\pp')\cdot\yy}\left[1-e^{- i\pp\cdot(\yy-\xx)}\right]       \\
    &\;\;\;\;\times \frac{2P^+ \, 2\pi\delta(P^+-K^+)  \, (2\pi)^2\delta^2(\underline{P} - \underline{K}- \pp - \pp')}{\braket{K|P}} \int [\dd x_i] \int [\dd^2 \qq_i] \, \frac{1}{x_1} \notag \\
    &\;\;\;\;\times \left[ \Phi^*(x_1,\qq_1+x_1\underline{P} - x_1\underline{K} -\pp-\pp';x_2,\qq_2 + x_2\underline{P} - x_2\underline{K};x_3,\qq_3 + x_3\underline{P} - x_3\underline{K}) \, \Phi(x_1,\qq_i) \right. \notag \\
    &\;\;\;\;\;\;\;\;- \left. \Phi^*(x_1,\qq_1+x_1\underline{P} - x_1\underline{K}-\pp';x_2,\qq_2+x_2\underline{P} - x_2\underline{K}-\pp;x_3,\qq_3 + x_3\underline{P} - x_3\underline{K}) \, \Phi(x_1,\qq_i) \right] , \notag
\end{align}
where we again had to include the contributions with the two gluon lines interacting with the same valence quark (Figs.~\ref{fig:F12_proj10_tg11} and \ref{fig:F12_proj11_tg11}) and distinct valence quarks (Figs.~\ref{fig:F12_proj10_tg12} and \ref{fig:F12_proj11_tg12}). Finally, we plug this result into Eq.~\eqref{Q_G} and obtain
\begin{align}\label{gl7}
    Q_f^{\text{G}}(r_{\perp},zs) = \frac{1}{3}\,\wg^{\text{G}}(r_{\perp},zs) &= \frac{16\pi^2\as^2}{9} \int\frac{\dd^2 \pp}{(2\pi)^2} \, \frac{1}{p^2_{\perp}} \left[1-e^{- i\pp\cdot\rr}\right]  \int [\dd x_i] \int [\dd^2 \qq_i] \, \frac{1}{x_1}  \\
    &\;\;\;\;\times \left[ |\Phi(x_1,\qq_i)|^2 - \Phi^*(x_1,\qq_1+\pp;x_2,\qq_2-\pp;x_3,\qq_3) \, \Phi(x_1,\qq_i) \right] , \notag
\end{align}
whose $r_{\perp}$-dependent terms proportional to $e^{-i\pp\cdot\rr}$ lead to the gluon-exchange contributions to Eqs.~\eqref{Q_IC} and \eqref{GT_IC}. A notable observation here is that the gluon-exchange term is flavor-independent, which makes physical sense because a gluon is expected to interact with light quarks the same way regardless of their flavors.

As for the flavor non-singlet term, we take the other linear combination of Eqs.~\eqref{gl1} and \eqref{gl1_cc}:
\begin{align}\label{gl8}
    \frac{zs}{2N_c} \left\langle\text{T\,tr}\left[V_{\xx}V_{\yy}^{\text{G}[1]\dagger}\right] - \text{T\,tr}\left[V_{\yy}^{\text{G}[1]}V_{\xx}^{\dagger}\right]\right\rangle &= - \frac{g^2P^+}{4N_c} \, \epsilon^{ij} \int\limits_{-\infty}^{\infty}\dd x_1^- \int\limits_{-\infty}^{\infty}\dd x_2^- \left\langle \left[ A^{+a}(x_2^-,\xx) , \frac{\partial}{\partial\yy^i}A^{ja}(x_1^-,\yy) \right]  \right\rangle  \\
    &\;\;\;\;+ \frac{g^2P^+}{4N_c} \, \epsilon^{ij} \int\limits_{-\infty}^{\infty}\dd x_1^- \int\limits_{-\infty}^{x_1^-}\dd x_2^- \left\langle \left[ A^{+a}(x_2^-,\yy) , \frac{\partial}{\partial\yy^i}A^{ja}(x_1^-,\yy) \right] \right\rangle \notag \\
    &\;\;\;\;- \frac{g^2P^+}{4N_c} \, \epsilon^{ij} \int\limits_{-\infty}^{\infty}\dd x_1^- \int\limits_{x_1^-}^{\infty}\dd x_2^- \left\langle \left[ A^{+a}(x_2^-,\yy) , \frac{\partial}{\partial\yy^i}A^{ja}(x_1^-,\yy) \right]   \right\rangle + \mathcal{O}(g^3) \, . \notag
\end{align}
Then, we follow the same steps outlined in Eqs.~\eqref{gl2} to \eqref{gl7}, obtain commutators of $\hat{b}^{\dagger}\hat{b}$ and eventually realize that the expression of Eq.~\eqref{gl8} vanishes. Thus, there is no gluon-exchange contribution to the flavor non-singlet dipole amplitudes, 
\begin{align}\label{gl9}
    Q_f^{\text{NS},\text{G}}(r_{\perp},zs) = 0 \, .
\end{align}
This result is physically consistent with the fact that gluon exchange should not be sensitive to the charge conjugation of the polarized (anti)quark in the dipole.

Finally, we turn our attention to the type-2 dipole amplitude, which only contains the gluon-exchange terms. Starting from Eqs.~\eqref{def_G2} and \eqref{ViG2}, we have that 
\begin{align}\label{gl10}
    G_2(r_{\perp},zs) &= \frac{zs}{2N_c}\, \frac{\epsilon^{ij}\rr^j}{r^2_{\perp}} \int \dd^2\left(\frac{\xx+\yy}{2}\right)   \left\langle\text{tr}\left[V_{\xx}^{\dagger}V_{\yy}^{i\text{G}[2]}\right] + \text{tr}\left[V_{\yy}^{i\text{G}[2]\dagger}V_{\xx} \right] \right\rangle \Big|_{r_\perp = |\yy-\xx|}  \\
    &= \frac{P^+}{4N_c}\, \frac{\epsilon^{ij}\rr^j}{r^2_{\perp}} \int \dd^2\left(\frac{\xx+\yy}{2}\right)  \int\limits_{-\infty}^{\infty}\dd x_1^- \left\langle\text{tr}\left[V_{\xx}^{\dagger} V_{\yy}[\infty,x_1^-] \left(\vec{D}^i(x_1^-,\yy) - \cev{D}^i(x_1^-,\yy)\right) V_{\yy}[x_1^-,-\infty] \right]  \right\rangle \Big|_{r_\perp = |\yy-\xx|}  \notag \\
    &\;\;\;\;- \frac{P^+}{4N_c}\, \frac{\epsilon^{ij}\rr^j}{r^2_{\perp}} \int \dd^2\left(\frac{\xx+\yy}{2}\right)  \int\limits_{-\infty}^{\infty}\dd x_1^- \left\langle \text{tr}\left[V_{\yy}[-\infty,x_1^-] \left(\vec{D}^i(x_1^-,\yy) - \cev{D}^i(x_1^-,\yy)\right) V_{\yy}[x_1^-,\infty] \, V_{\xx} \right] \right\rangle \Big|_{r_\perp = |\yy-\xx|} \, ,    \notag
\end{align}
where we recall that $\vec{D}^{\mu} = \vec{\partial}^{\mu} - igA^{\mu}$ and $\cev{D}^{\mu} = \cev{\partial}^{\mu} + igA^{\mu}$. To proceed, we first write down the transverse position derivative of a partial light-cone Wilson line,
\begin{align}\label{gl11}
    &\frac{\partial}{\partial \yy^i}V_{\yy}[a^-,b^-] = ig\int\limits_{b^-}^{a^-}\dd x_2^- \, V_{\yy}[a^-,x_2^-] \, F^{+i}(x_2^-,\yy) \, V_{\yy}[x_2^-,b^-] \, .
\end{align}
In Eq.~\eqref{gl11}, only the term $\sim \partial^iA^+$ is significant, while all the other terms would turn out to be sub-sub-eikonal and beyond the sub-eikonal order considered in small-$x$ helicity evolution~\cite{Cougoulic:2022gbk}. Thus, keeping only the relevant terms, we have that
\begin{align}\label{gl12}
    G_2(r_{\perp},zs) &= \frac{g^2P^+}{4N_c}\, \frac{\epsilon^{ij}\rr^j}{r^2_{\perp}} \int \dd^2\left(\frac{\xx+\yy}{2}\right)  \int\limits_{-\infty}^{\infty}\dd x_1^- \int\limits_{-\infty}^{\infty}\dd x_2^- \\
    &\;\;\;\;\;\times \left\langle \left\{ A^{ia}(x_1^-,\yy) - x_1^-\partial^iA^{+a}(x_1^-,\yy) , A^{+a}(x_2^-,\yy) - A^{+a}(x_2^-,\xx) \right\} \right\rangle \Big|_{r_\perp = |\yy-\xx|} \, , \notag   
\end{align}
where in each term we also expanded one of the eikonal Wilson lines to order $\mathcal{O}(g)$. Then, we write each gluon field in terms of the color current and subsequently the quark fields, 
\begin{subequations}\label{gl13}
\begin{align}
    -\partial^iA^{+a}(x_1^-,\yy) &= ig\sum_{f} \int\frac{\dd^2\pp}{(2\pi)^2} \, \frac{\pp^i}{p^2_{\perp}} \int \dd^2 \yy' \, e^{i\pp\cdot(\yy-\yy')} \bar{\psi}^{f}_{\ell\eta}(x_1^-,\yy') \, (\gamma^+)_{\eta\zeta}(t^a)_{\ell m}\psi^{f}_{m\zeta}(x_1^-,\yy') \, \\
    A^{ia}(x_1^-,\yy) &= g\sum_{f} \int\frac{\dd^2\pp}{(2\pi)^2} \, \frac{1}{p^2_{\perp}} \int \dd^2\yy' \, e^{i\pp\cdot(\yy-\yy')} \bar{\psi}^{f}_{\ell\eta}(x_1^-,\yy') \, (\gamma^i)_{\eta\zeta}(t^a)_{\ell m}\psi^{f}_{m\zeta}(x_1^-,\yy') \, . 
\end{align}
\end{subequations}
Plugging these fields together with Eq.~\eqref{qk24} into Eq.~\eqref{gl12}, we obtain
\begin{align}\label{gl14}
    &G_2(r_{\perp},zs) = - \frac{g^4P^+}{4N_c} \, (\gamma^+)_{\alpha\beta} \, (t^a)_{\ell m} \, (t^a)_{pq} \, \frac{\epsilon^{ij}\rr^i}{r^2_{\perp}} \int \dd^2\left(\frac{\xx+\yy}{2}\right)  \int\limits_{-\infty}^{\infty}\dd x_1^- \int\limits_{-\infty}^{\infty}\dd x_2^- \int\frac{\dd^2\pp}{(2\pi)^2} \int\frac{\dd^2\pp'}{(2\pi)^2} \, \frac{(\gamma^j + ix_1^-\pp^j\gamma^+)_{\eta\zeta}}{p^2_{\perp}p'^2_{\perp}} \notag \\
    &\;\;\;\;\times \sum_{f,f'}\left[\int \dd^2\xx' \int \dd^2\yy' \, e^{i\pp\cdot(\yy-\yy') + i\pp'\cdot(\yy-\xx')} \left\langle \left\{ \bar{\psi}^f_{\ell\eta}(x_1^-,\yy') \, \psi^f_{m\zeta}(x_1^-,\yy') , \bar{\psi}^{f'}_{p\alpha}(x_2^-,\xx') \, \psi^{f'}_{q\beta}(x_2^-,\xx') \right\} \right\rangle \right.  \\
    &\;\;\;\;\;\;\;\;\;- \left. \int \dd^2\xx' \int \dd^2\yy' \, e^{i\pp\cdot(\yy-\yy') + i\pp'\cdot(\xx-\xx')} \left\langle \left\{ \bar{\psi}^f_{\ell\eta}(x_1^-,\yy') \, \psi^f_{m\zeta}(x_1^-,\yy') , \bar{\psi}^{f'}_{p\alpha}(x_2^-,\xx') \, \psi^{f'}_{q\beta}(x_2^-,\xx') \right\} \right\rangle \right] \Big|_{r_\perp = |\yy-\xx|}  + \mathcal{O}(g^5) \, .  \notag
\end{align}

Consider the term proportional to $(\gamma^+)_{\eta\zeta}$ in Eq.~\eqref{gl14}. Upon plugging in the quark fields in terms of the creation and annihilation operators, we would obtain two powers of $\bar{u}_{S'}\gamma^+u_S \sim \delta_{SS'}$ and eventually end up with an expression $\sim\sum_{\mathcal{S}_L}\mathcal{S}_L = 0$. Thus, the term vanishes. This makes physical sense because the term is known to correspond to the color charge correlator, $\left\langle \hat{\rho}^a\hat{\rho}^a\right\rangle$, which is not sensitive to helicity. As for the term proportional to $(\gamma^j)_{\eta\zeta}$, it is remarkably similar to Eq.~\eqref{gl4}. This allows us to read off the result from Eq.~\eqref{gl6} and obtain
\begin{align}\label{gl15}
    &G_2(r_{\perp},zs) = \frac{16\pi^2\as^2}{9} \int \dd^2\left(\frac{\xx+\yy}{2}\right) \int\frac{\dd^2 \pp}{(2\pi)^2} \int\frac{\dd^2 \pp'}{(2\pi)^2} \, \frac{i(\pp'\cdot\rr)}{p^2_{\perp}p'^2_{\perp}r^2_{\perp}} \, e^{i(\pp+\pp')\cdot\yy}\left[1-e^{- i\pp\cdot \rr}\right]      \\
    &\;\;\;\;\times \lim_{K\to P}  \frac{2P^+ \, 2\pi\delta(P^+-K^+)  \, (2\pi)^2\delta^2(\underline{P} - \underline{K}- \pp - \pp')}{\braket{K|P}} \int [\dd x_i] \int [\dd^2 \qq_i] \, \frac{1}{x_1} \notag \\
    &\;\;\;\;\times \left[ \Phi^*(x_1,\qq_1+x_1\underline{P} - x_1\underline{K} -\pp-\pp';x_2,\qq_2 + x_2\underline{P} - x_2\underline{K};x_3,\qq_3 + x_3\underline{P} - x_3\underline{K}) \, \Phi(x_1,\qq_i) \right. \notag \\
    &\;\;\;\;\;\;\;\;- \left. \Phi^*(x_1,\qq_1+x_1\underline{P} - x_1\underline{K}-\pp';x_2,\qq_2+x_2\underline{P} - x_2\underline{K}-\pp;x_3,\qq_3 + x_3\underline{P} - x_3\underline{K}) \, \Phi(x_1,\qq_i) \right] \Big|_{r_\perp = |\yy-\xx|}  \notag \\
    &= - \frac{16\pi^2\as^2}{9}  \int\frac{\dd^2 \pp}{(2\pi)^2} \, \frac{i(\pp\cdot\rr)}{p^4_{\perp}r^2_{\perp}} \left[1-e^{- i\pp\cdot\rr}\right] \int [\dd x_i] \int [\dd^2 \qq_i] \, \frac{1}{x_1} \left[ |\Phi(x_1,\qq_i)|^2 - \Phi^*(x_1,\qq_1+\pp;x_2,\qq_2-\pp;x_3,\qq_3) \, \Phi(x_1,\qq_i) \right] . \notag 
\end{align}
This gives the moderate-$x$ expression written in Eq.~\eqref{G2_IC} for the type-2 polarized dipole amplitude according to the valence quark model. Although both terms appear to be sensitive to the dipole size, $r_{\perp}$, it appears that the term without the Fourier factor, $e^{-i\pp\cdot\rr}$, would vanish because the term, $\int [\dd x_i] \int [\dd^2 \qq_i] \, \frac{1}{x_1} \, \Phi^*(x_1,\qq_1+\pp;x_2,\qq_2-\pp;x_3,\qq_3) \, \Phi(x_1,\qq_i)$, only depends on $p^2_{\perp}$ and not the transverse direction of $\pp$ itself. The reason for this follows from the fact that the momentum-space wave function, $\Phi$, is totally symmetric under exchange of any pair of quarks.

\section{Analytic Parametrization of Polarized Dipole Amplitudes}
\label{sec:HO-LCwf}

In this Appendix, we determine the numerical values of polarized dipole amplitudes in Eqs.~\eqref{IC_summary} as a function of dipole size, $r_{\perp}$, starting from the Gaussian momentum-space wave function~\eqref{Phi} of the proton and the analytic expression~\eqref{IC2} of $F(x_i;p_{\perp})$. Essentially, there are two types of integrals to evaluate, one for Eqs.~\eqref{Q_IC}--\eqref{GT_IC} and another for Eq.~\eqref{G2_IC}. They differ in the functions of $\pp$ and $\rr$ in the integral over $\pp$. For the first type of integrals relevant to type-1 dipole amplitudes, we have
\begin{align}\label{IC4}
    &\int\frac{\dd^2 \pp}{(2\pi)^2} \, \frac{1}{p^2_{\perp}} \, e^{-i\pp\cdot\rr}  \, F(x_i;\pp)  \\
    &= \frac{\mathcal{N}^2 \beta^4}{32\pi^3}\,x_1x_2x_3\,\exp\left[- \frac{M^2}{\beta^2}\sum_{i=1}^3 \frac{1}{x_i}\right] \int \frac{\dd p_{\perp}}{p_{\perp}} \, J_0(p_{\perp}r_{\perp}) \left\{1- \exp\left[- \frac{p^2_{\perp}}{4\beta^2}\left(\frac{1}{x_1}+\frac{1}{x_2}\right)\right] \right\} \notag \\
    &= \frac{\mathcal{N}^2 \beta^4}{64\pi^3}\,x_1x_2x_3\,\exp\left[- \frac{M^2}{\beta^2}\sum_{i=1}^3 \frac{1}{x_i}\right] \Gamma\left( 0 , \frac{x_1x_2}{(x_1+x_2)} \, \beta^2r^2_{\perp} \right) , \notag
\end{align}
where we used the fact that
\begin{align}\label{IC5}
    &\int\limits_0^{2\pi}\dd\theta\, e^{-ia\cos\theta} = 2\pi\, J_0(a) \, ,
\end{align}
for some constant, $a > 0$. Here, $J_0$ is the Bessel function of the first kind, and $\Gamma$ is the incomplete gamma function, which is defined as
\begin{align}\label{IC6}
    &\Gamma(s,x) = \int\limits_x^{\infty}\dd t \; t^{s-1}e^{-t} \, .
\end{align}
Then, to compute Eqs.~\eqref{Q_IC}--\eqref{GT_IC}, we need to evaluate the integral, 
\begin{align}\label{IC7}
    I_1(r_{\perp}) &= \int [\dd x_i] \, \frac{1}{x_1} \int\frac{\dd^2 \pp}{(2\pi)^2} \, \frac{1}{p^2_{\perp}} \, e^{-i\pp\cdot\rr}  \, F(x_i;\pp) \\
    &= \frac{\mathcal{N}^2 \beta^4}{(4\pi)^5} \int\limits_0^1 \dd x_1 \int\limits_0^{1-x_1} \dd x_2 \, x_2(1-x_1-x_2)\,\exp\left\{- \frac{M^2}{\beta^2}\left[\frac{1}{x_1} + \frac{1}{x_2} + \frac{1}{(1-x_1-x_2)}\right]\right\} \Gamma\left( 0 , \frac{x_1x_2}{(x_1+x_2)} \, \beta^2r^2_{\perp} \right) , \notag
\end{align}
which need to be performed numerically. 

As for the integral over $\pp$ in Eq.~\eqref{G2_IC}, we obtain
\begin{align}\label{IC8}
    &\int\frac{\dd^2 \pp}{(2\pi)^2} \, \frac{i(\pp\cdot\rr)}{p^4_{\perp}r^2_{\perp}} \, e^{- i\pp\cdot\rr}  \, F(x_i;\pp) \\
    &= \frac{\mathcal{N}^2 \beta^4}{32\pi^3}\,x_1x_2x_3\,\exp\left[- \frac{M^2}{\beta^2}\sum_{i=1}^3 \frac{1}{x_i}\right] \frac{1}{r_{\perp}} \int\frac{\dd p_{\perp}}{p^2_{\perp}} \, J_1(p_{\perp}r_{\perp}) \left\{ 1-\exp\left[- \frac{p^2_{\perp}}{4\beta^2}\left(\frac{1}{x_1}+\frac{1}{x_2}\right)\right] \right\}       \notag \\
    &=  \frac{\mathcal{N}^2 \beta^4}{128\pi^3}\,x_1x_2x_3\,\exp\left[- \frac{M^2}{\beta^2}\sum_{i=1}^3 \frac{1}{x_i}\right] \left\{ \Gamma\left( 0 , \frac{x_1x_2}{(x_1+x_2)} \, \beta^2r^2_{\perp} \right) + \frac{1}{\beta^2r^2_{\perp}}\left(\frac{1}{x_1}+\frac{1}{x_2}\right)\left[1 - \exp\left(-\frac{x_1x_2}{(x_1+x_2)} \, \beta^2r^2_{\perp}\right)\right] \right\} ,     \notag
\end{align}
where along the way we used
\begin{align}\label{IC9}
    &\int\limits_0^{2\pi}\dd\theta\,i\cos\theta \, e^{-ia\cos\theta} = -\frac{\dd}{\dd a} \int\limits_0^{2\pi}\dd\theta\, e^{-ia\cos\theta} = 2\pi \, J_1(a) \, ,
\end{align}
for $a>0$. Then, the final result for Eq.~\eqref{G2_IC} will follow from the integral,
\begin{align}\label{IC10}
    I_2(r_{\perp}) &= \int [\dd x_i] \, \frac{1}{x_1}  \int\frac{\dd^2 \pp}{(2\pi)^2} \, \frac{i(\pp\cdot\rr)}{p^4_{\perp}r^2_{\perp}} \, e^{- i\pp\cdot\rr}  \, F(x_i;\pp) \\
    &=  \frac{\mathcal{N}^2 \beta^4}{2(4\pi)^5} \int\limits_0^1 \dd x_1 \int\limits_0^{1-x_1} \dd x_2 \, x_2(1-x_1-x_2) \,\exp\left\{- \frac{M^2}{\beta^2}\left[\frac{1}{x_1} + \frac{1}{x_2} + \frac{1}{(1-x_1-x_2)}\right]\right\}  \notag \\
    &\;\;\;\;\;\times \left\{ \Gamma\left( 0 , \frac{x_1x_2}{(x_1+x_2)} \, \beta^2r^2_{\perp} \right) + \frac{1}{\beta^2r^2_{\perp}}\left(\frac{1}{x_1}+\frac{1}{x_2}\right)\left[1 - \exp\left(-\frac{x_1x_2}{(x_1+x_2)} \, \beta^2r^2_{\perp}\right)\right] \right\} ,     \notag
\end{align}
which again has to be evaluated numerically.

Before we proceed, a necessary ingredient essential to the computation is the normalization constant, $\mathcal{N}$, for the wave function. Starting from Eq.~\eqref{psi_norm}, we have that
\begin{align}\label{phi_norm}
    1 &= \int[\dd x_i] \int[\dd^2\qq_i] \; |\Phi(x_i,\qq_i)|^2 \sum_{\{f_1,f_2,f_3\}=\{u,u,d\}} \sum_{\sigma_1,\sigma_2,\sigma_3} \left|S_{\mathcal{S}_L}(\sigma_1,f_1;\sigma_2,f_2;\sigma_3,f_3) \right|^2   \\
    &= \mathcal{N}^2 \int[\dd x_i] \int[\dd^2 \qq_i] \, \exp\left[-\frac{1}{\beta^2}\sum_{i=1}^3\frac{q^2_{i\perp}+M^2}{x_i}\right]      \notag \\
    &= \frac{\mathcal{N}^2\beta^4}{16\pi^2} \int\limits_0^1\dd x_1 \int\limits_0^{1-x_1}\dd x_2 \, x_1x_2(1-x_1-x_2) \, \exp\left\{- \frac{M^2}{\beta^2}\left[\frac{1}{x_1} + \frac{1}{x_2} + \frac{1}{(1-x_1-x_2)}\right]\right\} .  \notag
\end{align}
This integral can be evaluated numerically. With $M = 0.26$ GeV and $\beta = 0.55$ GeV as determined in~\cite{Brodsky:1994fz}, one obtains $\mathcal{N} = 845$ fm$^2$.

\begin{figure}[t!]
    \centering
    \begin{subfigure}[t]{0.49\textwidth}
        \centering
        \includegraphics[width = \textwidth]{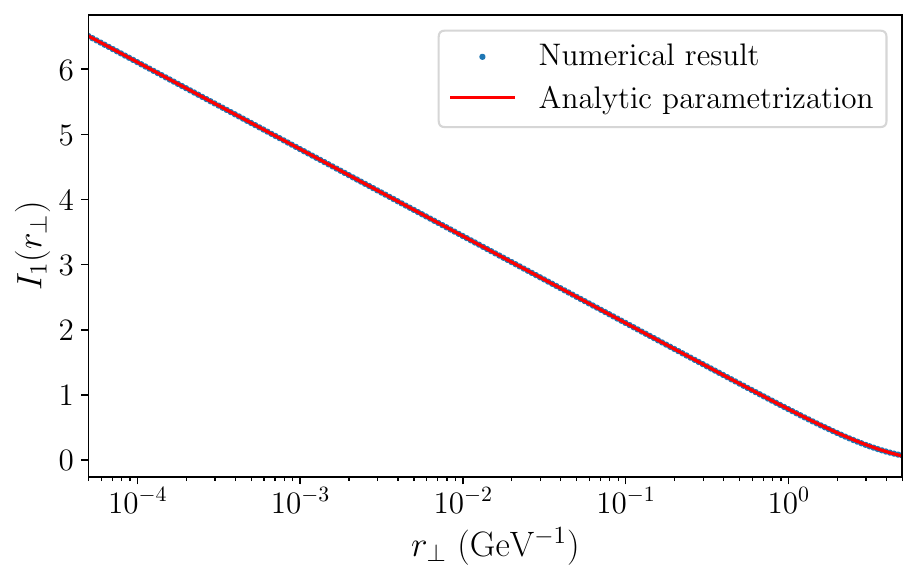}
        \caption{Full range}
        \label{fig:I1_res_model}
    \end{subfigure}
    \;
    \begin{subfigure}[t]{0.49\textwidth}
        \centering
        \includegraphics[width = \textwidth]{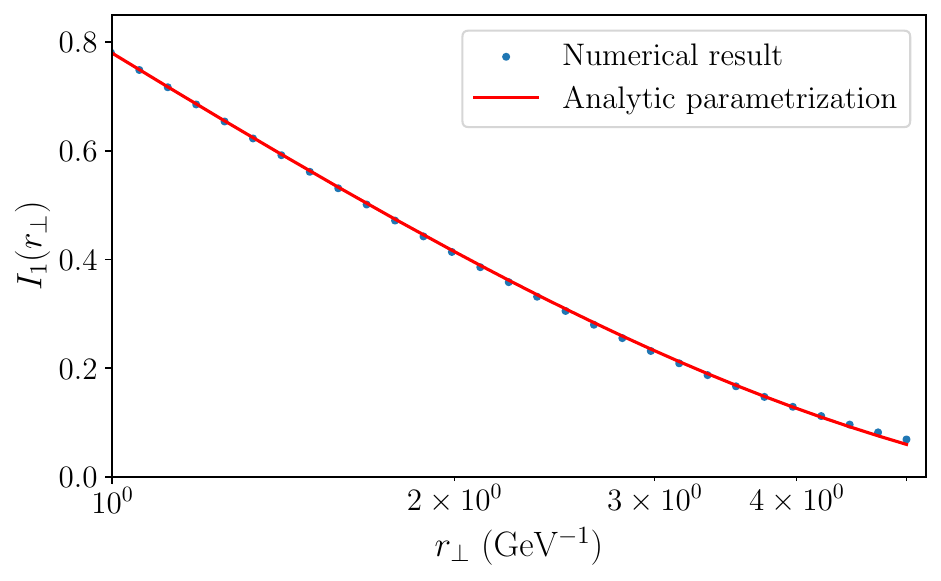}
        \caption{Zoomed in for 1 GeV$^{-1} \leq r_{\perp}\leq$ 5 GeV$^{-1}$}
        \label{fig:I1_res_model_zoomed}
    \end{subfigure}
    \caption{The integral $I_1(r_{\perp})$, obtained by numerically evaluating  Eq.~\eqref{IC7} (blue dots), as a function of dipole size $r_{\perp}$. The full result is compared to the analytical ansatz~\eqref{I1_ansatz} with parameters given in Eq.~\eqref{I1_params} (red lines).}
    \label{fig:I1_fit}
\end{figure}

Then, for a given value of $r_{\perp}$, we numerically evaluate Eq.~\eqref{IC7} using \texttt{scipy.integrate.dblquad}~\cite{2020SciPy-NMeth}. The results are shown as blue dots in Fig.~\ref{fig:I1_res_model} for 200 values of $r_{\perp}$ between 5 GeV$^{-1}$ ($\Lambda_{\text{QCD}}$ scale) and $5\times 10^{-5}$ GeV$^{-1}$ (LHC energy scale), equally spaced in the logarithmic scale. The plot displays a clear linear pattern that breaks down as the dipole size increases to approach the strong interaction scale, $\Lambda_{\text{QCD}}^{-1}$. This inspires the ansatz that $I_1(r_{\perp})$ is a transverse logarithm with an infrared regulator,
\begin{align}\label{I1_ansatz}
    &I_1(r_{\perp}) = a_1 \ln\left(\frac{1}{r^2_{\perp}\Lambda^2_1} + b_1\right) .
\end{align}
Then, we perform a fit using \texttt{scipy.optimize.curve\_fit}~\cite{2020SciPy-NMeth} and find that 
\begin{align}\label{I1_params}
    a_1 = 0.290\;,\;\;\;\;\Lambda_1 = 0.267~\text{GeV}\;\;\;\;\text{and}\;\;\;\;b_1 = 0.668,
\end{align}
with the uncertainties dominated by those coming from the available accuracies of $M$ and $\beta$. Indeed, $\Lambda_1$ is close to $\Lambda_{\text{QCD}}$. Here, it is reasonable to relate $a_1$ to $\overline{x^{-1}} = 3.64$, which has the physical meaning as the average of the inverse light-cone momentum fraction of a valence quark in the proton, c.f. Eq.~\eqref{x-1bar}. To our accuracy, we have $a_1=\overline{x^{-1}}/4\pi$. Qualitatively, Eq.~\eqref{I1_ansatz} is an excellent fit, as evident in Figs.~\ref{fig:I1_fit}, in which the fitted function is shown to reproduce the numerical results well, even in the large-dipole regime where the logarithmic behavior breaks down. 

\begin{figure}[t!]
    \centering
    \begin{subfigure}[t]{0.49\textwidth}
        \centering
        \includegraphics[width = \textwidth]{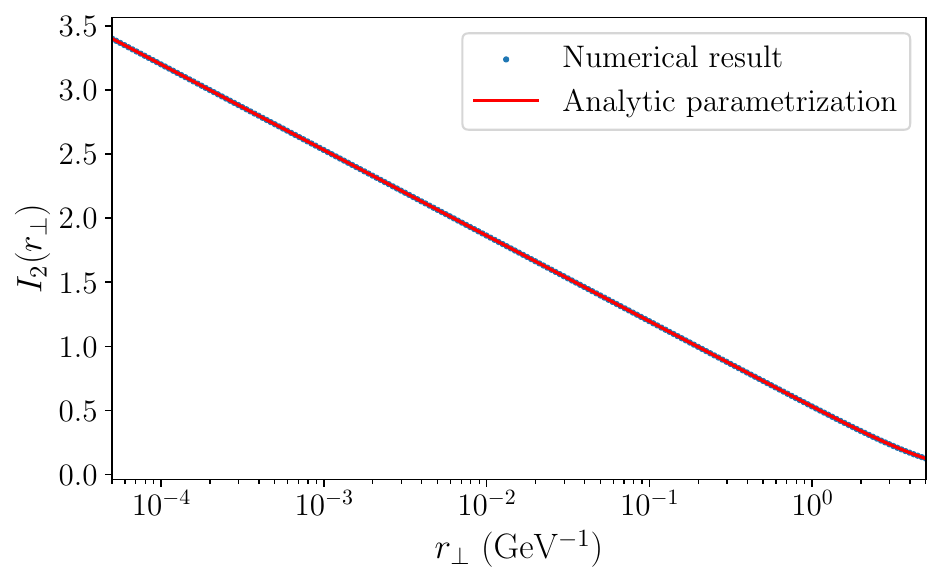}
        \caption{Full range}
        \label{fig:I2_res_model}
    \end{subfigure}
    \;
    \begin{subfigure}[t]{0.49\textwidth}
        \centering
        \includegraphics[width = \textwidth]{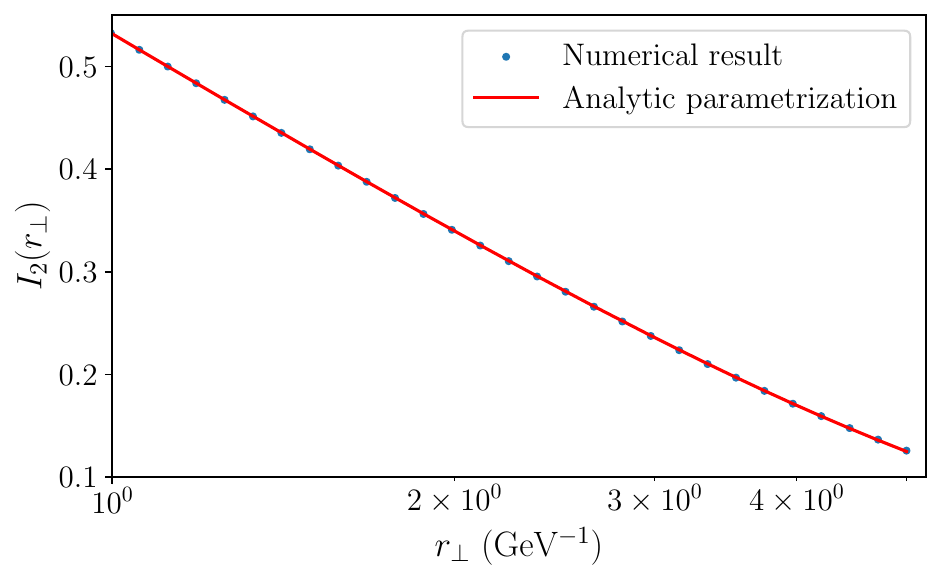}
        \caption{Zoomed in for 1 GeV$^{-1} \leq r_{\perp}\leq$ 5 GeV$^{-1}$}
        \label{fig:I2_res_model_zoomed}
    \end{subfigure}
    \caption{The integral $I_2(r_{\perp})$, obtained by numerically evaluating Eq.~\eqref{IC10} (blue dots), as a function of dipole size $r_\perp$. The full reuslt is compared to the analytical ansatz~\eqref{I2_ansatz} with parameters given in Eq.~\eqref{I2_params} (red lines).}
    \label{fig:I2_fit}
\end{figure}

Next, we repeat the process for $I_2(r_{\perp})$, starting from Eq.~\eqref{IC10}. The steps are similar, albeit with a slightly more complicated integrand. The results of the numerical integration are shown as blue dots in Fig.~\ref{fig:I2_res_model}. Again, we see a very similar logarithmic behavior that breaks down as $r_{\perp}$ increases and approaches the strong interaction scale, $\Lambda_{\text{QCD}}$. This leads to the similar ansatz,
\begin{align}\label{I2_ansatz}
    &I_2(r_{\perp}) = a_2\ln\left(\frac{1}{r^2_{\perp}\Lambda^2_2} + b_2\right) .
\end{align}
We again fit coefficients $a_2, \Lambda_2$ and $\beta_2$ to the numerically obtained results and find 
\begin{align}\label{I2_params}
    a_2 = 0.145\;,\;\;\;\;\Lambda_2 = 0.161~\text{GeV}\;\;\;\;\text{and}\;\;\;\;b_2 = 0.832,
\end{align}
with the uncertainties dominated by the available accuracies of $M$ and $\beta$. Similarly to the $I_1$ case, these parameters provide an excellent fit to the numerical integral, as shown in Figs.~\ref{fig:I2_fit}. Quantitatively, we see that $\Lambda_2$ is smaller than $\Lambda_1$ but remains close to the $\Lambda_{\text{QCD}}$ scale. Remarkably, to our accuracy (and a couple more digits), we see that $a_2 = \frac{a_1}{2}$. We believe that this relation is likely analytic, at least in the logarithmic regime with a not-too-large dipole. Similar to the $I_1$ case, we put $a_2=\overline{x^{-1}}/8\pi$. Eq.~\eqref{I2_params}, together with Eq.~\eqref{I1_params}, are the main results of this Appendix. They provide the central piece for the analytic parametrizations~\eqref{IC_summary_final} of all polarized dipole amplitudes relevant to small-$x$ helicity evolution.

\section{Terms Independent of the Dipole Size}
\label{sec:r-indep}

In this Section, we discuss the contributions independent of the dipole size, $r_{\perp}$. As discussed in the main text, these terms have possible issues that stem from their lack of external perturbative scale inherent to the setup. This Appendix discusses their results assuming that the perturbative calculation still holds. 

Similar to the $r_{\perp}$-dependent term, we start with the quark-exchange contributions. Again, the leading terms are at order $\mathcal{O}(\as^2)$ and come from two origins: (i) an eikonal gluon connecting the dipole with the target, and (ii) a gluon emitted and absorbed within the target. Note that gluons emitted and absorbed within the dipole should be included as parts of the small-$x$ evolution. 

For (i), as discussed in Appendix~\ref{sec:O-g4-gluon-Xchange}, the $r_{\perp}$-independent terms result from expanding the Wilson lines at $\yy$, including $V_{\yy}[-\infty,x_1^-]$, $U_{\yy}[x_1^-,x_2^-]$ and $V_{\yy}[x_2^-,\infty]$, to the linear order, $\mathcal{O}(g)$. Starting from Eqs.~\eqref{qk20_Vyminus}--\eqref{qk20_Vyplus}, we employ the Poisson equation~\eqref{qk21} on the gluon field, then subsequently write the color density in terms of the quark fields, c.f. Eq.~\eqref{qk23}, and at the end express all the quark fields in terms of quark creation and annihilation operators via Eq.~\eqref{qkfield}. Putting all the three terms together, we have that
\begin{align}\label{r1}
    &\frac{zs}{2N_c} \,\text{Re}\left\langle\text{T\,tr}\left[V_{\xx}V_{\yy}^{\text{q}[1]\dagger}\right]\right\rangle\Big|_{\text{dipole}} \\
    &= \frac{zs}{2N_c} \left\langle\text{T\,tr}\left[V_{\xx}V_{\yy}^{\text{q}[1]\dagger}\right]\right\rangle\Big|_{V_{\yy}[-\infty,x_1^-]} + \frac{zs}{2N_c} \left\langle\text{T\,tr}\left[V_{\xx}V_{\yy}^{\text{q}[1]\dagger}\right]\right\rangle\Big|_{U_{\yy}[x_1^-,x_2^-]} + \frac{zs}{2N_c} \left\langle\text{T\,tr}\left[V_{\xx}V_{\yy}^{\text{q}[1]\dagger}\right]\right\rangle\Big|_{V_{\yy}[x_2^-,\infty]} \notag \\
    &= \frac{2i \pi^2 \alpha_s^2 P^+}{N_c^2} \, (t^a)_{ji} \, (t^a)_{\ell m} \int\frac{\dd^2\pp}{(2\pi)^2}\,\frac{1}{p^2_{\perp}} \int d^2\xx'     \notag \\
    &\;\;\;\;\times \int\frac{\dd^2 \pp_1 \, \dd p_1^+}{(2\pi)^3\sqrt{2p_1^+}} \int\frac{\dd^2 \pp_2 \, \dd p_2^+}{(2\pi)^3\sqrt{2p_2^+}} \int\frac{\dd^2 \pp_3 \, \dd p_3^+}{(2\pi)^3\sqrt{2p_3^+}} \int\frac{\dd^2 \pp_4 \, \dd p_4^+}{(2\pi)^3\sqrt{2p_4^+}} \, e^{i(\pp + \pp_1-\pp_2)\cdot\yy - i(\pp-\pp_3+\pp_4)\cdot\xx'}   \notag \\
    &\;\;\;\;\times \left\{ \, \int\limits_{-\infty}^{\infty}\dd x_1^-\int\limits_{x_1^-}^{\infty}\dd x_2^- \int\limits_{-\infty}^{x_1^-}\dd x_3^- \, e^{-ip_1^+x_1^- + ip_2^+x_2^- - i(p_3^+ - p_4^+)\,x_3^-} \sum_{f'} \sum_{S,S'}S \left\langle \hat{b}^{f'\dagger}_{p_4,\ell,S'} \hat{b}^{f'}_{p_3,m,S'} \hat{b}^f_{p_1,i,S} \hat{b}^{f\dagger}_{p_2,j,S}  \right\rangle      \right.  \notag \\
    &\;\;\;\;\;\;\;\;+ N_c^2 \int\limits_{-\infty}^{\infty}\dd x_1^-\int\limits_{x_1^-}^{\infty}\dd x_2^-\int\limits_{x_1^-}^{x_2^-}\dd x_3^- \, e^{-ip_1^+x_1^- + ip_2^+x_2^- - i(p_3^+ - p_4^+)\,x_3^-} \sum_{f'} \sum_{S,S'}S \left\langle \hat{b}^f_{p_1,i,S} \hat{b}^{f'\dagger}_{p_4,\ell,S'} \hat{b}^{f'}_{p_3,m,S'}  \hat{b}^{f\dagger}_{p_2,j,S} \right\rangle   \notag \\
    &\;\;\;\;\;\;\;\;+ \left. \int\limits_{-\infty}^{\infty}\dd x_1^-\int\limits_{x_1^-}^{\infty}\dd x_2^- \int\limits_{x_2^-}^{\infty}\dd x_3^- \, e^{-ip_1^+x_1^- + ip_2^+x_2^- - i(p_3^+ - p_4^+)\,x_3^-} \sum_{f'} \sum_{S,S'}S \left\langle \hat{b}^f_{p_1,i,S} \hat{b}^{f\dagger}_{p_2,j,S} \hat{b}^{f'\dagger}_{p_4,\ell,S'} \hat{b}^{f'}_{p_3,m,S'} \right\rangle   \right\} . \notag
\end{align}
Then, with the proton state given in Section~\ref{sec:proton_model}, we write down the correlators of the creation and annihilation operators  in Eq.~\eqref{r1}. We see that all the contributions come from the terms with the eikonal gluon acting on a different quark than the one exchanging quarks with the dipole, similar to Fig.~\ref{fig:gqq_12}. Altogether, we obtain
\begin{align}\label{r2}
    &\frac{zs}{2N_c} \,\text{Re}\left\langle\text{T\,tr}\left[V_{\xx}V_{\yy}^{\text{q}[1]\dagger}\right]\right\rangle\Big|_{\text{dipole}} = -  \lim_{K\to P} \frac{8\pi^2 \alpha_s^2}{81} \left[4\delta^{f,u}-\delta^{f,d}\right] \frac{4\pi P^+\delta(P^+-K^+) }{\braket{K|P}} \;  e^{i(\underline{P}-\underline{K})\cdot\yy}  \int\frac{\dd^2\pp}{(2\pi)^2} \, \frac{1}{p^2_{\perp}}      \\  
    &\times \int[\dd x_i] \int[\dd^2\qq_i] \, \frac{1}{x_1} \, \Phi^*(x_1,\qq_1 - (1-x_1)\underline{P} + (1 - x_1)\underline{K} + \pp;x_2,\qq_2 + x_2\underline{P} - x_2\underline{K} - \pp;x_3,\qq_3 + x_3\underline{P} - x_3\underline{K}) \, \Phi(x_i,\qq_i) \, .  \notag
\end{align}
Similarly, starting from Eq.~\eqref{qk29}, we obtain
\begin{align}\label{r3}
    &\frac{zs}{2N_c} \sum_f \text{Re}\left\langle\text{T\,tr}\left[V_{\xx}W_{\yy}^{\text{q}[1]\dagger}\right]\right\rangle\Big|_{\text{dipole}} =   \lim_{K\to P} \frac{4\pi^2\alpha_s^2}{9}  \frac{4\pi P^+\delta(P^+-K^+) }{\braket{K|P}} \;  e^{i(\underline{P}-\underline{K})\cdot\yy}  \int\frac{\dd^2\pp}{(2\pi)^2} \, \frac{1}{p^2_{\perp}}      \\  
    &\times \int[\dd x_i] \int[\dd^2\qq_i] \, \frac{1}{x_1} \, \Phi^*(x_1,\qq_1 - (1-x_1)\underline{P} + (1 - x_1)\underline{K} + \pp;x_2,\qq_2 + x_2\underline{P} - x_2\underline{K} - \pp;x_3,\qq_3 + x_3\underline{P} - x_3\underline{K}) \, \Phi(x_i,\qq_i) \, .  \notag
\end{align}

\begin{figure}
    \centering
    \includegraphics[width = 0.35\textwidth]{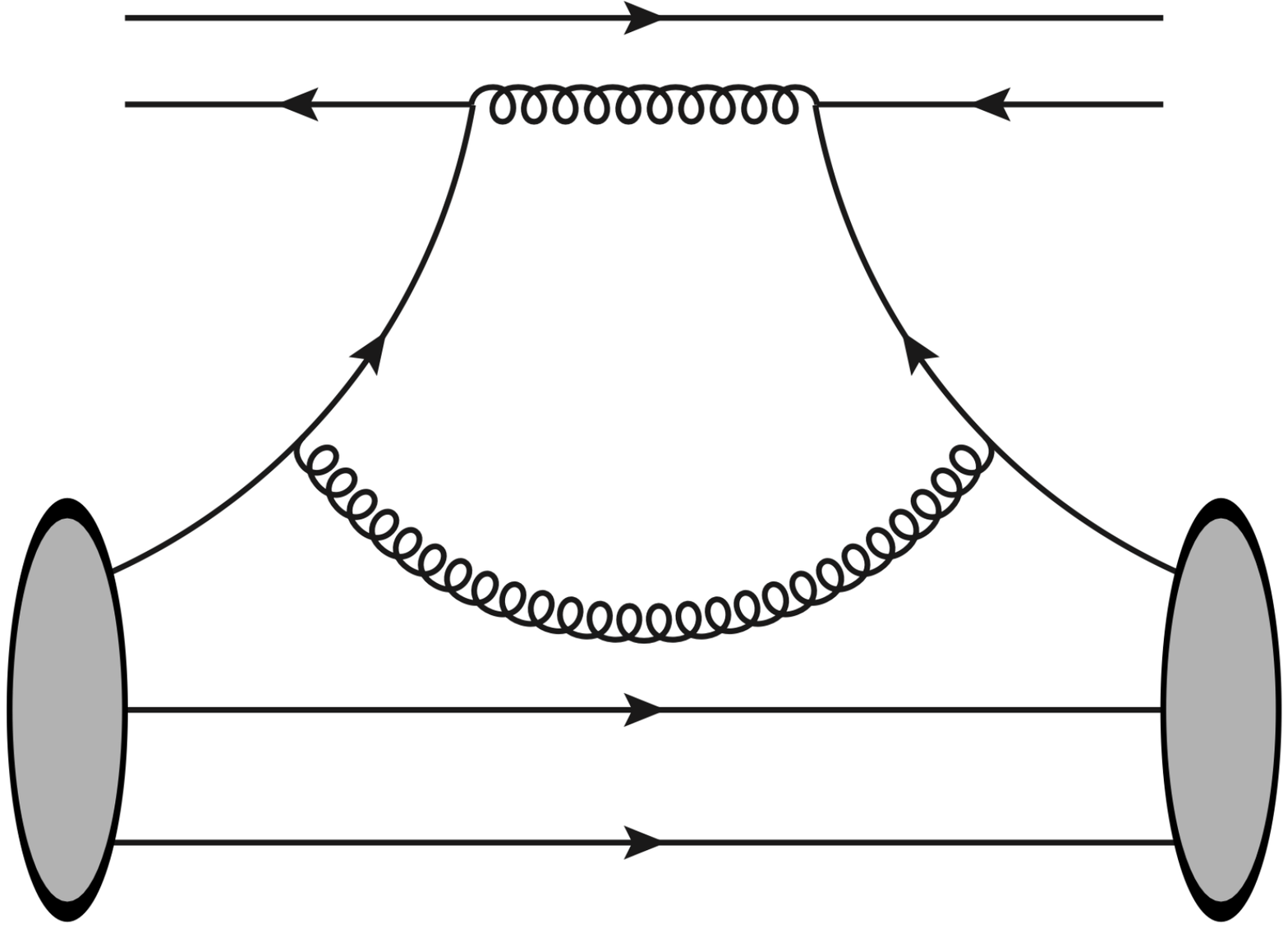}
    \caption{The only diagram with gluon emission and absorption inside the proton that contributes to the quark-exchange term of the polarized dipole amplitude}
    \label{fig:1a}
\end{figure}

Another contribution comes from an emission and an absorption of a gluon inside the proton. This contribution vanishes for the $r_{\perp}$-dependent term, as the only interactions with the dipole are the two exchanged quarks that take place at $\yy$ as parts of the polarized Wilson line, $V_{\yy}^{\text{q}[1]\dagger}$. The calculation of this term amounts to including $\mathcal{O}(g^2)$ corrections to the target states, as detailed in Ref.~\cite{Dumitru:2020gla}. Then, the resulting linear combination of $\ket{qqq}$ and $\ket{qqqg}$ states is employed to calculate the correlator in Eq.~\eqref{qk3}. As it turns out, most of the terms, including all of those with virtual gluon emissions, become proportional to the delta function of $x_i$ or $x'_i$, which are the longitudinal momentum fractions of one of the incoming or outgoing valence quarks. Then, the proton wave function dictates that most terms vanish~\cite{Schlumpf:1992vq,Brodsky:1994fz}. The only remaining term corresponds to the diagram in Fig.~\ref{fig:1a}, with the lower gluon contained within the proton target. To calculate this term, we first write down the valence quark state with $\mathcal{O}(g)$ correction term,
\begin{align}\label{r4}
    \prod_{n=1}^3 \ket{q(p_n,i_n,\sigma_n,f_n)} &= \prod_{n=1}^3 \ket{q(p_n,i_n,\sigma_n,f_n)}_0 - \sum_{n=1}^3 \, \sum_{\sigma,\lambda,j,a} 2g(t^a)_{ji_n} \int\frac{\dd k_g^+\,\dd^2 \kk_g}{2k_g^+(2\pi)^3} \, \frac{1}{2(p_n^+-k_g^+)} \, \hat{\psi}_{q\to qg}(p_n; p_n-k_g,k_g) \notag  \\
    &\;\;\;\;\;\times \ket{q(p_n-k_g,j,\sigma,f_n)}_0 \otimes \ket{g(k_g,a,\lambda)}_0 \otimes \prod_{m\neq n}\ket{q(p_m,i_m,\sigma_m,f_m)}_0 + \mathcal{O}(g^2) \, ,
\end{align}
where the $q\to qg$ splitting function is given by
\begin{align}\label{r5}
    &\hat{\psi}_{q\to qg}(p;k_q,k_g) = 2p^+\sqrt{1-z}\left[\left(1-\frac{z}{2}\right) \delta^{\ell m} + \frac{z}{2}\,i\sigma\epsilon^{\ell m} \right] \delta_{\sigma \sigma'} \, \frac{(\kk_g-z\pp)^{\ell}\ee_{\lambda}^{*m}}{|\kk_g-z\pp|^2} \, ,
\end{align}
with $z=\frac{k_g^+}{p^+}$ and $\sigma$($\sigma'$) being the helicity of the outgoing(incoming) quark. In the diagram in Fig.~\ref{fig:1a}, it is the second term of Eq.~\eqref{r4} that sandwiches the quark creation and annihilation operators in Eq.~\eqref{qk3}. Calculating the correlator and performing the integrals, we obtain
\begin{align}\label{r6}
    &\frac{zs}{2N_c} \,\text{Re}\left\langle\text{T\,tr}\left[V_{\xx}V_{\yy}^{\text{q}[1]\dagger}\right]\right\rangle\Big|_{\text{target}} = - \frac{64\pi^2\as^2}{81} \left[4\delta^{f,u} - \delta^{f,d} \right]\lim_{K\to P}  \frac{4\pi P^+\delta(P^+-K^+)}{\braket{K|P}} \, e^{i(\underline{P}-\underline{K})\cdot\yy} \int [\dd x_i] \int [\dd^2 \qq_i]    \\
    &\;\;\;\;\times \int\frac{dk_g^+\,d^2\kk_g}{2k_g^+(2\pi)^3} \; 2\pi\delta\left(x_1-\frac{k_g^+}{P^+}\right)  \frac{\left[\kk_g-\qq_1-x_1\underline{P}\right]\cdot\left[\kk_g-\qq_1+(1-x_1)\underline{P}-\underline{K}\right]}{\left|\kk_g-\qq_1-x_1\underline{P}\right|^2\left|\kk_g-\qq_1+(1-x_1)\underline{P}-\underline{K}\right|^2}     \notag \\
    &\;\;\;\;\times \Phi(x_i,\qq_i)\,\Phi^*(x_1,\qq_1-(1-x_1)\underline{P}+(1-x_1)\underline{K};x_2,\qq_2+x_2\underline{P}-x_2\underline{K};x_3,\qq_3+x_3\underline{P}-x_3\underline{K}) \, .  \notag
\end{align}
Here, physically, $k_g$ is the momentum of the gluon within the proton. Then, we integrate over $k_g^+$ and re-label $\kk_g\mapsto\pp+\qq_1+x_1\underline{P}$ to get
\begin{align}\label{r7}
    &\frac{zs}{2N_c} \,\text{Re}\left\langle\text{T\,tr}\left[V_{\xx}V_{\yy}^{\text{q}[1]\dagger}\right]\right\rangle\Big|_{\text{target}} = - \frac{32\pi^2\as^2}{81} \left[4\delta^{f,u} - \delta^{f,d} \right] \lim_{K\to P} \frac{4\pi P^+\delta(P^+-K^+)}{\braket{K|P}}  e^{i(\underline{P}-\underline{K})\cdot\yy} \int [\dd x_i] \int [\dd^2 \qq_i]  \frac{1}{x_1}    \\
    &\times \int\frac{\dd^2\pp}{(2\pi)^2} \, \frac{\pp\cdot(\pp+\underline{P}-\underline{K})}{p_{\perp}^2\left|\pp+\underline{P}-\underline{K}\right|^2}     \, \Phi(x_i,\qq_i)\,\Phi^*(x_1,\qq_1-(1-x_1)\underline{P}+(1-x_1)\underline{K};x_2,\qq_2+x_2\underline{P}-x_2\underline{K};x_3,\qq_3+x_3\underline{P}-x_3\underline{K}) \, .  \notag
\end{align}
Furthermore, by Eq.~\eqref{qk7}, we deduce from Eq.~\eqref{r7} that
\begin{align}\label{r8}
    &\frac{zs}{2N_c} \sum_f \text{Re}\left\langle\text{T\,tr}\left[V_{\xx}W_{\yy}^{\text{q}[1]\dagger}\right]\right\rangle\Big|_{\text{target}} = - \frac{2\pi^2\as^2}{9} \lim_{K\to P}  \frac{4\pi P^+\delta(P^+-K^+)}{\braket{K|P}} \, e^{i(\underline{P}-\underline{K})\cdot\yy} \int [\dd x_i] \int [\dd^2 \qq_i] \, \frac{1}{x_1}    \\
    &\times \int\frac{\dd^2\pp}{(2\pi)^2} \, \frac{\pp\cdot(\pp+\underline{P}-\underline{K})}{p_{\perp}^2\left|\pp+\underline{P}-\underline{K}\right|^2}     \, \Phi(x_i,\qq_i)\,\Phi^*(x_1,\qq_1-(1-x_1)\underline{P}+(1-x_1)\underline{K};x_2,\qq_2+x_2\underline{P}-x_2\underline{K};x_3,\qq_3+x_3\underline{P}-x_3\underline{K}) \, .  \notag
\end{align}

Then, adding together the ``dipole'' and ``target'' terms, we obtain the $r_{\perp}$-independent quark-exchange terms for each polarized dipole amplitude of type 1. Similar to the previous cases, the term with polarized quark contains no quark-exchange contribution because the proton in our valence quark model contains no antiquark. For $Q_f^{\text{q}}$ and $Q_f^{\text{NS},\text{q}}$, we add Eqs.~\eqref{r2} to \eqref{r7} together and obtain
\begin{align}\label{r9}
    &Q_f^{\text{q}}(r_{\perp},zs)\big|_{r_{\perp}-\text{indep}} = Q_f^{\text{NS},\text{q}}(r_{\perp},zs)\big|_{r_{\perp}-\text{indep}} = -  \frac{8\pi^2\alpha_s^2}{81} \left[4\delta^{f,u}-\delta^{f,d}\right] \int\frac{\dd^2 \pp}{(2\pi)^2} \, \frac{1}{p^2_{\perp}}       \\  
    &\;\;\;\;\;\times \int[\dd x_i] \int[\dd^2 \qq_i] \, \frac{1}{x_1} \left[ 4\,|\Phi(x_i,\qq_i)|^2 + \Phi^*(x_1,\qq_1 + \pp;x_2,\qq_2 - \pp;x_3,\qq_3) \, \Phi(x_i,\qq_i) \right] . \notag
\end{align}
As for the adjoint dipole amplitude, $\wg^{\text{q}}(r_{\perp},zs)$, we similarly have
\begin{align}\label{r10}
    &\wg^{\text{q}}(r_{\perp},zs)\big|_{r_{\perp}-\text{indep}} = -  \frac{2\pi^2\alpha_s^2}{9} \int\frac{\dd^2\pp}{(2\pi)^2} \, \frac{1}{p^2_{\perp}}       \\  
    &\;\;\;\;\;\times \int[\dd x_i] \int[\dd^2\qq_i] \, \frac{1}{x_1} \left[ |\Phi(x_i,\qq_i)|^2 - 2\Phi^*(x_1,\qq_1 + \pp;x_2,\qq_2 - \pp;x_3,\qq_3) \, \Phi(x_i,\qq_i) \right] , \notag
\end{align}
using Eqs.~\eqref{r3} and \eqref{r8}. Again, as discussed in Appendix~\ref{sec:matrix_element_details}, the type-2 polarized dipole amplitude, $G_2(r_{\perp},zs)$, contains no quark-exchange contribution.

The gluon-exchange contributions as calculated in Appendix~\ref{sec:O-g4-gluon-Xchange} already include $r_{\perp}$-inpendent terms, which can be read off directly from Eqs.~\eqref{gl7}, \eqref{gl9} and \eqref{gl15}. Together with the results written in Eqs.~\eqref{r9} and \eqref{r10}, we can write the complete $r_{\perp}$-independent term for all polarized dipole amplitudes as 
\begin{subequations}\label{r_summary}
\begin{align}
    Q_f(r_{\perp},zs)\big|_{r_{\perp}-\text{indep}} &=   \frac{16\pi^2\alpha_s^2}{81} \left[ \delta^{f,u} + 11\delta^{f,d} + 9\delta^{f,s}\right] \int\frac{\dd^2\pp}{(2\pi)^2} \, \frac{1}{p^2_{\perp}} \int[\dd x_i] \, \frac{1}{x_1} \, f_1(x_i)    \label{Q_r}   \\  
    &\;\;\;\;\;- \frac{8\pi^2\alpha_s^2}{81} \left[22\delta^{f,u}+17\delta^{f,d} + 18\delta^{f,s}\right] \int\frac{\dd^2\pp}{(2\pi)^2} \, \frac{1}{p^2_{\perp}} \int[\dd x_i] \, \frac{1}{x_1} \, f_2(x_i;\pp) \, , \notag \\
    Q^{\text{NS}}_f(r_{\perp},zs)\big|_{r_{\perp}-\text{indep}} &= - \frac{8\pi^2\alpha_s^2}{81} \left[4\delta^{f,u}-\delta^{f,d}\right] \int\frac{\dd^2\pp}{(2\pi)^2} \, \frac{1}{p^2_{\perp}} \int[\dd x_i]  \, \frac{1}{x_1} \left[4f_1(x_i) + f_2(x_i;\pp)\right] ,   \label{QNS_r}   \\  
    \wg(r_{\perp},zs)\big|_{r_{\perp}-\text{indep}} &=   \frac{2\pi^2\alpha_s^2}{9} \int\frac{\dd^2\pp}{(2\pi)^2} \, \frac{1}{p^2_{\perp}}\int[\dd x_i]  \, \frac{1}{x_1} \left[23f_1(x_i) - 22f_2(x_i;\pp) \right]  , \label{GT_r}     \\  
    G_2(r_{\perp},zs)\big|_{r_{\perp}-\text{indep}} &= - \frac{16\pi^2\as^2}{9}  \int\frac{\dd^2\pp}{(2\pi)^2} \, \frac{i(\pp\cdot\rr)}{p^4_{\perp}r^2_{\perp}}  \int [\dd x_i] \, \frac{1}{x_1} \left[f_1(x_i) - f_2(x_i;\pp)\right] , \label{G2_r} 
\end{align}
\end{subequations}
where $f_1(x_i)$ and $f_2(x_i;\pp)$ are defined as 
\begin{subequations}\label{r11}
\begin{align}
    f_1(x_i) &= \int [\dd^2\qq_i]  \,  |\Phi(x_i,\qq_i)|^2 \, , \label{r11_f1} \\
    f_2(x_i;\pp) &= \int [\dd^2\qq_i]  \, \Phi^*(x_1,\qq_1 + \pp;x_2,\qq_2 - \pp;x_3,\qq_3) \, \Phi(x_i,\qq_i) \, . \label{r11_f2} 
\end{align}
\end{subequations}
The next step is to evaluate the remaining integrals. First, we recall that $f_2(x_i;\pp)$ is independent of the azimuthal direction of $\pp$, c.f. Eq.~\eqref{IC3}. An important consequence is that the integrand of Eq.~\eqref{G2_r} is odd under $\pp\to -\pp$. Thus, Eq.~\eqref{G2_r} vanishes, and there is no $r_{\perp}$-independent contribution to $G_2(r_{\perp},zs)$. 

As for the other dipole amplitudes, we need to evaluate two types of integrals -- one involving $f_1(x_i)$ and another involving $f_2(x_i;\pp)$. To compute these integrals, we begin by writing down the analytic expressions for $f_1(x_i)$ and $f_2(x_i;\pp)$. The former is a straightforward generalization of Eq.~\eqref{phi_norm}, while the latter subsequently follows via Eq.~\eqref{IC3}. Explicitly, this gives
\begin{subequations}\label{r12}
\begin{align}
    f_1(x_i) &= \frac{\mathcal{N}^2 \beta^4}{16\pi^2}\, x_1x_2x_3\,\exp\left[- \frac{M^2}{\beta^2}\sum_{i=1}^3 \frac{1}{x_i}\right] ,  \label{r12_f1} \\
    f_2(x_i;\pp) &= \frac{\mathcal{N}^2 \beta^4}{16\pi^2}\,x_1x_2x_3\,\exp\left[- \frac{M^2}{\beta^2}\sum_{i=1}^3 \frac{1}{x_i}\right] \exp\left[- \frac{p^2_{\perp}}{4\beta^2}\left(\frac{1}{x_1}+\frac{1}{x_2}\right)\right] . \label{r12_f2}  
\end{align}
\end{subequations}
Then, the relevant integral involving $f_1(x_i)$ is relatively straightforward because the logarithmic integral over $\pp$ separates. However, the integral is both ultraviolet and infrared divergence, requiring the cutoffs on both ends. Explicitly, we have
\begin{align}\label{r13}
    \int\frac{\dd^2\pp}{(2\pi)^2} \, \frac{1}{p^2_{\perp}}\int[\dd x_i]  \, \frac{1}{x_1} \, f(x_i) &= \frac{\mathcal{N}^2 \beta^4}{(4\pi)^5} \, \ln\left(\frac{\Lambda^2_{\text{UV}}}{\Lambda^2_{\text{IR}}}\right) \int\limits_0^1\dd x_1 \int\limits_0^{1-x_1} \dd x_2 \, x_2(1-x_1-x_2) \\
    &\;\;\;\;\;\times \exp\left\{- \frac{M^2}{\beta^2}\left[\frac{1}{x_1}+\frac{1}{x_2}+\frac{1}{(1-x_1-x_2)}\right]\right\} . \notag
\end{align}
The integrals over $x_1$ and $x_2$ can now be evaluated numerically using \texttt{scipy.integrate.dblquad}~\cite{2020SciPy-NMeth}. This gives
\begin{align}\label{r14}
    \int\frac{\dd^2\pp}{(2\pi)^2} \, \frac{1}{p^2_{\perp}}\int[\dd x_i]  \, \frac{1}{x_1} \, f(x_i) &= \frac{\overline{x^{-1}}}{4\pi} \, \ln\left(\frac{\Lambda^2_{\text{UV}}}{\Lambda^2_{\text{IR}}}\right) ,  
\end{align}
where $\overline{x^{-1}} = 3.64$ is the same parameter that appeared in the $r_{\perp}$-dependent results in Section~\ref{sec:main_results}. For the integral involving $f_2(x_i;\pp)$, we similar start by evaluating the integral over $\pp$ analytically, obtaining
\begin{align}\label{r15}
    \int\frac{\dd^2\pp}{(2\pi)^2} \, \frac{1}{p^2_{\perp}}\int[\dd x_i]  \, \frac{1}{x_1} \, f_2(x_i;\pp) &= \frac{\mathcal{N}^2 \beta^4}{(4\pi)^5} \int\limits_0^1\dd x_1 \int\limits_0^{1-x_1} \dd x_2 \, x_2(1-x_1-x_2) \,   \Gamma\left(0,\frac{\Lambda^2_{\text{IR}}}{4\beta^2}\left(\frac{1}{x_1}+\frac{1}{x_2}\right)\right)  \\
    &\;\;\;\;\;\times \exp\left\{- \frac{M^2}{\beta^2}\left[\frac{1}{x_1}+\frac{1}{x_2}+\frac{1}{(1-x_1-x_2)}\right]\right\} .   \notag
\end{align}
Notice that we only require an infrared cutoff for this integral. In fact, if we take $\Lambda_{\text{IR}}$ to be much smaller than $\beta$, then the incomplete gamma function can be written as a power series such that
\begin{align}\label{r16}
    &\Gamma\left(0,\frac{\Lambda^2_{\text{IR}}}{4\beta^2}\left(\frac{1}{x_1}+\frac{1}{x_2}\right)\right) = - \gamma_E + \ln\left(\frac{\beta^2}{\Lambda^2_{\text{IR}}}\right) - \ln\left[\frac{1}{4}\left(\frac{1}{x_1}+\frac{1}{x_2}\right)\right] + \mathcal{O}\left(\frac{\Lambda^2_{\text{IR}}}{\beta^2}\right) ,
\end{align}
where $\gamma_E$ is the Euler–Mascheroni constant. Then, Eq.~\eqref{r15} becomes
\begin{align}\label{r17}
    \int\frac{\dd^2\pp}{(2\pi)^2} \, \frac{1}{p^2_{\perp}}\int[\dd x_i]  \, \frac{1}{x_1} \, f_2(x_i;\pp) &= \frac{\overline{x^{-1}}}{4\pi} \left[\ln\left(\frac{\beta^2}{\Lambda^2_{\text{IR}}}\right)-\gamma_E\right] - K \simeq  \frac{\overline{x^{-1}}}{4\pi} \, \ln\left(\frac{\beta^2}{\Lambda^2_{\text{IR}}}\right) ,        
\end{align}
where $K=0.186$. Here, we again employed \texttt{scipy.integrate.dblquad}~\cite{2020SciPy-NMeth} to evaluate the integrals over $x_1$ and $x_2$ numerically. In the final step, we simply kept the dominant logarithmic term.

Finally, plugging Eqs.~\eqref{r14} and \eqref{r17} into Eqs.~\eqref{r_summary}, we obtain
\begin{subequations}\label{r_final}
\begin{align}
    Q_f(r_{\perp},zs)\big|_{r_{\perp}-\text{indep}} &=   \frac{4\pi\alpha_s^2}{81} \left[ \delta^{f,u} + 11\delta^{f,d} + 9\delta^{f,s}\right] \overline{x^{-1}} \, \ln\left(\frac{\Lambda^2_{\text{UV}}}{\Lambda^2_{\text{IR}}}\right)   \label{Q_r_final}   \\  
    &\;\;\;\;\;- \frac{2\pi\alpha_s^2}{81} \left[22\delta^{f,u}+17\delta^{f,d} + 18\delta^{f,s}\right] \overline{x^{-1}} \, \ln\left(\frac{\beta^2}{\Lambda^2_{\text{IR}}}\right) , \notag    \\  
    Q^{\text{NS}}_f(r_{\perp},zs)\big|_{r_{\perp}-\text{indep}} &= - \frac{2\pi\alpha_s^2}{81} \left[4\delta^{f,u}-\delta^{f,d}\right] \overline{x^{-1}} \left[4\ln\left(\frac{\Lambda^2_{\text{UV}}}{\Lambda^2_{\text{IR}}}\right) + \ln\left(\frac{\beta^2}{\Lambda^2_{\text{IR}}}\right)\right] ,   \label{QNS_r_final}   \\  
    \wg(r_{\perp},zs)\big|_{r_{\perp}-\text{indep}} &=   \frac{\pi\alpha_s^2}{18} \, \overline{x^{-1}} \left[23\ln\left(\frac{\Lambda^2_{\text{UV}}}{\Lambda^2_{\text{IR}}}\right) - 22\ln\left(\frac{\beta^2}{\Lambda^2_{\text{IR}}}\right) \right]  , \label{GT_r_final}     
\end{align}
\end{subequations}
where we recall that $G_2(r_{\perp},zs)$ has no $r_{\perp}$-independent term. Eq.~\eqref{r_final} is the main result of this Appendix. 

The $r_{\perp}$-independent terms of all type-1 polarized dipole amplitudes contain both ultraviolet and infrared divergences. In small-$x$ dipole calculations, it is a common practice to employ the center-of-mass energy as the ultraviolet cutoff for the transverse momentum transfer, so that $p^2_{\perp}\ll zs$~\cite{Kovchegov:2015pbl,Kovchegov:2018znm,Cougoulic:2022gbk}.  Had we imposed such constraint on the momentum transfer, $p_{\perp}$, in our problem, the ultraviolet logarithms in Eqs.~\eqref{r_final} would have become $\ln(zs/\Lambda^2_{\text{IR}})$, which is the common logitudinal logarithms considered in~\cite{Kovchegov:2020hgb,Adamiak:2021ppq,Cougoulic:2022gbk,Borden:2023ugd,Adamiak:2023okq,Adamiak:2023yhz}.

On a more general note, since $\beta$ is taken to be 0.55 GeV throughout this work, it remains relatively close to $\Lambda_{\text{IR}}$. As a result, it might be justified in higher-energy settings to simply discard the second term in each of the square brackets in Eqs.~\eqref{r_final} as being negligible compared to the other logarithmic term. Furthermore, similar to the $r_{\perp}$-dependent case, constant terms could also be added to account for possible shifts in the cutoffs. Overall, the calculation performed in this Appendix fixes the coefficient of $\ln(\Lambda_{\text{UV}}^2/\Lambda_{\text{IR}}^2)$, which plays an important role in the moderate-$x$ initial condition for small-$x$ helicity evolution~\cite{Kovchegov:2015pbl,Kovchegov:2018znm,Cougoulic:2022gbk}.

Finally, if one proceeds to put $\Lambda_{\text{UV}}^2\to zs$ and neglect the terms proportional to $\ln(\beta^2/\Lambda^2_{\text{IR}})$, then the resulting coefficients of the ultraviolet logarithms can be compared to the Born-level calculation~\cite{Kovchegov:2016zex,Adamiak:2023okq}, c.f. Eqs.~\eqref{Born_IC}, and the recent JAM global analysis~\cite{Adamiak:2023yhz}. The results are qualitatively similar to those of the transverse logarithmic terms, which are discussed at the end of Section~\ref{sec:main_results}. Namely, the coefficients from Eqs.~\eqref{r_final} are greater in magnitude than the respective Born-level coefficients from Eqs.~\eqref{Born_IC} but smaller in magnitude than the JAM analysis results~\cite{Adamiak:2023yhz}, with the sign differences in a number of coefficients. Overall, as remarked in Section~\ref{sec:main_results}, the definite conclusion about the initial conditions from this work would require another global analysis based on the initial condition calculated in this work. In particular, the coefficients of the ultraviolet logarithm calculated in this Appendix could play a part in the future global analysis.


\bibliographystyle{JHEP}

\providecommand{\href}[2]{#2}\begingroup\raggedright\endgroup

\end{document}